\documentclass[aps,jcp,superscriptaddress,amsmath,amssymb]{revtex4-1}

\usepackage{comment}
\usepackage{tikz}
\usetikzlibrary{trees}
\usetikzlibrary{decorations.pathmorphing}
\usetikzlibrary{decorations.markings}
\usepackage{tikz-cd}
\usetikzlibrary{arrows,arrows.meta}
\usetikzlibrary{decorations.pathreplacing,snakes}
\tikzset{
    photon/.style={decorate, decoration={snake,segment length=1.5mm}, draw=black},
    coulomb/.style={dotted},
    electron/.style={draw=black, postaction={decorate},
        decoration={markings,mark=at position .55 with {\arrow[draw=black]{>}}}}, 
    gluon/.style={decorate, draw=magenta,
        decoration={coil,amplitude=4pt, segment length=5pt}},
    boundelectron/.style={thick, double},
    transverse/.style={dashed},
    marrow/.style={decoration={markings,mark=at position 0.5 with {\arrow{#1}}}, postaction=decorate}    
}
\usepackage{graphicx}% Include figure files
\usepackage{dcolumn}% Align table columns on decimal point
\usepackage{bm}% bold math
\usepackage{rotating}

\newcolumntype{.}{D{.}{.}{8}}

\usepackage[utf8]{inputenc}
\usepackage{physics}
\usepackage{amsmath, amssymb}
\usepackage{tabularx,booktabs,array,dcolumn}
\usepackage{setspace}
\usepackage{vmargin}
\usepackage{xcolor}
\usepackage{graphicx,psfrag,subfigure}

\usepackage{float}
\usepackage[all]{xy}

\usepackage{bm}
\usepackage{mathtools}
\usepackage{physics}
\usepackage[english]{babel}

\newcommand{\bos}[1]{\boldsymbol{#1}}
\newcommand{\mr}[1]{\mathrm{#1}}

\def\Eh{E_\mathrm{h}}

\def\iim{\mr{i}}
\def\eem{\mr{e}}

\def\nopair{\text{np}}
\def\withpair{\text{wp}}

\def\bur{\underline{\boldsymbol{r}}}
\def\bus{\underline{\boldsymbol{s}}}
\def\buA{\underline{\boldsymbol{A}}}

\def\kabs{k} % no four-vectors anymore

\def\dd{\mathrm{d}}

\def\bp{\boldsymbol{p}}
\def\br{\boldsymbol{r}}

\def\bs{\boldsymbol{s}}
\def\bk{\boldsymbol{k}}
\def\balpha{\boldsymbol{\alpha}}
\def\bsigma{\boldsymbol{\sigma}}

\def\epsi{\varepsilon}

\def\balpha{\boldsymbol{\alpha}}

\def\textC{\text{C}}

\def\tT{\text{T}}

\def\inst{\text{I}}
\def\epsi{\varepsilon}
\def\Vcal{{\cal{V}}}
\def\Kcal{{{K}}} 
\def\ahone{|h_1|}
\def\ahtwo{|h_2|}

\def\lnk0{\text{ln}k_0}

\usepackage[unicode]{hyperref}
\usepackage{soul}
\hypersetup{
   unicode=true,          % non-Latin characters in Acrobat??s bookmarks
   plainpages=false,
   colorlinks=true,       % false: boxed links; true: colored links
   linkcolor=black,          % color of internal links
   linkcolor=blue,          % color of internal links
   citecolor=blue,        % color of links to bibliography
   urlcolor=blue           % color of external links
}

\usepackage{url}

\begin{document}

\title{%
QED corrections to the correlated relativistic energy: one-photon processes \\
}

\author{\'Ad\'am Marg\'ocsy}
\author{Edit M\'atyus} 
\email{edit.matyus@ttk.elte.hu}
\affiliation{ELTE, Eötvös Loránd University, Institute of Chemistry, 
Pázmány Péter sétány 1/A, Budapest, H-1117, Hungary}

\date{\today}

\begin{abstract}
\noindent %
This work is a collection of initial calculations and formal considerations within the Salpeter--Sucher exact equal-time relativistic quantum electrodynamics framework.
The calculations are carried out as preparation for the computation of pair, retardation, and radiative corrections to the relativistic energy of correlated two-spin-1/2-fermion systems.  
In this work, particular attention is paid to the retardation and the `one-loop' self-energy corrections, which are known to be among the largest corrections to the correlated relativistic energy.
The theoretical development is supplemented with identifying formal connections to the non-relativistic quantum electrodynamics framework, which is based on a correlated but non-relativistic reference, as well as to the `$1/Z$ approach', which is built on a relativistic but {independent}-particle zeroth order. 
The two complementary directions currently provide the theoretical framework for light atomic-molecular precision spectroscopy and heavy-atom phenomena.
The present theoretical efforts pave the way for relativistic QED corrections to (explicitly) correlated relativistic computations. 
\end{abstract}

\maketitle

%\tableofcontents

%%%%%%%%%%%%%%%%%%%%%%%%%%%%%%%%%%%%%%%%%%%%%%%%%%%%%%%%
\section{Introduction\label{sec:intro}}
\noindent
Precision spectroscopy of light atoms and molecules, \emph{e.g.,} \cite{BeHoHuChSaEiUbJuMe19,SeJaCaMeScMe20,ClJaScAgScMe21,GuBaPRHoCa21,ShBaReHoCa23,ClScAgScMe23}, 
precision mass spectrometry of highly-charged heavy ions \cite{pentatrap20}, 
as well as wet-lab chemistry properties and reactions of compounds of heavy- and superheavy elements, \emph{e.g.,} \cite{St21}, 
require theoretical support that accounts for both special relativity and quantum mechanics. 
Although relativistic quantum electrodynamics (QED), unifying special relativity and quantum mechanics with electromagnetic forces, has been considered a mature field of physics by the mid-1950s, and a toy model for later developments in understanding nuclear forces, 
there are still open challenges in relation to the application of the relativistic QED theory to compute atomic and molecular bound states.

There are currently three distinct directions of research at the frontier of this field,
(a) non-relativistic QED (nrQED) \cite{Eides2001,JeAdBook22,Ye01,KoYe01,Pa06,YePaPa21,AdCaPR22}: an $\alpha$ (and $\ln(\alpha)$) series expansion of the full theory about a correlated non-relativistic reference; 
(b) the `$1/Z$ approach' 
\cite{MoPlSo98,Sh02,lindgrenRelativisticManyBodyTheory2011,YeMa20,ArShYePlSo05,KoMaGlShTu19,MaKoGlTuSh19}
an expansion of the full theory about an independent-particle relativistic reference; 
and 
(c) relativistic quantum chemistry
\cite{GoInDe87,BlJoLiSa89,PaGr90,DyFaBook07,ReWoBook15,Py12,Ku12,SmInNaPiSc23}: a practical combination of Dirac's wave equation with the many-particle correlation techniques of quantum chemistry without first-principles account of relativistic QED in most applications.

In recent work, we have identified the Bethe--Salpeter equation \cite{SaBe51} and its exact equal-time variant pioneered by Salpeter \cite{Sa52} and Sucher \cite{sucherPhD1958} that could serve as a formally correct and possibly useful starting equation for a variety of atomic-molecular applications by providing a correlated, relativistic reference.
First of all, a two-particle relativistic Hamiltonian emerges in the exact equal-time Salpeter--Sucher formalism, which establishes a connection to the relativistic quantum chemistry practice. 
Second, the first nrQED calculations \cite{sucherPhD1958,DoKr74,Zh96} have been carried out by expanding the (Bethe--)Salpeter--Sucher formalism in $\alpha$ (and $\ln\alpha$) series with respect to the non-relativistic reference.
Third, the Bethe--Salpeter (BS) equation is a formally correct relativistic QED equation, so direct (formal or numerical) comparison with the $1/Z$ approach implementing relativistic QED with respect to an independent-particle reference will be highly relevant.

In a series of recent work \cite{JeFeMa21,JeFeMa22,FeJeMa22,FeJeMa22b,JeMa22,FeMa23}, we have demonstrated that the no-pair part of the exact equal-time, Salpeter--Sucher wave equation, can be solved to high precision, comparable to the nrQED standards of precision, for two spin-1/2 fermion systems, and the numerical results are consistent with the available high-precision nrQED correction values. Although the equal-time Salpeter--Sucher equation formally includes `everything' (corresponding to the `effective' interaction kernel in the equation), the `QED' corrections to the no-pair energy have never been computed without resorting to a non-relativistic expansion. 

For these reasons, this work is a collection of elementary calculations in the Bethe--Salpeter--Sucher bound-state QED framework without resorting to any kind of non-relativistic expansion.
The formalism is developed with all quantities written in an abstract operator form, \emph{i.e.,} we do not assume an underlying single-particle (orbital) basis, as it is almost exclusively done in relativistic QED applications, \emph{e.g.,} \cite{MoPlSo98,lindgrenRelativisticManyBodyTheory2011}. 
The calculations serve as preparation for future computation of perturbative quantum electrodynamics corrections to a (high-precision) correlated relativistic reference. During the theoretical work, we also explore possibilities (or identify current limitations in the mathematical formalism) to formally resum (some of) the occurring series. 

The present work focuses on single-photon processes, \emph{i.e.,} the (single, transverse) photon exchange (retardation) and the `one-loop' self-energy contributions to the (effective) interaction kernel, and the corresponding (lowest-order) perturbative energy corrections. These processes are among the most important contributions to the no-pair energy for low-$Z$ systems, as it can be estimated from nrQED \cite{sucherPhD1958}. For medium to high-$Z$ systems, their importance can be assessed based on Ref.~\cite{LiPeSaLa95}. 

The present theoretical efforts have been initiated by a research program targeting the development of a \emph{relativistic} quantum electrodynamics counterpart to the well-established and successful \emph{non-relativistic} quantum electrodynamics framework \cite{sucherPhD1958,araki57,DoKr74,Zh96,KoYe01,Pa06,AdCaPR22} that currently provides the state-of-the-art computational results for precision spectroscopy. 
An \emph{a posteriori} unification of the nrQED and the $1/Z$ approaches have been proposed and successfully used in the past \cite{Dr88unified,YePaPuPa20,YePaPa22}. We have set out with the aim to develop an \emph{a priori}, first principles unified approach, \emph{i.e.,} a correlated relativistic QED computational framework.
Connections to nrQED have been explored from the beginning of this research direction, partly as a numerical test of the computational results \cite{JeFeMa21,JeFeMa22,FeJeMa22,FeJeMa22b,JeMa22,FeMa23}. Formal connections to nrQED, to the $1/Z$ and independent-particle approach, are elaborated throughout this work in order to check and better understand the formal results.

\section{Preliminaries to the Bethe--Salpeter equation: a formally exact wave equation for bound-state QED}
\noindent
In this section, the main steps are outlined to the equal-time formulation
of the Bethe--Salpeter equation~\cite{SaBe51} following  Salpeter \cite{Sa52} and Sucher \cite{sucherPhD1958}.
We also note that Mat{{}t}hews and Salam \cite{MaSa54} have shown that a renormalized Bethe--Salpeter equation can be
formulated using dressed and renormalized fermion and photon propagators and vertices following Dyson \cite{Dy49} (see also Ref.~\cite{JaRo76}).

The Bethe--Salpeter equation for the two-particle propagator {{}$S(x_1,x_2;x'_1,x'_2)$} is
{{}
\begin{align}
 S(x_1,x_2;x'_1,x'_2)=&\tilde{S}(x_1;x'_1)\tilde{S}(x_2;x'_2) \nonumber \\
 &-\iim\int\mathrm{d}^4y_1\mathrm{d}^4y_2\int\mathrm{d}^4y'_1\mathrm{d}^4y'_2
 \tilde{S}(x_1;y_1)\tilde{S}(x_2;y_2)
 K(y_1,y_2;y'_1,y'_2)S(y'_1,y'_2;x'_1,x'_2) \ ,
\end{align}
or with a condensed notation,
}
\begin{align}
  S_{12} = \tilde{S}_1 \tilde{S}_2 - {\iim}\tilde{S}_1 \tilde{S}_2 K S_{12} \ ,
  \label{2prop}
\end{align}
where $\tilde{S}_a=(S^{-1}_a-\Sigma_a)^{-1}$ is the dressed propagator of the spin-$1/2$ particle $a${{}, $\Sigma_a$ is the corresponding irreducible self-energy (containing radiative corrections),} and $K$ is the irreducible interaction kernel of the two particles.
$S_a$ is the Feynman propagator for particle $a$ in the field of the fixed nucleus (nuclei) \cite{Fe49a,Fu51}, 
\begin{align}
  [\iim\partial_{t_a}-{h}_a(\br_a)]S_{a}(x_a;x_a')
  =
  \iim\beta_a\delta(x_a-x_a') 
  \label{eq:greenSa}
\end{align}
with
\begin{align}
  h_a(\br_a)
  =
  \boldsymbol{\alpha}_a\cdot\boldsymbol{p}_a
  +\beta_a m_a
  +z_a e U_a 1_a
  \label{eq:defha}
\end{align}
including the $x=(\br,t)$ space-time coordinate, the $\boldsymbol{p}_a=-\iim\nabla_a$ momentum; $\boldsymbol{\alpha}_{a},\beta_a$ are the Dirac matrices, $1_a$ is the four-dimensional unit matrix, and 
$U_a$ labels the Coulomb potential of the fixed external charges (nuclei) interacting with the active particle of charge $z_ae$. 
Rationalized natural units, {$m=\hbar=c=\varepsilon_0=1$} and $\alpha=e^2/(4\pi)$ are used throughout this work.

%\newline
Using Eq. (\ref{2prop}), it can be shown that the $\Psi(x_1,x_2)$ wave function of a bound state satisfies the homogeneous equation {{}(Eq. (6.16) of Ref. \cite{GrReQEDBook09}, or the ideas explained in Ch. 12. of Ref. \cite{GrossQFTBook99})
\begin{align}
  \Psi(x_1,x_2)
  =
  -\iim 
  \int\mathrm{d}^4y_1\mathrm{d}^4y_2
  \int\mathrm{d}^4y'_1\mathrm{d}^4y'_2
  \tilde{S}_1(x_1;y_1) \tilde{S}_2(x_2;y_2) K(y'_1,y'_2;y''_1,y''_2) \Psi(y''_1,y''_2) \ ,
\end{align}
or}
\begin{align}
  \Psi
  =
  -\iim 
  \tilde{S}_1 \tilde{S}_2 K \Psi \ ,
\end{align}
which can be rearranged to
\begin{align}
  \Psi
  =
  -\iim 
  S_1 S_2 \tilde{K} \Psi \ .
  \label{eq:BSorig}
\end{align}
Here, the $\tilde{K}$ `effective' interaction kernel collects \cite{DoKr74,sucherPhD1958} both the 
$K$ interaction and the self-energy contributions from the dressed fermion propagators, 
\begin{align}
  \tilde{K}
  =
  K
  + S^{-1}_{1}(\iim\Sigma_2)
  + S^{-1}_{2}(\iim\Sigma_1)
  +\iim (\iim\Sigma_1)(\iim\Sigma_2) \ .
  \label{eq:Keff}
\end{align}
Next, we multiply Eq.~\eqref{eq:BSorig} from the left with $[\iim\partial_a-h_a]\ (i=1,2)$ and exploit that $S_a$ is the Green function, Eq.~\eqref{eq:greenSa}, 
\begin{align}
  [\iim\partial_{t_1}-{h}_1]
  [\iim\partial_{t_2}-{h}_2]
  \Psi(x_1,x_2)=
  \iim\beta_1\beta_2
  \int
  \mathrm{d}^4x'_1
  \mathrm{d}^4x'_2
  \tilde{K}(x_1,x_2;x'_1,x'_2)
  \Psi(x'_1,x'_2) \ .
  \label{eq:BSnew}
\end{align}
Bound, stationary states (no time-dependent external fields being present) can be written in the form {{}(Eq. (1.7) of Ref.~\citenum{sucherPhD1958} or Eqs. (2.20), (2.21) of Ref.~\citenum{DoKr74})} 
\begin{equation}
  \Psi(x_1,x_2) 
  =
  \eem^{-\iim E\frac{t_1+t_2}{2}}\psi(\boldsymbol{r}_1,\boldsymbol{r}_2,t_1-t_2) \ ,
\end{equation}
where $E$ is the energy of the state. 
This form motivates the coordinate transformation $t_1,t_2 \rightarrow T=(t_1+t_2)/2, \tau=t_1-t_2$, \emph{e.g.,} in Eq.~\eqref{eq:BSnew}.
Furthermore, it is useful to switch to momentum space to carry out calculations with the $S_a$ Feynman propagators and the $K$ interaction kernel. Then, the $T$ average and $\tau$ relative times are replaced by their conjugate variables, the $E$ total energy and the $\epsi$ relative energy variables. Correspondingly, Eq.~\eqref{eq:BSnew} turns into
\begin{equation}
  S_1^{-1}(\boldsymbol{p}_1,\varepsilon) S_2^{-1}(\boldsymbol{p}_2,-\varepsilon)
  \psi(\boldsymbol{p}_1,\boldsymbol{p}_2,\epsi)
  =
  {\tilde{K}} \psi(\boldsymbol{p}_1,\boldsymbol{p}_2,\epsi) \ ,
  \label{eq:BSalmostfinal}
\end{equation}
where the $S_1(\boldsymbol{p}_1,\varepsilon)$ and $S_2(\boldsymbol{p}_2,-\varepsilon)$ propagators carry Feynman's $\iim 0^+$ prescriptions \cite{Fe49a}
to describe the different (temporal) behaviour of the positive- and negative-energy states: 
\begin{align}
 S_1(\boldsymbol{p}_1,+\varepsilon)
 &=S_1^{(+)}(\boldsymbol{p}_1,+\varepsilon)+S_1^{(-)}(\boldsymbol{p}_1,+\varepsilon) \nonumber \\
 &=
 \frac{{L}_{1,+}}{\frac{E}{2}+\epsi-|h_1(\boldsymbol{p}_1)|+\iim 0^+}
 +
 \frac{{L}_{1,-}}{\frac{E}{2}+\epsi+|h_1(\boldsymbol{p}_1)|-\iim 0^+}
 \label{eq:propagator1} \\
 S_2(\boldsymbol{p}_2,-\varepsilon)
&=S_2^{(+)}(\boldsymbol{p}_1,-\varepsilon)+S_2^{(-)}(\boldsymbol{p}_1,-\varepsilon) \nonumber \\
 &=
 \frac{{L}_{2,+}}{\frac{E}{2}-\epsi-|h_2(\boldsymbol{p}_2)|+\iim 0^+}
 +
 \frac{{L}_{2,-}}{\frac{E}{2}-\epsi+|h_2(\boldsymbol{p}_2)|-\iim 0^+} \; .
 \label{eq:propagator2}
\end{align}
$L_{a,+}$ and $L_{a,-}$ are projectors onto the positive- and negative-energy space of particle $a$ {{}in the external field}, respectively,
and $|h_a(\bp_a)|$ labels the absolute value operator of the one-particle Hamiltonian. 
In this momentum-space representation, the action of the $K$ interaction kernel is understood as
\begin{equation}
  \tilde{K} \psi(\boldsymbol{p}_1,\boldsymbol{p}_2,\epsi)
  =
  \int\mathrm{d}^3 \bp'_1
    \mathrm{d}^3 \bp'_2
    \int_{-\infty}^{+\infty}\frac{\mathrm{d}\epsi'}{-2\pi \iim }
    \tilde{K}(\boldsymbol{p}_1,\boldsymbol{p}_2,\epsi;\boldsymbol{p}'_1,\boldsymbol{p}'_2,\epsi') %
    \psi(\boldsymbol{p}'_1,\boldsymbol{p}'_2,\epsi') \ .
\end{equation}
In general, the interaction and self-interaction contributions can be constructed according to Feynman's rules. For specific bound-state computations of two-particle systems, Sucher formulated the equivalent specific `rules' for constructing tree-level interactions, and we summarize them, with extension also for self-energy rules, in Appendix~\eqref{App:diagram}.

Salpeter \cite{Sa52} and Sucher \cite{sucherPhD1958} proposed to partition the interaction kernel to an instantaneous part and the rest:
\begin{equation}
  \tilde{K} = K_{\inst} + K_{\Delta} \ 
  \label{eq:kernelIDelta}
\end{equation}
This partitioning is useful, since the ${K}_{\inst}$ action is `trivial' in the relative-energy variable,
\begin{align}
 {{K}}_{\inst}
   \psi(\boldsymbol{p}_1,\boldsymbol{p}_2,\epsi)
 &=
 \int\mathrm{d}^3 \bk
 \int_{-\infty}^{+\infty}\frac{\mathrm{d}\omega}{-2\pi \iim}
    \kappa_{\inst}(\boldsymbol{k})T_\inst(\bk)
    \psi(\boldsymbol{p}_1-\boldsymbol{k},\boldsymbol{p}_2+\boldsymbol{k},\epsi-\omega) \nonumber\\
 &=
 \int\mathrm{d}^3 \bk\ 
   \kappa_{\inst}(\boldsymbol{k})T_\inst(\boldsymbol{k})
   \eta_1(\boldsymbol{k})
   \eta_2(-\boldsymbol{k})
 \int_{-\infty}^{+\infty}\frac{\mathrm{d}\omega}{-2\pi \iim}
   {\eta}_{{}\epsilon}(\omega)
   \psi(\boldsymbol{p}_1,\boldsymbol{p}_2,\epsi) \nonumber\\
 &=
 \frac{1}{-2\pi \iim}
 V_{\inst}\Phi(\boldsymbol{p}_1,\boldsymbol{p}_2)
 \ ,
 \label{eq:Kinst}
\end{align}
where the $\eta$ momentum- and energy-shift operators were implicitly defined (further details are in Appendix~\ref{App:eta}), and {{}$\kappa_\inst(\bk)$ and} $T(\boldsymbol{k})$ {{} are the kernel and} the `tensor part' of the interaction{{}, respectively}  \emph{(vide infra,} for the Coulomb or Breit options).
The trivial relative-energy dependence of the instantaneous kernel allows the definition of 
an (energy-independent) potential energy operator as 
\begin{equation} 
  V_{\inst} f(\bp_1,\bp_2)
  =
  \int\mathrm{d}^3\bk\ 
    \kappa_{\inst}(\boldsymbol{k})T_\inst(\boldsymbol{k})
    {\eta}_1(\boldsymbol{k})
  {\eta}_2(-\boldsymbol{k}) 
  f(\bp_1,\bp_2)
  \ ,
  \label{instint}
\end{equation}
and the `equal-time' wave function naturally emerges as
\begin{align}
  \Phi(\boldsymbol{p}_1,\boldsymbol{p}_2)
  &=
  \int_{-\infty}^{+\infty}\mathrm{d}\omega\ 
    {\eta}_{{}\epsilon}(\omega)
    \psi(\boldsymbol{p}_1,\boldsymbol{p}_2,\epsi)  
  =
  \int_{-\infty}^{+\infty}\mathrm{d}\omega\
  \psi(\boldsymbol{p}_1,\boldsymbol{p}_2,\omega)
    \ ,
\end{align}
where the $\epsi$-independence follows from the shift-invariance of the integral.

%eq:kernelIDelta
In atomic and molecular systems, the binding energy is dominated by the instantaneous part of the interaction and this fact motivates the Salpeter--Sucher partitioning, Eq.~\eqref{eq:kernelIDelta}. This partitioning becomes a natural choice in the Coulomb gauge representation of the photon propagator, and the interaction kernel corresponding to a single photon exchange naturally splits into the instantaneous Coulomb and the (retarded) transverse part (Sec.~\ref{sec:transverse} and Appendix~\ref{App:diagram}),
\begin{equation}
 {{K}}_{\textC}\psi(\boldsymbol{p}_1,\boldsymbol{p}_2,\epsi)
 =
 \frac{z_1z_2\alpha}{2\pi^2}
 \int\mathrm{d}^3 \bk\ 
   \frac{1}{{\kabs}^2}
   {\eta}_1(\boldsymbol{k})
   {\eta}_2(-\boldsymbol{k})%}
 \int_{-\infty}^{+\infty}\frac{\mathrm{d}\omega}{-2\pi \iim}
   {\eta}_{{}\epsilon}(\omega)
   \psi(\boldsymbol{p}_1,\boldsymbol{p}_2,\epsi) 
 \ ,
\end{equation}
\begin{equation}
 {{K}}_{\text{T}}\psi(\boldsymbol{p}_1,\boldsymbol{p}_2,\epsi)
 =
 \frac{z_1z_2\alpha}{2\pi^2}
 \int\mathrm{d}^3 \bk
 \int_{-\infty}^{+\infty}\frac{\mathrm{d}\omega}{-2\pi \iim}
   \frac{\tilde{\alpha}_{1,i}(\boldsymbol{k})\tilde{\alpha}_{2,i}(\boldsymbol{k})}{\omega^2-{\kabs}^2+\iim 0^+}
   {\eta}_1(\boldsymbol{k})
   {\eta}_2(-\boldsymbol{k})
   {\eta}_{{}\epsilon}(\omega)
   \psi(\boldsymbol{p}_1,\boldsymbol{p}_2,\epsi) 
 \ ,
 \label{eq:KT}
\end{equation}
where $\tilde{\alpha}_{a,i}(\boldsymbol{k})$ is the transverse component
of $\boldsymbol{\alpha}_a$, perpendicular to $\boldsymbol{k}$, 
\begin{equation}
 \tilde{\alpha}_{a,i}(\boldsymbol{k})
 =
 \delta^{\perp}_{ij}(\boldsymbol{k})\alpha_{a,j}
 =
 \left(\delta_{ij}-\frac{k_ik_j}{{\kabs}^2}\right)\alpha_{a,j} \ .
\end{equation}
Einstein's summation convention is used throughout this work unless we want to emphasize a particular aspect by making the summation explicit. 
For later convenience, the short notation for the Coulomb part, Eq. (\ref{eq:Kinst}), is
\begin{equation}
 \kappa_{\text{C}}(\boldsymbol{k})=+
 \frac{z_1z_2\alpha}{2\pi^2}
 \frac{1}{{\kabs}^2}
 \ \ \ , \ \ \ T_{\text{C}}(\boldsymbol{k})=I_1I_2
 \ ,
 \label{coulombkernel}
\end{equation}
and for the transverse part, Eq.~\eqref{eq:KT}
\begin{align}
  \kappa_\tT(\bk,\omega)
  &=
 \frac{z_1z_2\alpha}{2\pi^2}
  \frac{1}{\omega^2-{\kabs}^2+\iim 0^+} 
  \nonumber \\
  &= 
  \frac{z_1z_2\alpha}{2\pi^2}
  \frac{1}{2{\kabs}-\frac{\iim0^+}{{\kabs}}} 
  \left[%
    \frac{1}{\omega - {\kabs} + \frac{\iim0^+}{2{\kabs}}}
    -
    \frac{1}{\omega + {\kabs} - \frac{\iim0^+}{2{\kabs}}}
  \right]
  \; ,\quad T_{\text{T}}(\boldsymbol{k})=\tilde{\balpha}_1(\bk)\cdot\tilde{\balpha}_2(\bk) \;. 
  \label{eq:kappaT}
\end{align}
The instantaneous, $\omega=0$, approximation to the transverse interaction gives rise to the Breit term with
\begin{equation}
 \kappa_{\text{B}}(\boldsymbol{k})=-
 \frac{z_1z_2\alpha}{2\pi^2}
 \frac{1}{{\kabs}^2}
 \ \ \ , \ \ \ T_{\text{B}}(\boldsymbol{k})=\tilde{\alpha}_{1,i}(\boldsymbol{k})\tilde{\alpha}_{2,i}(\boldsymbol{k})
 \label{breitkernel}
 \ ,
\end{equation}
and to account for the sum of the instantaneous Coulomb--Breit contributions, we also define 
\begin{equation}
 \kappa_{\text{CB}}(\boldsymbol{k})=+
 \frac{z_1z_2\alpha}{2\pi^2}
 \frac{1}{{\kabs}^2}
 \ \ \ , \ \ \ T_{\text{CB}}(\boldsymbol{k})=
 I_1I_2-\tilde{\balpha}_{1}(\boldsymbol{k})\tilde{\balpha}_{2}(\boldsymbol{k})
 \ .
\end{equation}
The corresponding instantaneous potential energy operators, Eq.~\eqref{instint}, are the Coulomb, Breit, and Coulomb--Breit potential energy operators, ubiquitous in (relativistic) quantum chemistry, 
\begin{align}
  V_\text{C} 
  &=
  \int \dd^3\bk\ \kappa_\text{C}(\bk) T_\text{C} \eta_1(\bk) \eta_2(-\bk) 
  =
  \frac{z_1z_2\alpha}{2\pi^2} \int \dd^3\bk\ \frac{1}{{\kabs}^2} \eta_1(\bk) \eta_2(-\bk) \; ,\\
  V_\text{B} 
  &=
  \int \dd^3\bk\ \kappa_\text{B}(\bk) T_\text{B} \eta_1(\bk) \eta_2(-\bk) 
  =
  -\frac{z_1z_2\alpha}{2\pi^2} \int \dd^3\bk\ \frac{\tilde\balpha_1(\bk)\cdot\tilde\balpha_2(\bk)}{{\kabs}^2} \eta_1(\bk) \eta_2(-\bk)  \\
   V_\text{CB}
   &=
   V_\text{C} + V_\text{B} \; ,
\end{align}
respectively, written in a momentum-space form (and natural units).

\paragraph{Useful relations}
The energy-momentum shift operators appearing in the interaction kernels, Eqs.~\eqref{eq:Kinst} and \eqref{eq:KT}, are extensively used throughout this work, so we collect here their effect on the one-particle propagators {{}(Eqs. (\ref{eq:propagator1}), (\ref{eq:propagator2}))}:
\begin{align}
  S_1(\bp_1-\bk,\varepsilon-\omega) 
  = \eta_1(\bk)\eta_\epsilon(\omega) S_1(\bp_{{}1},\epsi) \eta_1(-\bk)\eta_\epsilon(-\omega) \ ,
  \label{eq:S1eta}
\end{align}
and
\begin{align}
  S_2(\bp_2{+}\bk,\omega-\varepsilon) 
  = \eta_2(-\bk)\eta_\epsilon(\omega) S_2(\bp_{{}2},-\epsi) \eta_2(\bk) \eta_\epsilon(-\omega)
  \label{eq:S2eta} \ .
\end{align}
Some further details and properties of the shift operators are made explicit in Appendix~\ref{App:eta}.
Note that we often suppress some or all arguments of the propagators for the sake of brevity. By $S_1$ and $S_2$,  we always mean $S_1(\bp_1,\varepsilon)$ and $S_2(\bp_2,-\varepsilon)$, respectively, with their `ordinary arguments'; the same applies when only their
energy arguments are shown explicitly, $S_1(\varepsilon)$ and $S_2(-\varepsilon)$.
Different 3-momentum arguments are always expressed with the help of the $\eta_a(\bk)$ shift operators.

Furthermore, we will often use the following relation $-$ based on the partial fraction decomposition $-$ for one-particle propagator products (see also Appendix~\ref{AppC}),
\begin{align}
  S_1 S_2 
  &=
  (S_1 + S_2) (S_1^{-1} + S_2^{-1})^{-1} = (S_1 + S_2) (E-h_1-h_2)^{-1} \nonumber \\
  &=
  (S_1^{-1} + S_2^{-1})^{-1} (S_1 + S_2) = (E-h_1-h_2)^{-1} (S_1 + S_2)  \; .
  \label{eq:partialFracS1S2}
\end{align}

\subsection{The Salpeter--Sucher exact equal-time equation \label{sec:SaSueq}}
\noindent %
Partitioning the $\tilde{K}$ interaction kernel into an instantaneous (Coulomb or Coulomb--Breit) part and the rest, makes it possible to rearrange Eq.~\eqref{eq:BSalmostfinal} to
\begin{align}
 \left[%
   S_1^{-1}(\boldsymbol{p}_1,\varepsilon)S_2^{-1}(\boldsymbol{p}_2,-\varepsilon)
   -{K}_{\Delta}
 \right]
 \psi(\boldsymbol{p}_1,\boldsymbol{p}_2,\epsi)
 &=
 \frac{1}{-2\pi \iim}
 {V}_{\inst}\Phi(\boldsymbol{p}_1,\boldsymbol{p}_2)  \; .
\end{align}
To arrive at a wave equation for the $\Phi$ equal-time wave function, we would like to have $\Phi$ on both sides of the equation, so the calculation continues as (we also adopt the general bra-ket notation which is equally valid with a $\bp_a$ and $\br_a$ representation)
\begin{align}
  |\Phi\rangle
  &=
  \int_{-\infty}^{+\infty} \dd \epsi\ |\psi(\epsi)\rangle
  \nonumber \\
  &=
  \int_{-\infty}^{+\infty}   
  \frac{\dd \epsi}{-2\pi \iim}
  \left[%
    S_1^{-1}S_{2}^{-1}-{\Kcal}_{\Delta}
  \right]^{-1}
  {V}_{\inst} |\Phi\rangle
  \nonumber \\
  &=
  \int_{-\infty}^{+\infty}   
  \frac{\dd \epsi}{-2\pi \iim}
  \left[%
    S_1 S_2
    +
    S_1S_2 \Kcal_{\Delta}
    \left(%
      S_1^{-1}S_2^{-1}
      -{\Kcal}_{\Delta}
    \right)^{-1}
  \right]
  {V}_{\inst} |\Phi\rangle
  \nonumber \\
  &=
  \left[%
  \frac{L_{++}-L_{--}}{E-h_1-h_2}
  +
  \left\lbrace%  
  \int_{-\infty}^{+\infty}   
  \frac{\dd \epsi}{-2\pi \iim}
    S_1S_2 \Kcal_{\Delta}
    \left(%
      S_1^{-1}S_2^{-1}
      -{\Kcal}_{\Delta}
    \right)^{-1}
  \right\rbrace
  \right]
  {V}_{\inst}
  |\Phi\rangle \; .
  \label{eq:SaSuEq0}  
\end{align}
Intermediate steps are elaborated in Appendix~\ref{AppC}, highlighting some useful relations in practice.
{{}The two-particle projection operators are defined as
\begin{equation}
 {L}_{\pm\pm}={L}_{1,\pm}{L}_{2,\pm} \ .
\end{equation}
}
Finally, a formally exact wave equation emerges, 
which we call the Salpeter--Sucher wave equation, in memory of its pioneers \cite{Sa52,sucherPhD1958},
\begin{align}
  \left[h_1+h_2+(L_{++}-L_{--}) V_\inst + \Vcal_{\epsilon}(E) \right]|\Phi\rangle
  =
  E|\Phi\rangle \ .
  \label{eq:SaSuEq1}
\end{align}
The `effective' Hamiltonian operator, which carries a non-trivial $E$ dependence in the ${\cal{V}}$ term,  can be partitioned in at least two natural ways.
The first option is 
\begin{equation}
  \left[H_{\nopair}+{\cal{V}}_{\text{QED}}(E)\right]|\Phi\rangle
  =
  E|\Phi\rangle \ ,
  \label{eq:SaSuEq}
\end{equation}
where $H_{\nopair}$ 
is the no-pair Hamiltonian with
the instantaneous interactions projected onto the positive-energy subspace,
\begin{equation}
  H_{\nopair}
  =
  {h}_1+{h}_2+{L}_{++}{{V}}_{\inst}{L}_{++} \ ,
  \label{Hnopair}
\end{equation}
the choices $V_\inst=V_\text{C}$ and $V_\inst=V_\text{CB}$ defining
the no-pair Dirac--Coulomb (DC) and Dirac--Coulomb--Breit (DCB) Hamiltonians, respectively;
 ${\cal{V}}_{\text{QED}}$ is essentially everything else, containing both radiative and non-radiative QED contributions:
\begin{align}
  {\cal{V}}_{\text{QED}}(E)
  =
  V_\delta + {\cal{V}_{{}\epsilon}}(E) \; .
\end{align}
$\Vcal_{\epsilon}(E)$ collects all retarded, including self-energy-type, contributions,
\begin{align}
  \Vcal_{\epsilon}(E)
  =
  (E-{h}_1-{h}_2)
  \int_{-\infty}^{+\infty}\frac{\mathrm{d}\epsi}{-2\pi \iim}
    %S_1S_2{K}_{\Delta}\left(S_1^{-1}S_2^{-1}-{\Kcal}_{\Delta}\right)^{-1}
    S_1S_2{\Kcal}_{\Delta}\left(1-S_1S_2 {\Kcal}_{\Delta}\right)^{-1} S_1 S_2
    V_{\inst} \ ,
  \label{eq:Vepsi}    
\end{align}
where $S_1,S_2,$ and $\Kcal$ are $E$-dependent factors.
At the same time, $V_\delta$ is  an energy-independent term and carries (non-crossed) Coulomb(--Breit) pair corrections, 
\begin{align}
  V_\delta = {L}_{++}{{V}}_{\inst}(1-{L}_{++})-{L}_{--}{{V}}_{\inst} \; .
  \label{eq:Vdelta}
\end{align}
So, in principle, it can be included in the reference Hamiltonian, and thereby, we have a (non-Hermitian) \emph{with-pair} Hamiltonian for instantaneous interactions,
\begin{align}
  H_\withpair = h_1+h_2+(L_{++}-L_{--})V_\inst \; .
  \label{eq:withpair}
\end{align}
Mathematical properties and numerical computations for $H_\withpair$  will be reported in Ref.~\cite{JeMa23CC}. 
With this partitioning, the full `QED' wave equation reads as
\begin{align}
  \left[H_{\withpair}+{\cal{V}}_{\epsilon}(E)\right]|\Phi\rangle
  =
  E|\Phi\rangle \ .
  \label{eq:SaSuEqWithPair}  
\end{align}

The $\Vcal_{\epsilon}(E)$ term, Eq.~\eqref{eq:Vepsi}, makes the eigenvalue equation non-linear in the energy. 
In principle, one could aim for an iterative solution of the Salpeter--Sucher equation, if
$\Vcal_{\epsilon}$ was written in a closed or at least, in some numerically, algorithmically tractable form.

\subsection{Calculations with $\cal{V}_{{}\epsilon}$ and resummation prospects \label{sec:IntroResum}}
A naive approach to $\Vcal_{\epsilon}$ starts with expanding the operator inverse in the integrand,
\begin{align}
 \Vcal_{\epsilon}(E)|\Phi\rangle
 &=
 (E-h_1-h_2)
 \int_{-\infty}^{+\infty}\frac{\mathrm{d}\epsi}{-2\pi \iim}
 S_1(\varepsilon)S_2(-\varepsilon)
 \sum_{n=1}^{\infty}\Big[K_\Delta S_1(\varepsilon)S_2(-\varepsilon)\Big]^nV_\inst|\Phi\rangle \ ,
 \label{eq:expansion}
\end{align}
followed by calculation of the $\epsi$ (and other) integrals, and then a possible resummation of the series. 
If it was possible to deal with $\Vcal_{\epsilon}$ in a compact form for the relevant interaction kernel, one could attempt an iterative solution of Eq.~\eqref{eq:SaSuEq}.

For a special, pedagogical case, this idea can be carried out easily; this will serve as an alternative derivation of the with-pair DCB equation. If $V_\inst$ contains the Coulomb interaction ($V_\inst=V_\text{C}$, Eqs. (\ref{instint}) and (\ref{coulombkernel})) 
and $K_\Delta$ is approximated by some instantaneous kernel ($K_\Delta\approx K_\inst'$, $\inst'\neq\text{C}$; actually, $K_\Delta\approx K_\text{T}\approx K_\text{B}$, Eq. (\ref{breitkernel}), will be used), then $\Vcal_{\epsilon}(E)$ can be obtained in a closed form.

First, we note that if $K_{\inst'} S_1(\varepsilon)S_2(\omega-\varepsilon)$ acts on some energy-independent ket state $|g\rangle$
(such as $V_\inst|\Phi\rangle$), then (using Appendix \ref{AppC}) we find
\begin{align}
 K_{\inst'} S_1(\varepsilon)S_2(\omega-\varepsilon)|g\rangle
 &=
 V_{\inst'}\int_{-\infty}^{+\infty}\frac{\mathrm{d}\nu}{{{}-}2\pi \iim}
 S_1(\varepsilon-\nu)S_2(\omega+\nu-\varepsilon)\eta_\epsilon(\nu)|g\rangle \nonumber \\
 &=
 V_{\inst'}\int_{-\infty}^{+\infty}\frac{\mathrm{d}\nu}{{{}-}2\pi \iim}
 S_1(\varepsilon-\nu)S_2(\omega+\nu-\varepsilon)|g\rangle \nonumber \\
 &=
 V_{\inst'}\int_{-\infty}^{+\infty}\frac{\mathrm{d}\nu}{{{}-}2\pi \iim}
 S_1(\nu)S_2(\omega-\nu)|g\rangle \nonumber \\
 &=V_{\inst'}{\cal{Q}}(\omega)|g\rangle \ ,
\end{align}
where
 \begin{align}
  {\cal{Q}}(\omega)
  &=
  {\cal{Q}}_{++}(\omega)+{\cal{Q}}_{--}(\omega) \nonumber \\
  &=
  \frac{L_{++}}{E-{h}_1-{h}_2+\omega+\iim 0^+}
  -
    \frac{L_{--}}{E-{h}_1-{h}_2+\omega-\iim 0^+} \nonumber \\
  &=
  \frac{L_{++}-L_{--}}{E-{h}_1-{h}_2+\omega+\iim 0^+(L_{++}-L_{--})}
   \ .
 \end{align}
The $\omega$ dependence was introduced only for the sake of later convenience, we will only need ${\cal{Q}}(0)$ in this section. 

The resulting state is again energy-independent, meaning
that we can calculate the action of $K_{\inst'} S_1(\varepsilon)S_2(\omega-\varepsilon)$ on it, and so on, leading to
\begin{align}
 \Large[K_{\inst'} S_1(\varepsilon)S_2(\omega-\varepsilon)\Large]^n|g\rangle
 =(V_{\inst'}{\cal{Q}}(\omega))^n|g\rangle \ ,
\end{align}
and
\begin{align}
 \Vcal_{\epsilon}(E)|\Phi\rangle
 &\approx
 (E-h_1-h_2)
 {\cal{Q}}(0)
 \sum_{n=1}^{\infty}(V_\text{B}{\cal{Q}}(0))^nV_\text{C}|\Phi\rangle \nonumber \\
 &=
(E-h_1-h_2)
 \sum_{n=1}^{\infty}({\cal{Q}}(0)V_\text{B})^n{\cal{Q}}(0)V_\text{C}|\Phi\rangle \nonumber \\ 
 &=
 (E-h_1-h_2)\left[\frac{1}{1-{\cal{Q}}(0)V_\text{B}}-1\right]{\cal{Q}}(0)V_\text{C}|\Phi\rangle \nonumber \\
 &=
 (E-h_1-h_2)\frac{1}{1-{\cal{Q}}(0)V_\text{B}}{\cal{Q}}(0)V_\text{C}|\Phi\rangle
 -(L_{++}-L_{--})V_\text{C}|\Phi\rangle \ .
\end{align}
Here we substituted the actual values $K_\inst=K_\text{B}$ and $\omega=0$.
Inserting this result in Eq. (\ref{eq:SaSuEqWithPair}) gives
\begin{align}
 (E-h_1-h_2)\frac{1}{1-{\cal{Q}}(0)V_\text{B}}{\cal{Q}}(0)V_\text{C}|\Phi\rangle=(E-h_1-h_2)|\Phi\rangle \ ,
\end{align}
which is rearranged to obtain
\begin{align}
 \left[h_1+h_2+(L_{++}-L_{--})V_{\text{CB}}\right]|\Phi\rangle=E|\Phi\rangle \ .
\end{align}
This is indeed the eigenvalue equation of the with-pair DCB Hamiltonian.

A more complicated approximate equation can be obtained if the leftmost
$K_\Delta$ factor is retained and only the other factors are approximated to be instantaneous (retaining more $K_\Delta$ factors in the first few terms could be realized along similar lines): 
\begin{align}
 \Vcal_{\epsilon}(E)|\Phi\rangle
 &\approx
 (E-h_1-h_2)
 \int_{-\infty}^{+\infty}\frac{\mathrm{d}\epsi}{-2\pi \iim}
 S_1(\varepsilon)S_2(-\varepsilon)K_\Delta S_1(\varepsilon)S_2(-\varepsilon)
  \sum_{n=0}^{\infty}\Big[K_{\inst'} S_1(\varepsilon)S_2(-\varepsilon)\Big]^nV_\inst|\Phi\rangle
  \nonumber \\
  &=
(E-h_1-h_2)
 \int_{-\infty}^{+\infty}\frac{\mathrm{d}\epsi}{-2\pi \iim}
 S_1(\varepsilon)S_2(-\varepsilon)K_\Delta S_1(\varepsilon)S_2(-\varepsilon)
 \frac{1}{1-V_{\inst'}{\cal{Q}}(0)}V_\inst|\Phi\rangle \nonumber \\
 &=
 v_\Delta{(E)} \frac{1}{E-h_1-h_2-(V_{\inst'}-\iim0^+)(L_{++}-L_{--})}V_\inst|\Phi\rangle
  \ .
\end{align}
where we used Eq. (\ref{eq:partialFracS1S2}) to introduce
\begin{align}
 v_\Delta{(E)}=
 \int_{-\infty}^{+\infty}\frac{\mathrm{d}\epsi}{-2\pi \iim}
 \large[S_1(\varepsilon)+S_2(-\varepsilon)\large]K_\Delta\large[S_1(\varepsilon)+S_2(-\varepsilon)\large] \ .
\end{align}
Now $\inst'$ may or may not be equal to $\inst$
(that is, the summed instantaneous interaction is not necessarily the same as the one that was originally separated).

In this case of approximating the higher-order factors with an instantaneous kernel, the Salpeter--Sucher equation can be written with a closed-form effective potential energy, 
\begin{align}
   \left[%
    h_1 + h_2 + (L_{++}-L_{--})V_\inst + 
    v_\Delta{(E)} 
    \frac{1}{E-[h_1+h_2+(V_{\inst'}-\iim0^+)(L_{++}-L_{--})]}
    V_\inst 
  \right]
  |\Phi\rangle=E|\Phi\rangle  \ ,
  \label{eq:resum}
\end{align}
and the equation may be included in iterative or perturbative computations.
For $\inst=\inst'$, the equation takes the form of a modified with-pair equation:
\begin{align}
   \left[%
    H_\text{wp} + 
    v_\Delta{(E)} 
    \frac{1}{E-H_\text{wp}^{\dagger}+\iim0^+(L_{++}-L_{--})]}
    V_\inst 
  \right]
  |\Phi\rangle=E|\Phi\rangle  \ .
  \label{eq:resumwp}
\end{align}
These equations contain a finite many crossed ladder and (or) retarded interactions, and the full expression is summed with approximating the (infinitely many) remaining factors with instantaneous interactions. This idea allowed us to arrive at a compact form, which might be tackled with algebraic methods. It is still necessary to evaluate the $v_\Delta$ integral, which is possible for finite many non-trivial (retarded, crossed, etc.) interaction terms and factors.

\subsection{Perturbative account of $\cal{V}_{\text{QED}}$}
\noindent%
Using recently developed numerical procedures \cite{JeFeMa21,JeFeMa22,FeJeMa22,FeJeMa22b,JeMa22}, we can converge the ground- (or excited)-state energy and wave function of the positive-energy block of the no-pair Hamiltonian, Eq.~\eqref{Hnopair}, to high precision, 
\begin{equation}
 H_{\nopair}|\Phi_{\nopair}\rangle=E_{\nopair}|\Phi_{\nopair}\rangle \ \ \ , \ \ \ 
 {L}_{++}|\Phi_{\nopair}\rangle=|\Phi_{\nopair}\rangle \; ,
\end{equation}
as briefly discussed in Appendix \ref{app:DCDCB}.

As a next step, we aim to perturbatively account for the effect of ${\cal{V}}_{\text{QED}}$ to the no-pair energy or perhaps even the effect of $\Vcal_{\epsilon}$ to the with-pair energy.
In this work, we will primarily focus on the (energy-dependent) $\Vcal_{\epsilon}(E)$ part of the QED potential energy. The instantaneous with-pair equation, Eq.~\eqref{eq:withpair}, is discussed separately, Ref.~\cite{JeMa23CC}. 

Brillouin--Wigner perturbation theory is used in this context \cite{sucherPhD1958,DoKr74,Zh96} {(see also Sec.~9.2 of Ref.~\citenum{LiMoBook} for an introduction)}, since it is applicable even if the perturbation operator itself is energy-dependent, 
\begin{align}
 \Delta E
 =E-E_{\nopair} 
 &=
 \langle\Phi_{\nopair}|
   {\Vcal}_{\text{QED}}
   (1-\Gamma(E){\Vcal}_{\text{QED}})^{-1}
 |\Phi_{\nopair}\rangle 
 \nonumber \\
 &=
 \sum_{n=1}^{\infty}
 \langle\Phi_{\nopair}|{\Vcal}_{\text{QED}}\left[\Gamma(E){\Vcal}_{\text{QED}}\right]^{n-1}|\Phi_{\nopair}\rangle \ .
 \label{eq:BWPT}
\end{align}
The $\Gamma(\omega)$ resolvent operator can be written as
\begin{align}
 \Gamma(\omega)
 &=
 (\omega-H_{\nopair})^{-1}(1-|\Phi_{\nopair}\rangle\langle \Phi_{\nopair}|) \\
 &=
 (\omega-H_{\nopair})^{-1}({L}_{++}-|\Phi_{\nopair}\rangle\langle \Phi_{\nopair}|)+(\omega-{h}_1-{h}_2)^{-1}(1-{L}_{++})
 \ ,
\end{align}
using the identities 
\begin{equation}
 (1-{L}_{++})\Gamma(\omega)=
 (1-{L}_{++})(\omega-{h}_1-{h}_2)^{-1}
 \ \ \ , \ \ \ 
 {L}_{++}\Gamma(\omega){L}_{--}=0 \ .
 \label{gammaidentity}
\end{equation}

\paragraph{First-order perturbative correction}
The first-order correction due to the $\Vcal_{\epsilon}$ term is
\begin{align}
  E^{(1)}_\epsi 
  &=
  \langle %
    \Phi_\nopair | \Vcal_{\epsilon}(E) | \Phi_\nopair 
  \rangle
  \nonumber \\
  &=
  \langle %
    \Phi_\nopair |
    (E-h_1-h_2) \int_{-\infty}^{+\infty}\frac{\dd\epsi}{-2\pi\iim} S_1 S_2 K_\Delta (1-S_1S_2K_\Delta)^{-1} S_1 S_2 V_\inst 
    | \Phi_\nopair
  \rangle \nonumber \\
  &=
  \langle %
    \Phi_\nopair |
    \int_{-\infty}^{+\infty} \frac{\dd\epsi}{-2\pi\iim} 
      (S_1+S_2) K_\Delta (1-S_1S_2 K_\Delta)^{-1} (S_1+S_2) (E-h_1-h_2)^{-1} V_\inst 
    | \Phi_\nopair
  \rangle  \nonumber \\
  &=
  \langle %
    \Phi_\nopair |
    \int_{-\infty}^{+\infty} \frac{\dd\epsi}{-2\pi\iim} 
      (S^{(+)}_1+S^{(+)}_2) K_\Delta (1-S_1S_2 K_\Delta)^{-1} (S_1+S_2) (E-h_1-h_2)^{-1} V_\inst L_{++}
    | \Phi_\nopair
  \rangle \; .
  \label{eq:PT1full}
\end{align}
We can start with the $(1-S_1 S_2K_\Delta)^{-1}\approx 1$ zeroth-order approximation (indicated by the `0' in the superscript)
\begin{align}
  E^{(1,0)}_\epsi 
  &=
  \langle %
    \Phi_\nopair |
    \int_{-\infty}^{+\infty} \frac{\dd\epsi}{-2\pi\iim} 
      (S^{(+)}_1+S^{(+)}_2) K_\Delta (S_1+S_2) (E-h_1-h_2)^{-1} V_\inst L_{++}
    {|}\Phi_\nopair
  \rangle
  \nonumber \\
  &\approx
  \langle %
    \Phi_\nopair |
    \int\frac{\dd\epsi}{-2\pi\iim} 
      (S^{(+)}_1+S^{(+)}_2) K_\Delta (S^{(+)}_1+S^{(+)}_2) 
    {|}\Phi_\nopair
  \rangle  \;,
  \label{eq:E10pt}
\end{align}
where, in the second step, we used two approximations ($1=L_{++}+L_{+-}+L_{-+}+L_{--}\approx L_{++}$ and $E\approx E_\nopair$) and the rearrangement
\begin{align}
  (E-h_1-h_2)^{-1} V_\inst L_{++} | \Phi_\nopair \rangle
  &=
  (E-h_1-h_2)^{-1} (L_{++}+L_{+-}+L_{-+}+L_{--}) V_\inst L_{++} | \Phi_\nopair \rangle
  \nonumber \\ 
  &\approx
  (E-h_1-h_2)^{-1} L_{++} V_\inst L_{++} | \Phi_\nopair \rangle  
  \nonumber \\ 
  &\approx
  (E_\nopair-h_1-h_2)^{-1} L_{++} V_\inst L_{++} | \Phi_\nopair \rangle  
  = | \Phi_\nopair \rangle  \; .
\end{align}
Retaining higher-orders in the $(1-S_1 S_2K_\Delta)^{-1}\approx 1 + S_1S_2 K_\Delta + \ldots$ expansion, will give rise to further $\Delta E_\epsi^{(1,n)}$ contributions.

We also note that $\langle\Phi_\nopair|{\cal{V}}_{\delta}|\Phi_\nopair\rangle=0$, \emph{i.e.,} the double-negative Coulomb(--Breit) ladder correction appears only at second order. Alternatively, the corresponding double-negative Coulomb(--Breit) ladder can be included in the wave equation thanks to its closed, algebraic form. 

\section{The simplest energy-dependent corrections: single transverse photon exchange and one-loop self-energy}
\subsection{Single transverse-photon exchange \label{sec:transverse}}
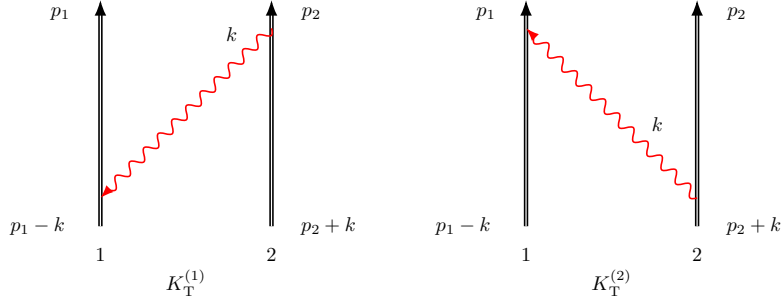
\begin{figure}
    \scalebox{0.75}{%
    \begin{tikzpicture}
    \draw[boundelectron,-Latex] (0.,0.) -- (0.,4.) ;
    \draw[boundelectron,-Latex] (3.,0.) -- (3.,4.) ;
    \draw[snake=coil, segment aspect=0,thick,red,-Latex]  (3.,3.5) -- (0.,0.5) ;
    \node (a0) at (0.,-0.5) {$1$};
    \node (a1) at (-0.7,3.75) {$p_1$};
    \node (a2) at (-1.1,0.0) {$p_1-k$};
    \node (b0) at (3.,-0.5) {$2$};
    \node (a1) at (3.7,3.75) {$p_2$};
    \node (a2) at (4.0,0.0) {$p_2+k$};
    \node (c) at (2.3,3.4) {$k$};
    \node (T,a) at (1.5,-1.0) {$K^{(1)}_{\text{T}}$};
    \end{tikzpicture} 
    \quad\quad\quad
    \begin{tikzpicture}
    \draw[boundelectron,-Latex] (0.,0.) -- (0.,4.) ;
    \draw[boundelectron,-Latex] (3.,0.) -- (3.,4.) ;
    \draw[snake=coil, segment aspect=0,thick,red,-Latex]  (3.,0.5) -- (0.,3.5) ;
    \node (a0) at (0.,-0.5) {$1$};
    \node (a1) at (-0.7,3.75) {$p_1$};
    \node (a2) at (-1.1,0.0) {$p_1-k$};
    \node (b0) at (3.,-0.5) {$2$};
    \node (a1) at (3.7,3.75) {$p_2$};
    \node (a2) at (4.0,0.0) {$p_2+k$};
    \node (c) at (2.3,1.8) {$k$};
    \node (T,a) at (1.5,-1.0) {$K^{(2)}_{\text{T}}$};    
    \end{tikzpicture} 
   }

  \caption{%
    Transverse interaction kernel, $K_{\text{T}}=K^{(1)}_{\text{T}}+K^{(2)}_{\text{T}}$ visualized in a time-ordered diagram.
    \label{fig:single}
  }
\end{figure}

A single transverse-photon exchange (Fig.~\ref{fig:single}) is described by the interaction kernel, Eq.~\eqref{eq:KT}, 
\begin{align}
  K_{\text{T}}
  =
  \frac{z_1z_2\alpha}{2\pi^2}
  \int\mathrm{d}^3 \bk\ 
  \int_{-\infty}^{+\infty}\frac{\mathrm{d}\omega}{-2\pi \iim}
    \frac{\tilde{\alpha}_{1,i}(\boldsymbol{k})\tilde{\alpha}_{2,i}(\boldsymbol{k})}{\omega^2-{\kabs}^2+\iim 0^+}
   {\eta}_1(\boldsymbol{k})
   {\eta}_2(-\boldsymbol{k})
   {\eta}_\epsilon(\omega) \ ,
  \label{eq:Ktransverse}
\end{align}
where $k=|\bk|$.
The lowest-order perturbative correction to the (retarded) transverse photon exchange can be calculated as (see also Appendix~\ref{AppC} or Ref.~\cite{MaFeJeMa23}): 
\begin{align}
  \Delta E_{\text{T}}&=
  \int_{-\infty}^{+\infty}\frac{\mathrm{d}\varepsilon}{-2\pi \iim}
  \langle\Phi_\nopair|[S^{(+)}_1(\varepsilon)+S^{(+)}_2(-\varepsilon)]K_{\text{T}}
  [S^{(+)}_1(\varepsilon)+S^{(+)}_2(-\varepsilon)]|\Phi_\nopair\rangle \nonumber \\
  &=
  \frac{z_1z_2 \alpha}{2\pi^2}
  \int \dd^3 \bk\  
    \frac{1}{2{\kabs}-\iim 0^+/{\kabs} }
    \langle\Phi_{\nopair}|
    \tilde{\boldsymbol{\alpha}}_2(\bk)\eta_2(-\bk)
    \frac{L_{++}}{E_\nopair-{h}_1-{h}_2-{\kabs} + \iim 0^+/(2{\kabs}) }
    \tilde{\boldsymbol{\alpha}}_1(\bk)\eta_1(\bk)
  |\Phi_{\nopair}\rangle \nonumber \\
   &+ \ (1\leftrightarrow2) \\
  &{=}
  \frac{z_1z_2 \alpha}{2\pi^2}
  \int \dd^3 \bk\  
    \frac{1}{2{\kabs}}
    \langle\Phi_{\nopair}|
    \tilde{\boldsymbol{\alpha}}_2(\bk)\eta_2(-\bk)
    \frac{L_{++}}{E_\nopair-{h}_1-{h}_2-{\kabs} + \iim 0^+ }
    \tilde{\boldsymbol{\alpha}}_1(\bk)\eta_1(\bk)
  |\Phi_{\nopair}\rangle %
  + \ (1\leftrightarrow2)
  \ ,
  \label{singletransverse}
\end{align}
where $(1\leftrightarrow2)$ denotes the same term with particle indices interchanged {(note that the $L_{++}$ factor could be neglected due to $\tilde{\boldsymbol{\alpha}}_2L_{++}\tilde{\boldsymbol{\alpha}}_1=L_{1+}\tilde{\boldsymbol{\alpha}}_2\tilde{\boldsymbol{\alpha}}_1L_{2+}$, but we keep it for later convenience)}.
If the zeroth-order solution, $(E_\nopair,\Phi_\nopair)$, was generated in the no-pair DC framework, then this result can be used without further manipulation.
However, if the zeroth-order is the no-pair DCB solution, then $E_\nopair$ already contains the instantaneous part of the transverse-photon exchange, \emph{i.e.,} the Breit term, so 
the perturbative correction must be computed only for the retarded part of the transverse correction to avoid double counting (Sec.~\ref{sec:Tresum} and the discussion at the beginning of Sec.~\ref{ch:unified}).

It is interesting to note in Eq.~\eqref{singletransverse} that the two-particle \emph{non-interacting} Hamiltonian is in the resolvent. 
As shown in the next section, it is possible to account for the effect of the full positive-energy projected Coulomb(--Breit) ladder in addition to the transverse (retarded) photon exchange, {(first three diagrams of Fig. \ref{fig:+-sum})}, and this will give rise to a two-particle \emph{interacting} Hamiltonian in the denominator.
This `slight' difference is also expected to allow a more consistent treatment with a (high-precision) interacting, relativistic reference state, $(E_\nopair,\Phi_\nopair)$, \emph{e.g.,} by ensuring that we can work with a positive-definite resolvent in (ground-state) computations.

Clear connections to nrQED can be made after the Coulomb ladder resummation (Sec.~\ref{sec:Tresum}) and combination with the one-loop self-energy correction (Secs.~\ref{sec:SE} and \ref{sec:SEresum}).

\vspace{0.25cm}
\paragraph{The special case of a non-interacting reference}
To make some connection with the independent-particle formulation of bound-state QED and relativistic quantum chemistry, let us evaluate the Coulomb- plus transverse-photon correction for a non-interacting reference
of two identical particles with charge number $z$ {{}(non-interacting only in the sense of interparticle interactions, the effect of the external potential being unchanged)}.
Then, 
\begin{equation}
 |\Phi_\nopair\rangle\approx
 |\Phi^{(0)}\rangle=
 \frac{1}{\sqrt{2}}
 \left(|\phi_p\phi_q\rangle-|\phi_q\phi_p\rangle\right)
 \ ,
 \label{phi0}
\end{equation}
\begin{equation}
 E_{\nopair}\approx E^{(0)}=\varepsilon_p+\varepsilon_q \ 
 \label{e0}
\end{equation}
in Eq. (\ref{singletransverse}) and in the analogous formula for the Coulomb part, 
\begin{align}
  \Delta E_{\text{C}}&=
  \int_{-\infty}^{+\infty}\frac{\mathrm{d}\varepsilon}{-2\pi \iim}
  \langle\Phi^{(0)}|[S^{(+)}_1(\varepsilon)+S^{(+)}_2(-\varepsilon)]K_{\text{C}}
  [S^{(+)}_1(\varepsilon)+S^{(+)}_2(-\varepsilon)]|\Phi^{(0)}\rangle \nonumber \\
  &=
  \langle\Phi^{(0)}|V_{\text{C}}|\Phi^{(0)}\rangle
  \ .  
  \label{singlecoulomb}
\end{align}
Note that the energy integration in the Coulomb case must be performed with some care (Appendix~{\ref{app:PoBe}}).
The orbitals are solutions of $h|\phi_p\rangle=\varepsilon_p|\phi_p\rangle$, and we obtain for the transverse photon interaction
\begin{align}
 \Delta E_{\text{T}}\approx&
  \frac{z^2 \alpha}{2\pi^2}
  \int \dd^3 \bk 
    \frac{1}{2{\kabs}-\frac{\iim0^+}{{\kabs}} }
    \Bigg[
    2\frac{\langle\phi_p\phi_q|\tilde{\boldsymbol{\alpha}}_1\eta_1(\bk)\tilde{\boldsymbol{\alpha}}_2\eta_2(-\bk)|\phi_p\phi_q\rangle}{-{\kabs}+ \frac{\iim0^+}{2{\kabs}}}
     \nonumber \\
    &-\langle\phi_p\phi_q|\tilde{\boldsymbol{\alpha}}_1\eta_1(\bk)\tilde{\boldsymbol{\alpha}}_2\eta_2(-\bk)|\phi_p\phi_q\rangle
    \left(\frac{1}{\varepsilon_p-\varepsilon_q-{\kabs}+ \frac{\iim0^+}{2{\kabs}}}+\frac{1}{\varepsilon_q-\varepsilon_p-{\kabs}+ \frac{\iim0^+}{2{\kabs}}}\right)
    \Bigg] \ .
\end{align}
The eight terms arising from expanding the determinant in both terms of the first equality of Eq.~(\ref{singletransverse}) could be simplified because $\eta_1(-\bk)\eta_2(\bk)$
 can be replaced by $\eta_1(\bk)\eta_2(-\bk)$ within the integral. The first term is just the expectation value of the Breit interaction,
while the second can be rewritten using Eq.~\eqref{eq:kappaT}, 
\begin{equation}
 \frac{1}{2{\kabs}-\frac{\iim0^+}{{\kabs}} }
 \left(\frac{1}{\varepsilon_p-\varepsilon_q-{\kabs}+ \frac{\iim0^+}{2{\kabs}}}-\frac{1}{\varepsilon_p-\varepsilon_q+{\kabs}- \frac{\iim0^+}{2{\kabs}}}\right)=
 \frac{1}{(\varepsilon_p-\varepsilon_q)^2-{\kabs}^2+\iim0^+} \ .
\end{equation}
Combining the transverse term with the similarly obtained Coulomb term, Eq.~(\ref{singlecoulomb}),
\begin{equation}
 \Delta E_\text{C}=
 \langle\phi_p\phi_q|V_{\text{C}}|\phi_p\phi_q\rangle-
 \langle\phi_p\phi_q|V_{\text{C}}|\phi_q\phi_p\rangle \ ,
\end{equation}
gives
\begin{equation}
 E_{\text{int}}\approx
 \langle\phi_p\phi_q|{\cal{V}}{_\text{1-photon}^{(0)}}(0)|\phi_p\phi_q\rangle-
 \langle\phi_p\phi_q|{\cal{V}}{_\text{1-photon}^{(0)}}(\varepsilon_p-\varepsilon_q)|\phi_q\phi_p\rangle \ ,
\end{equation}
where 
\begin{equation}
 {\cal{V}}{_\text{1-photon}^{(0)}}(\omega)=
 \frac{z^2\alpha}{2\pi^2}
 \int\mathrm{d}^3 \bk
 \left[
 \frac{1}{{\kabs}^2}+
  \frac{\tilde{\alpha}_{1}(\boldsymbol{k})\cdot\tilde{\alpha}_{2}(\boldsymbol{k})}{\omega^2-{\kabs}^2+\iim 0^+}
  \right]
   {\eta}_1(\boldsymbol{k})
   {\eta}_2(-\boldsymbol{k})
\end{equation}
is the one-photon exchange potential energy 
familiar from relativistic many-body theory \cite{lindgrenRelativisticManyBodyTheory2011,GrantBook07,DyFaBook07,Liu16}.
Note that ${\cal{V}}{_\text{1-photon}^{(0)}}(0)=V_{\text{CB}}$.

\subsection{Resummation for the instantaneous interaction ladder in the transverse-photon exchange correction\label{sec:Tresum}}
\noindent
It is possible to formally add the full instantaneous (either Coulomb or Coulomb--Breit) ladder of positive- and negative-energy \emph{in-ladder} states to the single transverse photon exchange (Fig. \ref{fig:+-sum}).
\emph{In-ladder} states {mean states represented by} the internal fermion lines \emph{between} the rungs of the ladder.
At the same time, the two \emph{in} and \emph{out} states of the ladder are forced to be of positive energy; this is trivially true when these states coincide with \emph{in}/\emph{out} states of the complete diagram (the external legs) because of the no-pair reference, but is a non-trivial restricition when they belong to internal lines connected to the emitting/absorbing vertex of the transverse photon. Besides derivational/computational convenience, the reason for this restriction is to treat all \emph{in}/\emph{out} states of the instantaneous ladder on the same footing.

Our result is a slight generalization of that of Sucher (in particular, Eqs.~5.22--5.25 of Ref.~\cite{sucherPhD1958}), who accomplished the inclusion of the positive-energy Coulomb ladder to the transverse photon exchange. 
The resummation is formally achieved for the general positive+negative energy case, but we will continue using the equations for the positive-energy case in the rest of the paper.
Sucher also calculated perturbative results to $\mathcal{O}(\alpha^3)$ order of nrQED corresponding to other processes involving a transverse photon and negative energy intermediate states (Fig.~\ref{fig:tr+neg}), which are, however, outside of the scope of the present paper.
    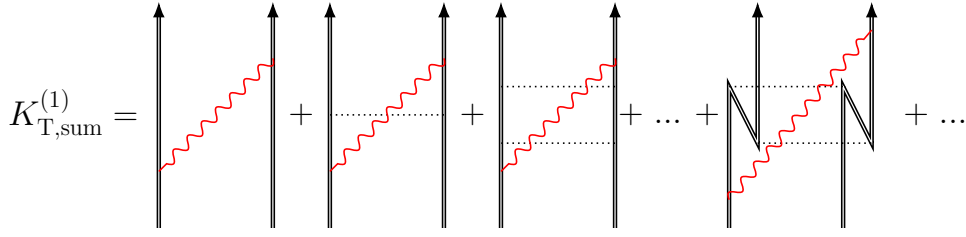
\begin{figure}
    \vspace{-0.175cm}
    \scalebox{0.75}{%
    \begin{tikzpicture}
    \node (T,a) at (-6.5,2.0) {\mbox{\LARGE$K^{(1)}_{\text{T,sum}}=$}};
    \draw[boundelectron,-Latex] (-5.,0.) -- (-5.,4.) ;
    \draw[boundelectron,-Latex] (-3.,0.) -- (-3.,4.) ;
    \draw[snake=coil, segment aspect=0,thick,red] (-3.,3.0) -- (-5.,1.0) ;
    {\color{black}
    \node (T,a) at (-2.5,2.0) {\mbox{\LARGE$+$}};
    \draw[boundelectron,-Latex] (-2.,0.) -- (-2.,4.) ;
    \draw[boundelectron,-Latex] (0.,0.) -- (0.,4.) ;
    \draw[coulomb, segment aspect=0,thick] (0.,2.0) -- (-2.,2.0) ;
    \draw[snake=coil, segment aspect=0,thick,red] (0.,3.0) -- (-2.,1.0) ;
    \node (T,a) at (0.5,2.0) {\mbox{\LARGE$+$}};
    \draw[boundelectron,-Latex] (1.,0.) -- (1.,4.) ;
    \draw[boundelectron,-Latex] (3.,0.) -- (3.,4.) ;
    \draw[coulomb, segment aspect=0,thick] (3.,1.5) -- (1.,1.5) ;
    \draw[coulomb, segment aspect=0,thick] (3.,2.5) -- (1.,2.5) ;
    \draw[snake=coil, segment aspect=0,thick,red] (3.,3.0) -- (1.,1.0) ;    
    \node (T,a) at (3.5,2.0) {\mbox{\ \ \LARGE$+ \  ...$}};}
    \node (T,a) at (4.5,2.0) {\mbox{\ \ \LARGE$+$}};
    \draw[boundelectron,-Latex] (5.0,0.) -- (5.0,2.5) -- (5.5,1.5) -- (5.5,4.) ;
    \draw[boundelectron,-Latex] (7.0,0.) -- (7.0,2.5) -- (7.5,1.5) -- (7.5,4.) ;
    \draw[coulomb,thick] (5.0,2.5) -- (7.0,2.5) ;
    \draw[coulomb,thick] (5.5,1.5) -- (7.5,1.5) ;
    \draw[snake=coil, segment aspect=0,thick,red] (5.,0.5) -- (7.5,3.5) ; 
    \node (T,a) at (8.5,2.0) {\mbox{\ \ \LARGE$+ \  ...$}};
        \end{tikzpicture} 
    \quad\quad\quad
   } 
   \caption{Transverse photon exchange on the background of an infinite number of instantaneous interactions involving both positive (second and third diagrams) and negative (last diagram) intermediate states \label{fig:+-sum}}
\end{figure}
    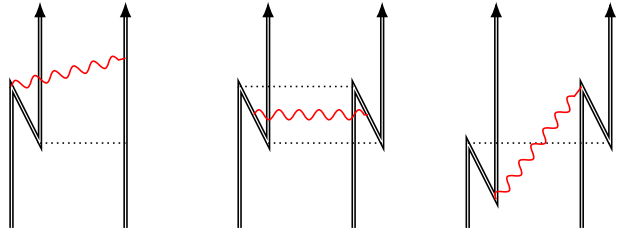
\begin{figure}
    \vspace{-0.175cm}
    \scalebox{0.75}{%
    \begin{tikzpicture}
    \draw[boundelectron,-Latex] (1.0,0.) -- (1.0,2.5) -- (1.5,1.5) -- (1.5,4.) ;
    \draw[boundelectron,-Latex] (3.0,0.) -- (3.0,4.) ;
    \draw[snake=coil, segment aspect=0,thick,red] (1.0,2.5) -- (3.0,3.0) ;
    \draw[coulomb,thick] (1.5,1.5) -- (3.0,1.5) ;
    \draw[boundelectron,-Latex] (5.0,0.) -- (5.0,2.5) -- (5.5,1.5) -- (5.5,4.) ;
    \draw[boundelectron,-Latex] (7.0,0.) -- (7.0,2.5) -- (7.5,1.5) -- (7.5,4.) ;
    \draw[coulomb,thick] (5.0,2.5) -- (7.0,2.5) ;
    \draw[coulomb,thick] (5.5,1.5) -- (7.5,1.5) ;
    \draw[snake=coil, segment aspect=0,thick,red] (5.25,2.0) -- (7.25,2.0) ; 
    \draw[boundelectron,-Latex] (9.0,0.) -- (9.0,1.5) -- (9.5,0.5) -- (9.5,4.) ;
    \draw[boundelectron,-Latex] (11.0,0.) -- (11.0,2.5) -- (11.5,1.5) -- (11.5,4.) ;
    \draw[coulomb,thick] (9.0,1.5) -- (11.5,1.5) ;
    \draw[snake=coil, segment aspect=0,thick,red] (9.5,0.5) -- (11.,2.5) ; 
    \end{tikzpicture} 
    \quad\quad\quad
   } 
   \caption{Examples of processes involving a transverse photon and negative-energy intermediate states, which are not considered in this paper.\label{fig:tr+neg}}
\end{figure}

To sum the $n$ instantaneous interaction steps in 
\begin{align}
  \Delta E_{\tau++}
  &=\sum_{n=0}^{\infty}  
  \int_{-\infty}^{+\infty}\frac{\mathrm{d}\varepsilon}{-2\pi \iim}
  \langle\Phi_\nopair|[S^{(+)}_1(\varepsilon)+S^{(+)}_2(-\varepsilon)]
  K_{\text{T};n}(\varepsilon)[S^{(+)}_1(\varepsilon)+S^{(+)}_2(-\varepsilon)]|\Phi_\nopair\rangle
  \ ,
  \label{eq:ET++}
 \end{align}
let us start by investigating the irreducible interaction kernel $K_{\text{T};n}$ %%%%%%%%%%
describing the summed effect of the \emph{with-pair interaction ladder}
 in addition to the effect of a single transverse photon. Whether these instantaneous interactions correspond to Coulomb or Coulomb-Breit interactions ($\inst=\text{C}$ or $\inst=\text{CB}$) depends on the no-pair wave function being $\nopair=\text{DC}$ or $\nopair=\text{DCB}$.

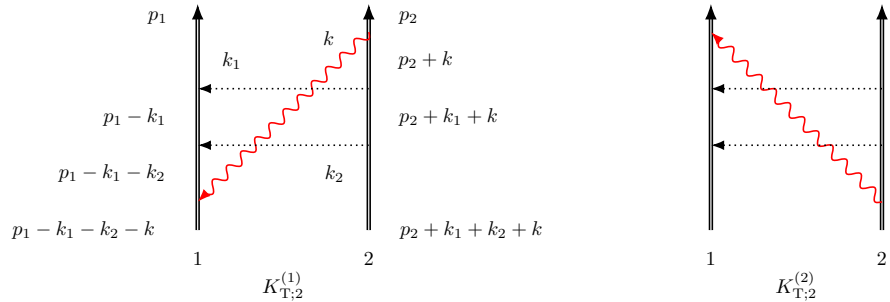
\begin{figure}
  \centering
  \quad\quad
  \scalebox{0.75}{%
    \begin{tikzpicture}
    \draw[boundelectron,-Latex] (0.,0.) -- (0.,4.) ;
    \draw[boundelectron,-Latex] (3.,0.) -- (3.,4.) ;
    \draw[snake=coil, segment aspect=0,thick,red,-Latex]  (3.,3.5) -- (0.,0.5) ;
    \draw[coulomb,thick,-Latex] (3.,1.5) -- (0.,1.5) ;
    \draw[coulomb,thick,-Latex] (3.,2.5) -- (0.,2.5) ;
    \node (a0) at (0.,-0.5) {$1$};
    \node (a1) at (-0.7,3.75) {$p_1$};
    \node (a2) at (-1.1,2.0) {$p_1-k_1$};
    \node (a3) at (-1.5,1.) {$p_1-k_1-k_2$};
    \node (a4) at (-2.,0.) {$p_1-k_1-k_2-k$};
    \node (b0) at (3.,-0.5) {$2$};
    \node (a1) at (3.7,3.75) {$p_2$};
    \node (a2) at (4.0,3.0) {$p_2+k$};
    \node (a3) at (4.4,2.) {$p_2+k_1+k$};
    \node (a4) at (4.8,0.) {$p_2+k_1+k_2+k$};
    \node (c) at (2.3,3.4) {$k$};
    \node (d) at (0.6,3.) {$k_1$};
    \node (e) at (2.4,1.) {$k_2$};
    \node (TxC,a) at (1.5,-1.0) {$K^{(1)}_{\text{T};2}$};
    \draw[boundelectron,-Latex] (9.,0.) -- (9.,4.) ;
    \draw[boundelectron,-Latex] (12.,0.) -- (12.,4.) ;
    \draw[snake=coil, segment aspect=0,thick,red,-Latex]  (12.,0.5) -- (9.,3.5) ;
    \draw[coulomb,thick,-Latex] (12.,1.5) -- (9.,1.5) ;
    \draw[coulomb,thick,-Latex] (12.,2.5) -- (9.,2.5) ;
    \node (a0) at (9.,-0.5) {$1$};
    \node (b0) at (12.,-0.5) {$2$};
    \node (TxC,a) at (10.5,-1.0) {$K^{(2)}_{\text{T};2}$};
    \end{tikzpicture} 
    }    
  \caption{%
   Example of a single transverse-photon exchange with two instantaneous photon exchanges, $K_{\text{T};n}=K^{(1)}_{\text{T};n}+K^{(2)}_{\text{T};n}$ for $n=2$.
   The upward oriented internal lines correspond to propagators restricted to the positive energy subspace. The diagram of $K^{(1)}$ shows the assignment of momenta.
   \label{fig:2instantaneouspic}
  }
\end{figure} 

We can write down the general $K_{\text{T};n}$ interaction kernel using the diagram rules 
of Appendix~\ref{App:diagram}
(an example for {{}the positive energy part of the} $n=2$ {{}case} is shown in Fig.~\ref{fig:2instantaneouspic}) as 
\begin{align}
  &K^{(1)}_{\text{T};n}(\varepsilon)\nonumber \\
  &=
  \int\mathrm{d}^3\bk\int_{-\infty}^{+\infty}\frac{\mathrm{d}\omega}{-2\pi \iim}
  \int\mathrm{d}^3\bk_n\int_{-\infty}^{+\infty}\frac{\mathrm{d}\omega_n}{-2\pi \iim}
  \eta_2(-\bk)\tilde{\boldsymbol{\alpha}}_2(\bk)
  S^{(+)}_2(\omega-\varepsilon)
  {{\cal{S}}^{(1..n-1)}_\text{inter}}(\varepsilon,\omega)
  \kappa_\text{T}(\bk,\omega) \nonumber \\
   &\quad\quad 
  \left\lbrace%
    \eta_1(\bk_n)\eta_2(-\bk_n)\eta_\epsilon(\omega_n)
    \kappa_\inst(\bk_n)T_\inst(\bk_n) 
  \right\rbrace
  S^{(+)}_1(\varepsilon)
    \tilde{\boldsymbol{\alpha}}_1(\bk)
  \eta_1(\bk)\eta_\epsilon(\omega) \nonumber \\
  &=
  \int\mathrm{d}^3\bk\int_{-\infty}^{+\infty}\frac{\mathrm{d}\omega}{-2\pi \iim}
  \eta_2(-\bk)\tilde{\boldsymbol{\alpha}}_2(\bk)
  S^{(+)}_2(\omega-\varepsilon)
  {{\cal{S}}^{(1..n-1)}_\text{inter}}(\varepsilon,\omega)
  \kappa_\text{T}(\bk,\omega)%  \tau_1(\bk_n)\tau_2(\bk_n)
  { K_\inst} S^{(+)}_1(\varepsilon)
  \tilde{\boldsymbol{\alpha}}_1(\bk)
  \eta_1(\bk)\eta_\epsilon(\omega)
   \ ,
  \label{T++}
\end{align}
where we defined the cumulative momenta and associated shift operators
\begin{align}
  \bk^{[p]}
  &=
  \sum_{q=1}^p\bk_q
  \; ,\quad 
  \eta_a(\bk^{[p]})
  =
  \prod_{q=1}^p\eta_a(\bk_q) \; , \\
  \omega^{[p]}
  &=
  \sum_{q=1}^p\omega_q \; , \quad
  \eta_\epsilon(\omega^{[p]})
  =
  \prod_{q=1}^p\eta_\epsilon(\omega_q) \; .
\end{align}
The $n$-th (`lowest') rung of the ladder has been written out explicitly,
and all the other rungs are collected in the expression (the product is understood left to right with increasing $p$, \emph{i.e.,} in the order of events),
\begin{align}
  &{{\cal{S}}^{(1..n-1)}_\text{inter}}(\varepsilon,\omega)
  \nonumber \\
  &=
  \Bigg[
 \prod_{p=1}^{n-1}
 \int\mathrm{d}^3\bk_p\int_{-\infty}^{+\infty}\frac{\mathrm{d}\omega_p}{-2\pi \iim}
  {\kappa_\inst(\bk_p)}
  T_\inst(\bk_p)
  %\right.
  \nonumber \\
  &\quad\quad
  %\left.
   \left\lbrace
   \eta_1(\bk^{[p]})
   S_1(\varepsilon-\omega^{[p]})
   \eta_1(-\bk^{[p]})
   \eta_2(-\bk^{[p]})
   S_2(\omega^{[p]}+\omega-\varepsilon)
   \eta_2(\bk^{[p]})
   \right\rbrace
   \Bigg]
  \nonumber \\
  &\quad\quad
   \eta_1(\bk^{[n-1]})\eta_2(-\bk^{[n-1]})\eta_\epsilon(\omega^{[n-1]})
  \label{eq:Sprod} \\
  &=
  \Bigg[%
 \prod_{p=1}^{n-1}
 \int\mathrm{d}^3\bk_p\int_{-\infty}^{+\infty}\frac{\mathrm{d}\omega_p}{-2\pi \iim}
  {\kappa_\inst(\bk_p)}
  T_\inst(\bk_p)
  %\right.
  \nonumber \\
  &\quad\quad
  %\left.
   \left\lbrace
   \eta_1(\bk^{[p]})
   \eta_2(-\bk^{[p]})   
   S_1(\varepsilon-\omega^{[p]})
   S_2(\omega^{[p]}+\omega-\varepsilon)
   \eta_1(-\bk^{[p]})   
   \eta_2(\bk^{[p]})
   \right\rbrace
   \Bigg]
  \nonumber \\
  &\quad\quad
   \eta_1(\bk^{[n-1]})\eta_2(-\bk^{[n-1]})\eta_\epsilon(\omega^{[n-1]}) \nonumber \\
  &=
  \prod_{p=1}^{n-1}
  \int\mathrm{d}^3\bk_p\int_{-\infty}^{+\infty}\frac{\mathrm{d}\omega_p}{-2\pi \iim}
  \left[%  
   {\kappa_\inst(\bk_p)}
   T_\inst(\bk_p)
    \eta_1(\bk_p)
    \eta_2(-\bk_p)   
    \eta_\epsilon(\omega_p)
    S_1(\varepsilon)
    S_2(\omega-\varepsilon)\right] \nonumber \\
    &=
    [K_\inst S_1(\varepsilon)S_2(\omega-\varepsilon)]^{n-1}
  \ .
\end{align}
Note that the leftmost and rightmost propagators of $K_\text{T};n$ were restricted to the positive energy subspace, in accordance with our beginning remarks.

In order to evaluate Eq.~\eqref{eq:ET++}, we consider the factors of the general $n$-th term (for $n\geq1$) from right to left.
The rightmost part reads
(with short notations $\eta_a=\eta_a(\bk)$, $\tilde{\boldsymbol{\alpha}}_a=\tilde{\boldsymbol{\alpha}}_a(\bk)$ and  using Appendix~\ref{AppC})
 \begin{align}
  |g\rangle
  &=
  \left[{{K}}_{\inst}
  S_1^{(+)}(\varepsilon)\tilde{\boldsymbol{\alpha}}_1\eta_1\eta_\epsilon(\omega)\right]
  \Big[S^{(+)}_1(\varepsilon)+S^{(+)}_2(-\varepsilon)\Big]|\Phi_{\nopair}\rangle \nonumber \\
  &=
  {{K}}_{\inst}S_1^{(+)}(\varepsilon)
  S_2^{(+)}(\omega-\varepsilon)\tilde{\boldsymbol{\alpha}}_1\eta_1|\Phi_{\nopair}\rangle \nonumber \\
  &=
  {{V}}_{\inst}
  \int_{-\infty}^{+\infty}\frac{\mathrm{d}\nu}{-2\pi\iim}
  S_1^{(+)}(\nu)S^{(+)}_2(\omega-\nu)\tilde{\boldsymbol{\alpha}}_1\eta_1|\Phi_{\nopair}\rangle \nonumber \\
  &=
  [V_{\inst}{\cal{Q}}_{++}(\omega)]
  \tilde{\boldsymbol{\alpha}}_1\eta_1|\Phi_{\nopair}\rangle \nonumber \\
  &=
  [V_{\inst}{\cal{Q}}(\omega)]
  \tilde{\boldsymbol{\alpha}}_1\eta_1|\Phi_{\nopair}\rangle \ .
  \label{gket}
 \end{align}
 In the last equality, ${\cal{Q}}_{++}$
 could be replaced by ${\cal{Q}}={\cal{Q}}_{++}+{\cal{Q}}_{--}$ due to the positive energy projected nature of $|\Phi_\nopair\rangle$.

 The action of the $\mathcal{S}_\text{inter}^{(1..n-1)}$ instantaneous ladder on this $|g\rangle$ can be evaluated as in Sec.~\ref{sec:IntroResum}
 after realizing that
 $\eta_\epsilon$ has no effect on $|g\rangle${{}:}
\begin{align}
  {{\cal{S}}^{(1..n-1)}_\text{inter}}(\varepsilon,\omega)|g\rangle
  &=
  \left[%
    K_\inst S_1(\varepsilon) S_2(\omega-\varepsilon)
  \right]^{n-1}|g\rangle \nonumber \\ 
  &=
  \left[%
    V_\inst {\cal{Q}}(\omega)
  \right]^{n-1}|g\rangle
 \ .
  \label{gketaction}
 \end{align}
 The result does not depend on $\varepsilon$, so the final, leftmost factor of the $n$-th
 energy term can be evaluated independently:
\begin{align}
\langle g'|
&=
\int_{-\infty}^{+\infty}\frac{\mathrm{d}\varepsilon}{-2\pi \iim} \langle\Phi_{\nopair}|(S^{(+)}_1(\varepsilon)+S^{(+)}_2(-\varepsilon))
 \eta_2^{\dagger}
\tilde{\boldsymbol{\alpha}}_2S_2^{(+)}(\omega-\varepsilon)
\nonumber \\
&=
\langle\Phi_{\nopair}|\eta_2^{\dagger}\tilde{\boldsymbol{\alpha}}_2
{\cal{Q}}_{++}(\omega) \nonumber \\
&=
\langle\Phi_{\nopair}|\eta_2^\dagger\tilde{\boldsymbol{\alpha}}_2
{\cal{Q}}(\omega) \ ,
\label{gbra}
\end{align}
Combining Eqs.~(\ref{gket}), (\ref{gketaction}) and (\ref{gbra}), we are left with
 \begin{equation}
 \Delta E^{(1)}_{{\text{T}};n}=
  \int\mathrm{d}^3\bk\int_{-\infty}^{+\infty}\frac{\mathrm{d}\omega}{-2\pi \iim}
  \langle\Phi_{\nopair}|
  \eta_2^{\dagger}\tilde{\boldsymbol{\alpha}}_2
  \kappa_{\text{T}}(\omega){\cal{Q}}(\omega)\left [V_{\inst}{\cal{Q}}(\omega)\right]^{n}
  \tilde{\boldsymbol{\alpha}}_1\eta_1
  |\Phi_{\nopair}\rangle \ .
  \label{eq:Tn1}
 \end{equation}
 To have the complete $n\geq1$ Coulomb photon contribution,
 the process of emission from electron $2$ also must be taken
 into account ($n=2$ example in the right subfigure of Fig.~\ref{fig:2instantaneouspic}).
 The result is the same as in Eq.~\eqref{eq:Tn1} with the
 $\tilde{\boldsymbol{\alpha}}_1(\bk)\leftrightarrow
 \tilde{\boldsymbol{\alpha}}_2(\bk)$,
 $\eta_1(\bk)\leftrightarrow\eta_2(\bk)$ replacements.
 The sum of the two contributions is
 \begin{align}
 \Delta E_{{\text{T}};n}=&
  \int\mathrm{d}^3\bk\int_{-\infty}^{+\infty}\frac{\mathrm{d}\omega}{-2\pi \iim}
  \langle\Phi_{\nopair}|
  \tilde{\boldsymbol{\alpha}}_2\eta_2^{\dagger}
  \kappa_{\text{T}}(\omega){\cal{Q}}(\omega)\left[V_{\inst}{\cal{Q}}(\omega)\right]^{n}
  \tilde{\boldsymbol{\alpha}}_1\eta_1
  |\Phi_{\nopair}\rangle 
  +
\ (1\leftrightarrow2)
   \label{poles}
  \ .
 \end{align}
 By adding the $n=0$ term, Eq. (\ref{singletransverse}), the ladder sum is obtained,
 \begin{align}
 \Delta E_{\text{T}}&=\sum_{n=0}^{\infty}\Delta E_{{\text{T}};n} \nonumber \\
  &=
  \int\mathrm{d}^3\bk\int_{-\infty}^{+\infty}\frac{\mathrm{d}\omega}{-2\pi \iim}
  \langle\Phi_{\nopair}|
  \tilde{\boldsymbol{\alpha}}_2\eta_2^\dagger
  \kappa_{\text{T}}(\omega){\cal{Q}}(\omega)\frac{1}{1-V_{\inst}{\cal{Q}}(\omega)}
  \tilde{\boldsymbol{\alpha}}_1\eta_1
  |\Phi_{\nopair}\rangle 
  + \ (1\leftrightarrow2)
  \ ,
  \label{ETsummed}
 \end{align}
 where the inner operator can be brought to the form (Appendix \ref{AppC})
 \begin{align}
  {\cal{Q}}(\omega)\frac{1}{1-V_{\inst}{\cal{Q}}(\omega)}
  &=
  \frac{L_{++}-L_{--}}{E_\nopair-h_1-h_2+\omega+\iim0^+(L_{++}-L_{--})}
  \frac{1}{1-V_{\inst}\frac{L_{++}-L_{--}}{E-h_1-h_2+\omega+\iim0^+(L_{++}-L_{--})}} \nonumber \\
  &=
  (L_{++}-L_{--})
  \frac{1}{E_\nopair-h_1-h_2+\omega-(V_{\inst}-\iim0^+)(L_{++}-L_{--})} \nonumber \\
  &=
  (L_{++}-L_{--})
  \frac{1}{E_\nopair-H^{\dagger}_{\text{wp}}+\omega+\iim0^+(L_{++}-L_{--})} \ .
  \label{generalQ}
 \end{align}
 In the last line, we recognized the adjoint of the with-pair Hamiltonian, Eq.~(\ref{eq:withpair}) (also Eq.~(\ref{eq:resumwp})). 
The $\omega$ integration can be formally performed by multiplying Eq.~(\ref{generalQ})
with
\begin{equation}
 1=\sum_n|\psi_n\rangle\langle\tilde{\phi}_n|
\end{equation}
from the right, where $\{|\psi_n\rangle\}$ are the right eigenvectors of the operator
\begin{equation}
 H_\Lambda=H_\text{wp}^{\dagger}-\iim0^+(L_{++}-L_{--})
\end{equation}
with eigenvalues $\lambda_n=\lambda_n'+\iim\lambda_n''\in\mathbb{C}$,
and $\{\langle\tilde{\phi}_n  |\}$ are elements of the dual space satisfying $\langle\tilde{\phi}_m|\psi_n\rangle=\delta_{mn}$ (left eigenvectors biorthogonalized).
Eq. (\ref{generalQ}) can then be split into two parts based on the sign of $\lambda_n''$:
\begin{equation}
 {\cal{Q}}(\omega)\frac{1}{1-V_{\inst}{\cal{Q}}(\omega)}=
 \left[
 \sum_{\substack{n \\ \lambda''_n>0}}
 \frac{L_{++}-L_{--}}{E_{\text{np}}+\omega-\lambda_n'-\iim\lambda_n''}
 +
 \sum_{\substack{n \\ \lambda_n''<0}}
 \frac{L_{++}-L_{--}}{E_{\text{np}}+\omega-\lambda_n'+\iim|\lambda_n''|}
 \right]|\psi_n\rangle\langle\tilde{\phi}_n|
 \ ,
\end{equation}
and the $\omega$ integration is then easily performed with Eq. (\ref{eq:kappaT}):
 \begin{equation}
 \int_{-\infty}^{+\infty}\frac{\mathrm{d}\omega}{-2\pi \iim}
 \kappa_{\text{T}}(\omega)
 {\cal{Q}}(\omega)\frac{1}{1-V_{\inst}{\cal{Q}}(\omega)}
 =\frac{L_{++}-L_{--}}{2k}
 \left[
 \frac{1}{E_\nopair+k-H_\Lambda}{\cal{P}}_{+}+
 \frac{1}{E_\nopair-k-H_\Lambda}{\cal{P}}_{-}
 \right] \ ,
 \label{combinedeff}
 \end{equation}
 with the`pseudo-projectors'
 \begin{equation}
  {\cal{P}}_{\mp}=
  \sum_{\substack{n \\ \lambda_n''\lessgtr0}}|\psi_n\rangle\langle\tilde{\phi}_n| \ .
  \label{pseudoproj}
 \end{equation}
 The above expressions make sense with
 an appropriate limiting procedure
 (that is, treating $H_\Lambda=H_\text{wp}^{\dagger}-\iim\delta(L_{++}-L_{--})$, for some finite $\delta>0$, and eventually letting $\delta\rightarrow0^+$).

 Eq. (\ref{combinedeff}) formally contains the summed effect of positive- and negative-energy intermediate states between instantaneous interactions.
 In this work, we continue only with the positive-energy ladder, and $L_{--}$ will be neglected in
 Eqs.~(\ref{generalQ})--(\ref{pseudoproj}).
 A more transparent way to arrive
 at the final result without $L_{--}$ is to turn to Eq. (\ref{generalQ}) again:
 \begin{align}
  {\cal{Q}}_{++}(\omega)\frac{1}{1-V_{\inst}{\cal{Q}}_{++}(\omega)}
  &=
  L_{++}
  \frac{1}{E_\nopair-h_1-h_2-V_\inst L_{++}+\omega+\iim0^+L_{++}}  \nonumber \\
  &=
  \frac{L_{++}}{E_\nopair-h_1-h_2-L_{++}V_\inst L_{++}+\omega+\iim0^+} \nonumber \\
  &=
  \frac{L_{++}}{E_\nopair-H_\nopair+\omega+\iim0^+} \ .
 \end{align}
 The equivalence of the first and second lines can be simply checked by expanding the denominator.
 Inserting this result in Eq. (\ref{ETsummed})
 and
 using the decomposition of $\kappa_{\tT}$ from Eq. (\ref{eq:kappaT}) leaves us with an even simpler $\omega$ integral, leading to
 \begin{align}
  \Delta E_{\text{T}++}&=
  \frac{z_1z_2\alpha}{2\pi^2}
  \int\mathrm{d}^3\bk\frac{1}{2{{\kabs}}}
  \langle\Phi_{\nopair}|
  \tilde{\boldsymbol{\alpha}}_2(\bk)\eta_2(-\bk)
  \frac{L_{++}}{E_{\nopair}-H_{\nopair}-{{\kabs}}+\iim0^+}
  \tilde{\boldsymbol{\alpha}}_1(\bk)\eta_1(\bk)
  |\Phi_{\nopair}\rangle 
  + \ (1\leftrightarrow2)  \ .  
  \label{transversepart}
 \end{align} 
 One must be careful if a no-pair DCB reference is used.
 The $n=0$ term in the expansion of Eq.~(\ref{eq:ET++})
 corresponds to the effect of a single transverse
 photon without any Coulomb or Breit photons, and
 its instantaneous part
 is already included in $E_{\text{DCB}}$;
 it must be subtracted from $\Delta E_{{{}\text{T}}++}$
 in order to avoid double counting. 
 
 The final one-photon exchange corrections for cases
 $\text{np}=\text{DC}$ and $\text{np}=\text{DCB}$ are thus
 \begin{align}
  \Delta E_{\text{T}++}^{\text{(DC)}}
  &=
  \frac{z_1z_2\alpha}{2\pi^2}
  \int\mathrm{d}^3\bk \frac{1}{2{\kabs}}
    \langle\Phi_{\text{DC}}|
    \tilde{\boldsymbol{\alpha}}_2(\bk)\eta_2(-\bk)
    %{L}_{++}
    \frac{{L}_{++}}{E_{\text{DC}}-H_{\text{DC}}-{{\kabs}}+\iim0^+}
    %{L}_{++}
    \tilde{\boldsymbol{\alpha}}_1(\bk)\eta_1(\bk)
  |\Phi_{\text{DC}}\rangle + \ (1\leftrightarrow2)
  \label{eq:trcorr}
 \end{align}
 \begin{align}
  \Delta E'_{\text{T}++} 
  &=
  \Delta E^{\text{(DCB)}}_{\text{T}++}-\langle\Phi_{\text{DCB}}|V_\text{B}|\Phi_{\text{DCB}}\rangle
  \nonumber
  \\
  &=
  \frac{z_1z_2\alpha}{2\pi^2}
  \int\mathrm{d}^3\bk 
  \langle\Phi_{\text{DCB}}|
  \tilde{\boldsymbol{\alpha}}_2(\bk)\eta_2(-\bk)
  \nonumber \\
  &\quad\quad%
  {L}_{++}
  \left[%
  \frac{1}{2{\kabs}}
  \frac{1}{E_{\text{DCB}}-H_{\text{DCB}}-{{\kabs}}+\iim0^+}
  +
  \frac{1}{2{\kabs}^2}
  \right]
  {L}_{++}
  \tilde{\boldsymbol{\alpha}}_1(\bk)\eta_1(\bk)
  |\Phi_{\text{DCB}}\rangle + (1\leftrightarrow 2) \nonumber \\
  &=
  \frac{z_1z_2\alpha}{2\pi^2}
  \int\mathrm{d}^3\bk 
  \frac{1}{2{\kabs}^2}
  \langle\Phi_{\text{DCB}}|
  \tilde{\boldsymbol{\alpha}}_2(\bk)\eta_2(-\bk)
  \nonumber \\
  &\quad\quad%
  {L}_{++} 
  \frac{E_{\text{DCB}}-H_{\text{DCB}}}{E_{\text{DCB}}-H_{\text{DCB}}-{{\kabs}}+\iim 0^+}  {L}_{++}
  \tilde{\boldsymbol{\alpha}}_1(\bk)\eta_1(\bk)
  |\Phi_{\text{DCB}}\rangle + (1\leftrightarrow 2) \; .
  \label{eq:retcorr}
 \end{align}

\subsection{One-loop self-energy \label{sec:SE}}

 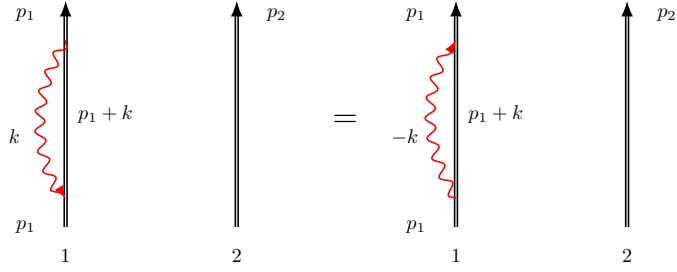
\begin{figure}
  \centering
  \quad\quad
  \scalebox{0.75}{%
    \begin{tikzpicture}
    \draw[boundelectron,-Latex] (0.,0.) -- (0.,4.) ;
    \draw[boundelectron,-Latex] (3.,0.) -- (3.,4.) ;
    \draw[decorate,decoration=snake,thick,red,-Latex]  (0.,3.3) to [bend right] (0.,0.5) ;
    \node (a0) at (0.,-0.5) {$1$};
    \node (a1) at (-0.7,3.75) {$p_1$};
    \node (a2) at (0.7,2.0) {$p_1+k$};
    \node (a4) at (-0.7,0.) {$p_1$};
    \node (b0) at (3.,-0.5) {$2$};
    \node (a1) at (3.7,3.75) {$p_2$};
    \node (c) at (-0.9,1.6) {$k$};
    %\node (TxC,a) at (1.5,-1.0) {$K^{(1)}_{\text{SE}\times 2\text{C};++}$};
    \end{tikzpicture} 
    \quad %
    \raisebox{2.5cm}{\mbox{\ \ \LARGE$=$}} %
    \quad %
    \begin{tikzpicture}
    \draw[boundelectron,-Latex] (0.,0.) -- (0.,4.) ;
    \draw[boundelectron,-Latex] (3.,0.) -- (3.,4.) ;
    \draw[decorate,decoration=snake,thick,red,-Latex] (0.,0.5) to [bend left] (0.,3.3);
    \node (a0) at (0.,-0.5) {$1$};
    \node (a1) at (-0.7,3.75) {$p_1$};
    \node (a2) at (0.7,2.0) {$p_1+k$};
    \node (a4) at (-0.7,0.) {$p_1$};
    \node (b0) at (3.,-0.5) {$2$};
    \node (a1) at (3.7,3.75) {$p_2$};
    \node (c) at (-0.9,1.6) {$-k$};
    \end{tikzpicture} 
    }    
  \caption{%
   The transverse part of the one-loop self-energy diagram. 
   \label{fig:oneSE}
  }
\end{figure} 

  Let us now turn to one-photon self-energy effects.
  In the Coulomb gauge, the one-loop self-energy can be partitioned to a Coulomb (C) and a transverse part (T),
  \begin{align}
    \Sigma_a
    =
    \Sigma^{\text{T}}_a+\Sigma^{\text{C}}_a \ .
  \end{align}
  Unlike in the case of photon exchange, the Coulomb part must be treated as well, since it is not included
  in the no-pair solution.
  The corresponding mathematical expressions
  follow from the diagram rules of Appendix \ref{App:diagram} when applied to a single particle
  (and noting that self-contracting interaction lines do not carry any $\eta$ shift operator).
  The transverse part reads (Fig. \ref{fig:oneSE})
  \begin{align}
   {\iim}  
   \Sigma^{\text{T}}_1(\bp_1,\varepsilon)
   &=
   \int\mathrm{d}^3\bk
   \int_{-\infty}^{+\infty}\frac{\mathrm{d}\omega}{-2\pi \iim}
     \tilde{\boldsymbol{\alpha}}_1(\bk)
     S_1(\bp_1+\bk,\varepsilon+\omega)
     \tilde{\boldsymbol{\alpha}}_1(\bk) 
     \kappa_\tT(\bk,\omega)
   \nonumber \\
   &=
   \int\mathrm{d}^3\bk
   \int_{-\infty}^{+\infty}\frac{\mathrm{d}\omega}{-2\pi \iim}
     \tilde{\boldsymbol{\alpha}}_1(\bk)
     \eta_1(-\bk)
     S_1(\bp_1,\varepsilon+\omega)
     \eta_1(\bk)
     \tilde{\boldsymbol{\alpha}}_1 (\bk)
     \kappa_\tT(\bk,\omega)
   \ ,
  \end{align}
  \begin{align}  
   {\iim}
   \Sigma^{\text{T}}_2(\bp_2,{-}\varepsilon)
   &=
   \int\mathrm{d}^3 \bk
   \int_{-\infty}^{+\infty}\frac{\mathrm{d}\omega}{-2\pi \iim}
     \tilde{\boldsymbol{\alpha}}_2(\bk)
     S_2(\bp_2+\bk,-\varepsilon+\omega)
     \tilde{\boldsymbol{\alpha}}_2(\bk)
     \kappa_\tT(\bk,\omega)
     \nonumber \\
   &=
   \int\mathrm{d}^3 \bk
   \int_{-\infty}^{+\infty}\frac{\mathrm{d}\omega}{-2\pi \iim}
     \tilde{\boldsymbol{\alpha}}_2(\bk)
     \eta_2(-\bk)
     S_2(\bp_2,-\varepsilon+\omega)
     \eta_2(\bk)
     \tilde{\boldsymbol{\alpha}}_2(\bk)
     \kappa_\tT(\bk,\omega)
   \ ,
  \end{align}
  whereas the Coulombic part is
  \begin{align}
   {\iim}  
   \Sigma^{\text{C}}_1(\bp_1,\varepsilon)
   &=
   \int\mathrm{d}^3\bk
   \int_{-\infty}^{+\infty}\frac{\mathrm{d}\omega}{-2\pi \iim}
   \eta_1(-\bk)
     S_1(\bp_1,\varepsilon+\omega)
     \eta_1(\bk)
     \kappa_\text{C}(\bk)
  \end{align}
  \begin{align}
   {\iim}  
   \Sigma^{\text{C}}_2(\bp_2,{-}\varepsilon)
   &=
   \int\mathrm{d}^3 \bk
   \int_{-\infty}^{+\infty}\frac{\mathrm{d}\omega}{-2\pi \iim}
    \eta_2(-\bk)
     S_2(\bp_2,-\varepsilon+\omega)
     \eta_2(\bk)
     \kappa_\text{C}(\bk) \; .
  \end{align}  
   Furthermore, let us partition both terms into positive- and negative-energy parts by
   using $S_a=S_a^{(+)}+S_a^{(-)}$: 
 \begin{align}
   {\iim}
   \Sigma^{{\text{T}}(\pm)}_1(\varepsilon)
   &=
   \alpha \frac{z_1^2}{2\pi^2}
   \int\mathrm{d}^3 \bk\ 
   \int_{-\infty}^{+\infty}\frac{\mathrm{d}\omega}{-2\pi \iim}
     \tilde{\boldsymbol{\alpha}}_1\eta_1^{\dagger}
     S^{(\pm)}_1(\varepsilon+\omega)
     \tilde{\boldsymbol{\alpha}}_1\eta_1
    \frac{1}{\omega^2-\bk^2+\iim 0^+} 
%    }
    \nonumber \\
   &=
   \alpha \frac{z_1^2}{2\pi^2}
   \int\mathrm{d}^3\bk\ \frac{1}{2{{\kabs}}}
     \tilde{\boldsymbol{\alpha}}_1\eta_1^{\dagger}
     S^{(\pm)}_1(\varepsilon\mp{{\kabs}})
     \tilde{\boldsymbol{\alpha}}_1\eta_1 \ ,
  \end{align}
  and similarly for particle 2,
  \begin{align}
   {\iim}  
   \Sigma^{{\text{T}}(\pm)}_2({-}\varepsilon)
   &=
   \alpha\frac{z_2^2}{2\pi^2}
   \int\mathrm{d}^3 \bk\ 
   \int_{-\infty}^{+\infty}\frac{\mathrm{d}\omega}{-2\pi \iim}
     \tilde{\boldsymbol{\alpha}}_2\eta_2^{\dagger}
     S^{(\pm)}_2(-\varepsilon+\omega)
     \tilde{\boldsymbol{\alpha}}_2\eta_2
     \frac{1}{\omega^2-\bk^2+\iim 0^+}
   \nonumber \\   
   &=
   \alpha \frac{z_2^2}{2\pi^2}
   \int\mathrm{d}^3\bk\ \frac{1}{2{{\kabs}}}
     \tilde{\boldsymbol{\alpha}}_2\eta_2^\dagger
     S^{(\pm)}_2(-\varepsilon\mp{{\kabs}})
     \tilde{\boldsymbol{\alpha}}_2\eta_2 \ .
  \end{align}
  A similar $(\pm)$ partitioning is applied to $\Sigma^{\text{C}}_a$, but the frequency integration requires special care due to the singular distributions, which is discussed in Appendix~\ref{app:PoBe}.

  The first-order perturbative energy correction, Eq.~\eqref{eq:PT1full}, is obtained by considering the second and third terms in the $\tilde{K}$ effective interaction kernel, Eq.~\eqref{eq:Keff} (the last term can be dropped at the one-loop level),  
\begin{align}
  &E'_\text{SE} \nonumber \\
  &=
  \int_{-\infty}^{+\infty}\frac{\dd\epsi}{-2\pi\iim}
    \langle \Phi_\nopair | 
      (E_\nopair-h_1-h_2) S_1 S_2 (S_1^{-1} {\iim}\Sigma_2 + {\iim}\Sigma_1 S_2^{-1}) S_1 S_2 V_\inst 
    \Phi_\nopair \rangle
    \nonumber \\
  &=
  \int_{-\infty}^{+\infty}\frac{\dd\epsi}{-2\pi\iim}
    \langle \Phi_\nopair | 
      (E_\nopair-h_1-h_2) \left[%
      S_2 {\iim}\Sigma_2 
      +
      S_1 {\iim}\Sigma_1 
      \right] S_1 S_2 V_\inst
    \Phi_\nopair \rangle
    \nonumber \\
  &=
  \int_{-\infty}^{+\infty}\frac{\dd\epsi}{-2\pi\iim}
    \langle \Phi_\nopair | 
      D
      \left[%
      S_2 {\iim}\Sigma_2 
      +
      S_1 {\iim}\Sigma_1 
      \right] (S_1+S_2) (E-h_1-h_2)^{-1} V_\inst
    \Phi_\nopair \rangle
    \nonumber \\
  &\approx
  \int_{-\infty}^{+\infty}\frac{\dd\epsi}{-2\pi\iim}
    \langle \Phi_\nopair | 
      D
      \left[%
      S_2 {\iim}\Sigma_2 
      +
      S_1 {\iim}\Sigma_1 
      \right] 
      \left[%
        S_1^{(+)}+S_2^{(+)}
      \right]
    \Phi_\nopair \rangle
    \nonumber \\
  &=
  \int_{-\infty}^{+\infty}\frac{\dd\epsi}{-2\pi\iim}
    \langle \Phi_\nopair | 
      D
      \left[%
      S^{(+)}_2 {\iim}\Sigma_2 
      +
      S^{(+)}_1 {\iim}\Sigma_1 
      \right] 
      \left[%
        S^{(+)}_1 + S^{(+)}_2
      \right]
    \Phi_\nopair \rangle
    \nonumber \\
  &=
    \langle \Phi_\nopair | 
      \int_{-\infty}^{+\infty}\frac{\dd\epsi}{-2\pi\iim}
      D
      \left[%
      S^{(+)}_2 {\iim}\Sigma^{(+)}_2 S^{(+)}_1 + S^{(+)}_2 {\iim}\Sigma^{(+)}_2 S^{(+)}_2 
      +
      S^{(+)}_1 {\iim}\Sigma^{(+)}_1 S^{(+)}_1 + S^{(+)}_1 {\iim}\Sigma^{(+)}_1 S^{(+)}_2
      \right] 
    \Phi_\nopair \rangle
    \nonumber \\
  &\ +
    \langle \Phi_\nopair | 
      \int_{-\infty}^{+\infty}\frac{\dd\epsi}{-2\pi\iim}
      D
      \left[%
      S^{(+)}_2 {\iim}\Sigma^{(-)}_2 S^{(+)}_1 + S^{(+)}_2 {\iim}\Sigma^{(-)}_2 S^{(+)}_2 
      +
      S^{(+)}_1 {\iim}\Sigma^{(-)}_1 S^{(+)}_1 + S^{(+)}_1 {\iim}\Sigma^{(-)}_1 S^{(+)}_2
      \right] 
    \Phi_\nopair \rangle    
    \nonumber \\
  &=
    \langle \Phi_\nopair | 
      D    
      \int_{-\infty}^{+\infty}\frac{\dd\epsi}{-2\pi\iim}
      S^{(+)}_1 S^{(+)}_2 
      \left[%
      {\iim}\Sigma_1 + {\iim}\Sigma_2
      \right] 
    \Phi_\nopair \rangle
    +
    \langle \Phi_\nopair | 
      D
      \int_{-\infty}^{+\infty}\frac{\dd\epsi}{-2\pi\iim}
      \left[%
      S^{(+)}_2 {\iim}\Sigma^{(-)}_2 S^{(+)}_2 
      +
      S^{(+)}_1 {\iim}\Sigma^{(-)}_1 S^{(+)}_1 
      \right] 
    \Phi_\nopair \rangle    
    \nonumber \\    
  &=
    \langle \Phi_\nopair | 
      \int_{-\infty}^{+\infty}\frac{\dd\epsi}{-2\pi\iim}
      [S^{(+)}_1 + S^{(+)}_2]
      \left[%
      {\iim}\Sigma_1 + {\iim}\Sigma_2
      \right] 
    | \Phi \rangle
    +
    \langle \Phi_\nopair | 
      D
      \int_{-\infty}^{+\infty}\frac{\dd\epsi}{-2\pi\iim}
      \left[%
      S^{(+)}_2 {\iim}\Sigma^{(-)}_2 S^{(+)}_2 
      +
      S^{(+)}_1 {\iim}\Sigma^{(-)}_1 S^{(+)}_1 
      \right] 
    | \Phi_\nopair \rangle    
    \nonumber \\
  &=
    \langle \Phi_\nopair | 
      \int_{-\infty}^{+\infty}\frac{\dd\epsi}{-2\pi\iim}
      \left[%
        S^{(+)}_1 {\iim}\Sigma^{(+)}_2 + S^{(+)}_2 {\iim}\Sigma^{(+)}_1
       +S^{(+)}_1 {\iim}\Sigma^{(-)}_1 + S^{(+)}_2 {\iim}\Sigma^{(-)}_2
      \right]
    | \Phi_\nopair \rangle
    \nonumber \\
    &\quad +
    \langle \Phi_\nopair | 
      (E_\nopair-h_1-h_2)
      \int_{-\infty}^{+\infty}\frac{\dd\epsi}{-2\pi\iim}
      \left[%
      S^{(+)}_2 {\iim}\Sigma^{(-)}_2 S^{(+)}_2 
      +
      S^{(+)}_1 {\iim}\Sigma^{(-)}_1 S^{(+)}_1 
      \right] 
    | \Phi_\nopair \rangle  \ ,  
    \label{eq:SEpt1}
\end{align}
where we dropped terms that give zero contribution to the $\epsi$ integrals (due to having all the poles on the same {{}complex} half-plane), and introduced the shorthand $D=E_\nopair-h_1-h_2$ notation in the intermediate steps.
The expression can be evaluated in an appropriate basis representation.

 In what follows, we focus on the $\Sigma_a^{(+)}$ positive-energy terms, which give the dominant contribution  (at least for small-$Z$ systems) and can be further simplified.
 Then, the transverse part of the one-loop self-energy correction reads (using Eq.~\eqref{eq:S1S2int}, Appendix \ref{AppC})
 \begin{align}
  E^{\text{T}(+)}_{{\text{SE}}}
  =&
  \int_{-\infty}^{+\infty}\frac{\mathrm{d}\varepsilon}{-2\pi \iim}
  \langle\Phi_{\nopair}|
  S_1^{(+)}(\varepsilon){\iim}\Sigma_2^{{\text{T}}(+)}(-\varepsilon)+{\iim}\Sigma_1^{{\text{T}}(+)}(\varepsilon)S_2^{(+)}(-\varepsilon)
  |\Phi_{\nopair}\rangle
  \nonumber \\
  =&
\frac{z_1^2\alpha}{2\pi^2}
  \int\mathrm{d}^3\bk\frac{1}{2{{\kabs}}}
  \langle\Phi_{\nopair}|
  \tilde{\boldsymbol{\alpha}}_1(\bk)\eta_1(-\bk)
  \frac{L_{1,+}}{E_\nopair-{h}_1-{h}_2-{\kabs}+\iim0^+}
  \tilde{\boldsymbol{\alpha}}_1(\bk)\eta_1(\bk)
  |\Phi_{\nopair}\rangle %\nonumber \\
  + \ (1\leftrightarrow2)
  \ .
  \label{SE0_PES}
 \end{align}  
Similarly, for the Coulomb part (Appendix~\ref{app:PoBe}), 
\begin{align}
  E^{\text{C}(+)}_{{\text{SE}}}
  =&
  \int_{-\infty}^{+\infty}\frac{\mathrm{d}\varepsilon}{-2\pi \iim}
  \langle\Phi_{\nopair}|
  S_1^{(+)}(\varepsilon) {\iim}\Sigma_2^{{\text{C}}(+)}(-\varepsilon)
  +
  {\iim}\Sigma_1^{{\text{C}}(+)}(\varepsilon)S_2^{(+)}(-\varepsilon)
  |\Phi_{\nopair}\rangle
  \nonumber \\
  =&
\frac{z_1^2\alpha}{2\pi^2}
  \int\mathrm{d}^3\bk\frac{1}{{{\kabs}}^2}
  \int_{-\infty}^{+\infty}\frac{\mathrm{d}\omega}{-2\pi \iim}
  \langle\Phi_{\nopair}|
  \eta_1(-\bk)
  \frac{L_{1,+}}{E_{\nopair}-h_1-h_2+\omega+\iim0^+}
  \eta_1(\bk)
  |\Phi_{\nopair}\rangle 
  + \ (1\leftrightarrow2) \nonumber \\
  =&
\frac{z_1^2\alpha}{2\pi^2}
  \int\mathrm{d}^3\bk\frac{1}{2{{\kabs}}^2}
  \langle\Phi_{\nopair}|
  \eta_1(-\bk)
  L_{1,+}\eta_1(\bk)
  |\Phi_{\nopair}\rangle 
+ \ (1\leftrightarrow2)
  \ .
  \label{SECoulomb}
 \end{align}  

\paragraph{The special case of a non-interacting reference}
Based on the structure of the diagrams (Fig.~\ref{fig:oneSE}), one may expect
that the self-energy correction is an inherent one-body quantity;
the previous equations, Eqs. (\ref{SE0_PES}) and (\ref{SECoulomb}), however, defy this view, as both terms
have ${h}_1+{h}_2$ in the denominator,
giving the energy contribution a two-body character.
This is a consequence of using an interacting two-particle state
as the zeroth-order solution.
In the relativistic QED literature, non-interactive reference states have been used, so we consider this choice as a special case and test of our formalism.
Let us consider the non-interacting approximation in Eqs. (\ref{phi0}) and (\ref{e0}) again. 
Our previous expressions can be simplified along the following lines:
\begin{align}
 &\frac{L_{1,+}}{E^{(0)}-{h}_1-{h}_2-{\kabs}+\iim0^+}
  \tilde{\boldsymbol{\alpha}}_1(\bk)\eta_1(\bk)
  |\phi_p\phi_q\rangle \nonumber \\
  &=
  \frac{1}{\varepsilon_p-{h}_1-{\kabs}+\varepsilon_q-{h}_2+\iim0^+}
  \left[L_{1,+}
  \tilde{\boldsymbol{\alpha}}_1(\bk)\eta_1(\bk)|\phi_p\rangle\right]
  |\phi_q\rangle  \nonumber \\
  &=
  \frac{1}{\varepsilon_p-{h}_1-{\kabs}+\iim0^+}
  \left[L_{1,+}
  \tilde{\boldsymbol{\alpha}}_1(\bk)\eta_1(\bk)|\phi_p\rangle\right]
  |\phi_q\rangle  
  \ ,
\end{align}
and similarly for $|\phi_q\phi_p\rangle$. 
Then, the positive-energy, transverse part of the self-energy correction simplifies to
\begin{align}
  E^{\text{T}(+)}_{\text{SE}}
  =\ &
  \frac{z^2\alpha}{2\pi^2}
  \int\mathrm{d}^3\bk\ \frac{1}{2{{\kabs}}}
    \langle\phi_p|
      \tilde{\boldsymbol{\alpha}}(\bk)\eta(-\bk)
      \frac{L_{+}}{\varepsilon_p-{h}-{\kabs}+\iim0^+}
      \tilde{\boldsymbol{\alpha}}(\bk)\eta(\bk)
    |\phi_p\rangle % \nonumber \\
    + \ {{}(p\leftrightarrow q)}
\ ,
\end{align}
and a similar Coulombic part;
particle indices were dropped, since now 
we have explicitly one-body quantities, and there is no more ambiguity which operators act on which particles' Hilbert space.
Using the above non-interacting approximation, we can also proceed with the evaluation of the negative-energy contributions in Eq.~\eqref{eq:SEpt1}. 
In this special case, the last two terms of Eq.~\eqref{eq:SEpt1} vanish,
\begin{align}
   \langle \Phi^{(0)} | 
      (E^{(0)}-h_1-h_2)
      \int_{-\infty}^{+\infty}\frac{\dd\epsi}{-2\pi\iim}
      \left[%
      S^{(+)}_2 {\iim}\Sigma^{(-)}_2 S^{(+)}_2 
      +
      S^{(+)}_1 {\iim}\Sigma^{(-)}_1 S^{(+)}_1 
      \right] 
    | \Phi^{(0)} \rangle    
    = 
    0 \; ,
\end{align}
and we are left with
\begin{align}
  E^{\text{T}(-)}_{\text{SE}}
  =&
    \int_{-\infty}^{+\infty}\frac{\mathrm{d}\varepsilon}{-2\pi \iim}
    \langle\Phi^{(0)}|
    {\iim}\Sigma_1^{(-)}S_1^{(+)}+{\iim}\Sigma_2^{(-)}S_2^{(+)}
    |\Phi^{(0)}\rangle \nonumber \\
  =&
  \frac{1}{2}
    \int_{-\infty}^{+\infty}\frac{\mathrm{d}\varepsilon}{-2\pi \iim}
    \left[%
    \langle\phi_p| {\iim}\Sigma_1^{(-)}S_1^{(+)} |\phi_p\rangle
    +
    \langle\phi_q| {\iim}\Sigma_1^{(-)}S_1^{(+)} |\phi_q\rangle
    +    
    \langle\phi_p| {\iim}\Sigma_2^{(-)}S_2^{(+)} |\phi_p\rangle
    +
    \langle\phi_q| {\iim}\Sigma_2^{(-)}S_2^{(+)} |\phi_q\rangle
    \right]
   \ .  
\end{align}
We write out the evaluation of the first term in detail; the other three are calculated similarly:
\begin{equation}
   S_1^{(+)}(\varepsilon)|\phi_p\rangle
   =
   \frac{1}{\frac{1}{2}[\varepsilon_q-\varepsilon_p]+\varepsilon+\iim 0^+}|\phi_p\rangle \ ,
\end{equation}
and 
\begin{align}
  \int_{-\infty}^{+\infty}\frac{\mathrm{d}\varepsilon}{-2\pi \iim}
    \frac{L_{1,-}}{\frac{1}{2}[\varepsilon_p+\varepsilon_q]+\varepsilon+{\kabs}+|{h}_1|-\iim 0^+}
    \frac{1}{\frac{1}{2}[\varepsilon_q-\varepsilon_p]+\varepsilon+\iim 0^+}
  =
  \frac{L_{1,-}}{\varepsilon_p-{h}_1+{\kabs}-\iim0^+} \ .
\end{align}
So, as a result, we obtain for the transverse part 
\begin{align}
 E^{\text{T}(-)}_{\text{SE}}
 =&\ 
 \frac{z^2\alpha}{2\pi^2}
  \int\mathrm{d}^3\bk\frac{1}{2{{\kabs}}}
    \langle\phi_p|
      \tilde{\boldsymbol{\alpha}}(\bk)\eta(-\bk)
      \frac{L_{-}}{\varepsilon_p-{h}+{\kabs}-\iim0^+}
      \tilde{\boldsymbol{\alpha}}(\bk)\eta(\bk)
    |\phi_p\rangle 
 + \ {{}(p\leftrightarrow q)}
%  \nonumber \\
%  &
%  +  \frac{z^2\alpha}{2\pi^2}
%  \int\mathrm{d}^3\bk\frac{1}{2{{\kabs}}}
%    \langle\phi_q|
%      \tilde{\boldsymbol{\alpha}}(\bk)\eta(-\bk)
%      \frac{L_{-}}{\varepsilon_q-{h}+{\kabs}-\iim0^+}
%      \tilde{\boldsymbol{\alpha}}(\bk)\eta(\bk)
%    |\phi_q\rangle \ .
  \label{SE0_NES}
\end{align} 
Hence, for a non-interacting reference, 
the sum of  the positive- and negative-energy transverse self-energy contributions is
\begin{align}
 {{}E^{\text{T}}_{\text{SE}}}=&
  E^{\text{T}(+)}_{\text{SE}}
  +
  E^{\text{T}(-)}_{\text{SE}} \nonumber \\
  =&\ 
  \frac{z^2\alpha}{2\pi^2}
  \int\mathrm{d}^3\bk\ \frac{1}{2{{\kabs}}}
    \langle\phi_p|
      \tilde{\boldsymbol{\alpha}}(\bk)\eta(-\bk)
      \frac{1}{\varepsilon_p-{h}-({\kabs}-\iim0^+)(L_+-L_-)}
      \tilde{\boldsymbol{\alpha}}(\bk)\eta(\bk)
    |\phi_p\rangle
    {{} + \ (p\leftrightarrow q)}
    \nonumber \\
% +&
%  \frac{z^2\alpha}{2\pi^2}
%  \int\mathrm{d}^3\bk\ \frac{1}{2{{\kabs}}}
%    \langle\phi_q|
%      \tilde{\boldsymbol{\alpha}}(\bk)\eta(-\bk)
%      \frac{1}{\varepsilon_q-{h}-({\kabs}-\iim0^+)(L_+-L_-)}
%      \tilde{\boldsymbol{\alpha}}(\bk)\eta(\bk)
%    |\phi_q\rangle 
  \label{eq:SEnointTr}    
\end{align}
Similarly, we obtain for the Coulomb part, 
\begin{align}
 {{}E^{\text{C}}_{\text{SE}}}=&
 E^{\text{C}(+)}_{\text{SE}}
 +
 E^{\text{C}(-)}_{\text{SE}} \nonumber \\ =
 &\ 
 \frac{z^2\alpha}{2\pi^2}
  \int\mathrm{d}^3\bk\ \frac{1}{{{\kabs}}^2}
    \langle\phi_p|
      \eta(-\bk)
      \frac{L_{+}-L_-}{2}
      \eta(\bk)
    |\phi_p\rangle 
    {{} + \ (p\leftrightarrow q)}
%    \nonumber \\
%+& \frac{z^2\alpha}{2\pi^2}
%  \int\mathrm{d}^3\bk\frac{1}{{{\kabs}}^2}
%    \langle\phi_q|
%      \eta(-\bk)
%      \frac{L_{+}-L_-}{2}
%      \eta(\bk)
%    |\phi_q\rangle \ .
\label{eq:SEnointC}
\end{align} 
Eqs.~\eqref{eq:SEnointTr} and \eqref{eq:SEnointC} recover the results known from the independent-particle approaches to bound state QED \cite{LiPeSaYn93,PeLiSa93,lindgrenRelativisticManyBodyTheory2011,GrQu22}.

\subsection{Resummation for the instantaneous interaction ladder in the one-loop self-energy correction \label{sec:SEresum}}
 \noindent
 Instantaneous interactions can be summed for the self-energy in a similar fashion to
 the case of photon exchange (Fig.~\ref{fig:+-sumSE}).
 A formula similar to the $K_{\text{T},n}$ in Eq.~\eqref{transversepart} is obtained for the self-energy contribution
 with either 
 $\tilde{\boldsymbol{\alpha}}_1\eta_1(\bk)$ or
 $\tilde{\boldsymbol{\alpha}}_2\eta_2(\bk)$
 on both sides.
    \begin{figure}
    \vspace{-0.175cm}
    \scalebox{0.75}{%
    \begin{tikzpicture}
         \node (T,a) at (3.5,2.0) {\mbox{\LARGE$K^{(1)}_{\text{SE,sum}}=$}};
    \draw[boundelectron,-Latex] (5.5,0.) -- (5.5,4.) ;
    \draw[boundelectron,-Latex] (7.5,0.) -- (7.5,4.) ;
    \draw[decorate,decoration=snake,thick,red]  (5.5,3.3) to [bend right] (5.5,0.5) ;
    {\color{black}
         \node (T,a) at (8.0,2.0) {\mbox{\LARGE$+$}};
    \draw[boundelectron,-Latex] (9.0,0.) -- (9.0,4.) ;
    \draw[boundelectron,-Latex] (11.,0.) -- (11.,4.) ;
    \draw[coulomb, segment aspect=0,thick] (11.,2.0) -- (9.,2.0) ;
    \draw[decorate,decoration=snake,thick,red]  (9.,3.3) to [bend right] (9.,0.5) ;
         \node (T,a) at (11.5,2.0) {\mbox{\LARGE$+$}};
    \draw[boundelectron,-Latex] (12.5,0.) -- (12.5,4.) ;
    \draw[boundelectron,-Latex] (14.5,0.) -- (14.5,4.) ;
    \draw[coulomb, segment aspect=0,thick] (14.5,1.5) -- (12.5,1.5) ;
    \draw[coulomb, segment aspect=0,thick] (14.5,2.5) -- (12.5,2.5) ;
    \draw[decorate,decoration=snake,thick,red]  (12.5,3.3) to [bend right] (12.5,0.5) ;
    \node (T,a) at (15.0,2.0) {\mbox{\LARGE$ \ \ + \ ...$}};
    \node (T,a) at (16.0,2.0) {\mbox{\ \ \LARGE$+$}};
    \draw[boundelectron,-Latex] (17.0,0.) -- (17.0,2.5) -- (17.5,1.5) -- (17.5,4.) ;
    \draw[boundelectron,-Latex] (19.0,0.) -- (19.0,2.5) -- (19.5,1.5) -- (19.5,4.) ;
    \draw[coulomb,thick] (17.0,2.5) -- (19.0,2.5) ;
    \draw[coulomb,thick] (17.5,1.5) -- (19.5,1.5) ;
    \draw[decorate,decoration=snake,thick,red]  (17.0,0.4) to [bend left] (17.5,3.6) ;
    \node (T,a) at (20.3,2.0) {\mbox{\ \ \LARGE$+ \  ...$}};
    }
        \end{tikzpicture} 
    \quad\quad\quad
   } 
  \caption{One-loop self-energy on the background of an infinite number of instantaneous interactions involving both positive (second and third diagrams) and negative (last diagram) intermediate states \label{fig:+-sumSE}}
\end{figure}
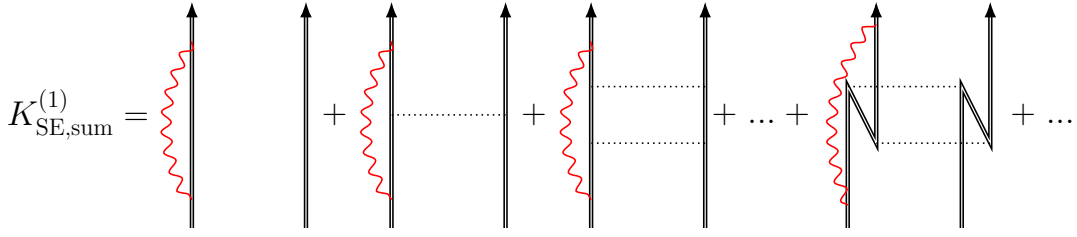
 \begin{figure}
  \centering
  \quad\quad
  \scalebox{0.75}{%
    \begin{tikzpicture}
    \draw[boundelectron,-Latex] (0.,0.) -- (0.,4.) ;
    \draw[boundelectron,-Latex] (3.,0.) -- (3.,4.) ;
\draw[decorate,decoration=snake,thick,red,-Latex]  (0.,3.3) to [bend right] (0.,0.5) ;
    \draw[coulomb,thick,-Latex] (3.,1.5) -- (0.,1.5) ;
    \draw[coulomb,thick,-Latex] (3.,2.5) -- (0.,2.5) ;
    \node (a0) at (0.,-0.5) {$1$};
    \node (a1) at (-0.7,3.75) {$p_1$};
    \node (a2) at (-1.8,2.8) {$p_1+k$};
    \node (a2) at (-2.0,2.0) {$p_1-k_1+k$};
    \node (a3) at (-2.0,1.) {$p_1-k_1-k_2+k$};
    \node (a4) at (-2.,0.) {$p_1-k_1-k_2$};
    \node (b0) at (3.,-0.5) {$2$};
    \node (a1) at (3.7,3.75) {$p_2$};
    \node (a3) at (4.1,2.) {$p_2+k_1$};
    \node (a4) at (4.5,0.) {$p_2+k_1+k_2$};
    \node (c) at (-0.7,1.6) {$k$};
    \node (d) at (1.5,3.) {$k_1$};
    \node (e) at (1.5,1.) {$k_2$};
    \node (TxC,a) at (1.5,-1.0) {$K^{(1)}_{\text{SE}\times 2\text{C};++}$};
    \end{tikzpicture} 
    }    
  \caption{%
   An example of the self-energy contribution for $n=2$. 
  }
\end{figure}
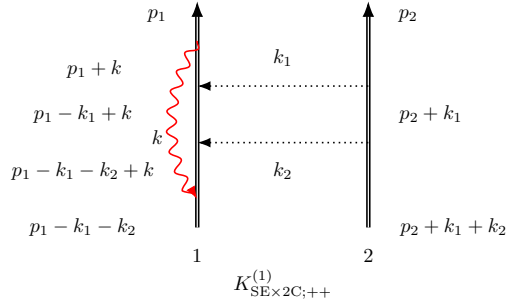 
The kernel $K^{(1)}_{\text{SE};n}$ can be constructed
similarly to that of the transverse photon, the only difference
being that the self-interacting transverse photon line does not invoke shift operators (see also Appendix \ref{App:diagram}).
Processing the diagram rules analogously to the 
case of the transverse photon, we find 
 \begin{align}
  K^{(1)}_{\text{SE};n}=&
  \int\mathrm{d}^3\bk\int_{-\infty}^{+\infty}\frac{\mathrm{d}\omega}{-2\pi \iim}
  \eta_1(-\bk)
  \tilde{\boldsymbol{\alpha}}_1(\bk)
  S_1(\varepsilon+\omega)
  {{\cal{S}}^{(1..n-1)'}_\text{inter}}
  \kappa_\text{T}(\bk,\omega)
  {K}_\inst
  S_1(\varepsilon+\omega)  
\tilde{\boldsymbol{\alpha}}_1(\bk) %\nonumber \\
  \eta_1(\bk)
  ,
  \label{SE++}
 \end{align}
\begin{align}
{{\cal{S}}^{(1..n-1)'}_\text{inter}}(\varepsilon,\omega)
 &=
\prod_{p=1}^{n-1}
\left[
\int\mathrm{d}^3\bk_p\int_{-\infty}^{+\infty}\frac{\mathrm{d}\omega_p}{-2\pi \iim}
 {\kappa_\inst(\bk_p)}
 T_\inst(\bk_p) 
\right. 
 \nonumber \\
 &
\left. 
 \left\lbrace
 \eta_1(\bk^{[p]})S_1(\varepsilon-\omega^{[p]}+\omega)
 \eta_1(-\bk^{[p]})
 \eta_2(-\bk^{[p]})
 S_2(-\varepsilon+\omega^{[p]})
 \eta_2(\bk^{[p]})
 \right\rbrace
 \right]
 \nonumber \\
 &\eta_1(\bk^{[n-1]})\eta_2(-\bk^{[n-1]})\eta_\epsilon(\omega^{[n-1]})
 \nonumber \\
 &=[K_\inst S_1(\varepsilon+\omega)S_2(-\varepsilon)]^{n-1} \nonumber \\
 &=
 \eta_\epsilon(-\omega){{\cal{S}}^{(1..n-1)}_\text{inter}}(\varepsilon,\omega)\eta_\epsilon(\omega)
 \ ,
\end{align}
where {${\cal{S}}^{(1..n-1)}_\text{inter}$} is the same as in Eq.~(\ref{eq:Sprod}).
Then, 
 \begin{align}
  K^{(1)}_{\text{SE};n}=&
  \int\mathrm{d}^3\bk\int_{-\infty}^{+\infty}\frac{\mathrm{d}\omega}{-2\pi \iim}
  \eta_\epsilon(-\omega)
  \eta_1^\dagger
  \tilde{\boldsymbol{\alpha}}_1
  S_1(\varepsilon)
  {{\cal{S}}^{(1..n-1)}_\text{inter}}(\varepsilon,\omega)
  \kappa_\text{T}(\omega)
  {K}_\inst
  S_1(\varepsilon)  
\tilde{\boldsymbol{\alpha}}_1 %\nonumber \\
  \eta_1\eta_\epsilon(\omega)
  ,
 \end{align}
all notations being the same as for the transverse photon term (Sec.~\ref{sec:Tresum}).
From this point, everything goes just like previously,
using Appendix~\ref{AppC}
 and $\langle\Phi_{\nopair}|\eta_\epsilon(-\omega)=\langle\Phi_{\nopair}|$,
leading to
 \begin{align}
  \Delta E_{\text{SE}}&=\sum_{n=0}^{\infty} \Delta E_{{\text{SE}};n} \nonumber \\
  &=
  \int{\mathrm{d}^3\bk}\int_{-\infty}^{+\infty}\frac{\mathrm{d}\omega}{-2\pi \iim}
  \langle\Phi_{\nopair}|
  \tilde{\boldsymbol{\alpha}}_1\eta_1^{\dagger}
  \kappa_{\text{T}}(\omega){\cal{Q}}(\omega)\frac{1}{1-V_{\inst}{\cal{Q}}(\omega)}
  \tilde{\boldsymbol{\alpha}}_1\eta_1
  |\Phi_{\nopair}\rangle
  + \ (1\leftrightarrow2)
  \ ,
  \label{SEsummed}
 \end{align}
The frequency integration could be again formally performed for combined positive and negative energy intermediate states, as was done in Eq. (\ref{combinedeff}).
Resorting to positive-energy intermediate states gives
\begin{align}
  \Delta E_{\text{SE}++}
  &=
  \int{\mathrm{d}^3\bk}\int_{-\infty}^{+\infty}\frac{\mathrm{d}\omega}{-2\pi \iim}
  \langle\Phi_{\nopair}|
  \tilde{\boldsymbol{\alpha}}_1\eta_1^{\dagger}
  \kappa_{\text{T}}(\omega){\cal{Q}}_{++}(\omega)\frac{1}{1-V_{\inst}{\cal{Q}}_{++}(\omega)}
  \tilde{\boldsymbol{\alpha}}_1\eta_1
  |\Phi_{\nopair}\rangle + \ (1\leftrightarrow2) \nonumber \\
  &=
  \frac{{z_1^2}\alpha}{2\pi^2}
  \int\mathrm{d}^3\bk\frac{1}{2{\kabs}}
  \langle\Phi_\nopair|
  \tilde{\boldsymbol{\alpha}}_1(\bk)\eta_1(-\bk)
  \frac{{L}_{++}}{E_{\nopair}-H_{\nopair}-{{\kabs}}+\iim0^+}
   \tilde{\boldsymbol{\alpha}}_1(\bk)\eta_1(\bk)
  |\Phi_\nopair\rangle + \ (1\leftrightarrow2) %\nonumber \\
  \ .  
  \label{SEpart}
\end{align}

\section{A unified treatment of transverse photon exchange and self-energy\label{ch:unified}}

\noindent
%In this section, atomic units are used, and factors of $c=1/\alpha$ are thus restored.
\newline
Let us specify our reference to be $\nopair=\text{DC}$ from now on,
with shorthand notation $\langle(...)\rangle=\langle\Phi_\text{DC}|(...)|\Phi_\text{DC}\rangle$.
The energy shifts associated with the transverse photon exchange and self-energy, Eqs.~(\ref{transversepart}) and (\ref{SEpart}), (\ref{SECoulomb}),
can be conveniently grouped together, 
\begin{align}
 \Delta E
 =&
 E^{\text{C}(+)}_{\text{SE}}+
 \frac{\alpha^3}{2\pi^2}
  \int\mathrm{d}^3\bk\frac{1}{2{\kabs}}
  \left\langle
    \boldsymbol{j}^{\dagger}(\bk)
    \frac{1}{E_{\text{DC}}-H_{\text{DC}}-{{\kabs}}+\iim0^+} 
    \boldsymbol{j}(\bk)
  \right\rangle  
   \nonumber \\
 =&
 E^{\text{C}(+)}_{\text{SE}}+\Delta E_{\text{SE}}^{0(+)}+\Delta E_{\text{T}}^{0(+)} \nonumber \\
 &+
 \frac{\alpha^3}{2\pi^2}
  \int\mathrm{d}^3\bk\frac{1}{2{\kabs}^2}\left\langle
    \boldsymbol{j}^{\dagger}(\bk)
    \left[%
    \frac{E_{\text{DC}}-H_{\text{DC}}}{E_{\text{DC}}-H_{\text{DC}}-{{\kabs}}+\iim0^+}
    \right]
    \boldsymbol{j}(\bk)
   \right\rangle 
  \ ,
  \label{eq:dEretinst}
\end{align}
where
\begin{equation}
 \boldsymbol{j}(\bk)=
 {\frac{1}{\alpha}}
 L_{++}
 \left[
 {z_1}\eta_1(\bk)\tilde{\boldsymbol{\alpha}}_1(\bk)+{z_2}\eta_2(\bk)\tilde{\boldsymbol{\alpha}}_2(\bk)
 \right]L_{++}
 \ ,
 \label{velocityop}
\end{equation}
and we used the simple identity
\begin{align}
 \frac{1}{2{\kabs}^2}\frac{{\kabs}}{E_{\nopair}-H_{\nopair}-{{\kabs}}+\iim0^+}
 &=
 -\frac{1}{2{\kabs}^2}
 +
 \frac{1}{2{\kabs}^2}
 \left[% 
   \frac{{\kabs}}{E_{\nopair}-H_{\nopair}-{{\kabs}}+\iim0^+} + 1
 \right] \nonumber
 \\
 &=
 -\frac{1}{2{\kabs}^2}+\frac{1}{2{\kabs}^2}\frac{E_{\nopair}-H_{\nopair}}{E_{\nopair}-H_{\nopair}-{{\kabs}}+\iim0^+} 
 \label{simpleid}
\end{align}
to separate the terms $\Delta E_{\text{T}}^{0(+)}=\langle V_\text{B}\rangle$,
and
\begin{equation}
\Delta E_{\text{SE}}^{0(+)}=-
    \frac{z_1^2\alpha}{2\pi^2}
  \int\mathrm{d}^3\bk\frac{1}{2{{\kabs}^2}}
  \left\langle 
  \tilde{\boldsymbol{\alpha}}_1(\bk)\eta_1(-\bk)
  L_{++}
  \tilde{\boldsymbol{\alpha}}_1(\bk)\eta_1(\bk)
  \right\rangle  +  \ (1\leftrightarrow2) \ .
\end{equation}
Since we are interested in retardation effects in this paper, it would be natural to omit
the Breit term from now on
(in case of $\nopair=\text{DCB}$ this would be actually necessary in order to avoid double counting, as the Breit interaction
is already treated variationally in the zeroth-order solution).
Nevertheless, we keep the Breit term for the time being to simplify the developments of the next section.

The divergent term $\Delta E_{\text{SE}}^{0(+)}$ is to be treated in mass renormalization (see Secs.~\ref{ch:relBl} and \ref{ch:renorm}).
\newline
{The factor of $1/\alpha$ was introduced in Eq. (\ref{velocityop}) to make the the leading ${\cal{O}}(m\alpha^5)={\cal{O}}(\alpha^3E_h)$ scaling of the retardation+self-energy correction manifest whenever an $\alpha$ expansion is meaningful (in the non-relativistic limit, particularly); see Appendix \ref{app:AU} for further details.}

\subsection{Energy shift in the dipole approximation and a relativistic Bethe logarithm? \label{ch:relBl}}
\noindent
In the nrQED evaluation of the energy shift, Eq. (\ref{eq:dEretinst}) is split into high- and low-photon momentum parts. Different approximations are
possible in the different momentum ranges; the parts of the energy shift
are to be calculated using these approximations, with the obvious requirement
that the final result cannot depend on the arbitrary separation of the energy ranges.
The low-energy part of the shift is usually processed
in the non-relativistic approximation (understood in the Foldy--Wouthuysen sense, with the exact non-relativistic energy eigenstate) combined with the dipole approximation $\eta_a(\bk)=\eem^{\iim\bk\cdot\br_a}\approx1$, and hence
$j_i(\bk)\approx\delta^{\perp}_{il}(\bk)(\boldsymbol{j}_0)_{l}$, where
\begin{equation}
  \boldsymbol{j}_0=
  {\frac{1}{\alpha}}
  L_{++}\left[
  {z_1}{\boldsymbol{\alpha}}_1+{z_2}{\boldsymbol{\alpha}}_2\right]L_{++} \ .
 \label{eq:j0def}
\end{equation}
Then, we can easily integrate out the angular $\bk$ dependence,
\begin{align}
 \Delta E_K
 &=
  \frac{\alpha^3}{2\pi^2}
  \int_{k<K}\mathrm{d}^3\bk\frac{1}{2{\kabs}}
  \left\langle 
(\boldsymbol{j}_0)_i
\frac{1}
    {E_{\text{DC}}-H_{\text{DC}}-{{\kabs}}
     }
    (\boldsymbol{j}_0)_l
  \right\rangle \nonumber\delta_{il}^{\perp}(\bk) \\
 &=
  \frac{\alpha^3}{4\pi^2}
  \int_0^K\mathrm{d}{\kabs}\ {\kabs}
  \left\langle 
(\boldsymbol{j}_0)_i
\frac{1}
    {E_{\text{DC}}-H_{\text{DC}}-{{\kabs}}
     }
    (\boldsymbol{j}_0)_l
  \right\rangle
  \int\mathrm{d}{\Omega_{\bk}}
  \delta_{il}^{\perp}(\bk) \nonumber \\
  &=
  \frac{2\alpha^3}{3\pi }
  \int_0^K\mathrm{d}{\kabs}\ {\kabs}
  \left\langle 
\boldsymbol{j}_0
\frac{1}
    {E_{\text{DC}}-H_{\text{DC}}-{{\kabs}}
     }
    \boldsymbol{j}_0
  \right\rangle  \nonumber \\
  &=
  -\frac{2\alpha^3}{3\pi }
  \int_0^K\mathrm{d}{\kabs}\ {\kabs}J({\kabs})
  \ ,
  \label{Eorig0}
\end{align}
where
\begin{equation}
 J(\kabs)=\left\langle 
\boldsymbol{j}_0
\frac{1}
    {H_{\text{DC}}-E_{\text{DC}}+{{\kabs}}
     }
    \boldsymbol{j}_0
  \right\rangle \ .
\end{equation}
In these derivations, $|\Phi_{\text{DC}}\rangle$ is assumed to be the ground state of the no-pair DC Hamiltonian, and for this reason, the $\iim0^+$ prescription can be omitted.

Just like in the case of the analogous non-relativistic energy formula, $\Delta E_K$ has a linearly and logarithmically growing part as $K\rightarrow\infty$, as it can be seen from
\begin{align}
 \frac{{\kabs}}{H_\text{DC}-E_\text{DC}+{\kabs}} 
 =
 1-\frac{H_\text{DC}-E_\text{DC}}{{\kabs}}
 +\frac{1}{{\kabs}}\frac{(H_\text{DC}-E_\text{DC})^2}{H_\text{DC}-E_\text{DC}+{\kabs}} \ ,
\end{align}
\begin{align}
 kJ(k)
 =
 \left\langle 
\boldsymbol{j}_0
\cdot
    \boldsymbol{j}_0
  \right\rangle
 -\frac{\left\langle 
\boldsymbol{j}_0
(H_\text{DC}-E_\text{DC})
    \boldsymbol{j}_0
  \right\rangle}{{\kabs}}
 +\frac{1}{{\kabs}}\left\langle {{}\boldsymbol{j}_0}\frac{(H_\text{DC}-E_\text{DC})^2}{H_\text{DC}-E_\text{DC}+{\kabs}}{{}\boldsymbol{j}_0}\right\rangle \ ,
\end{align}
and it must be renormalized. 
It is interesting to note that the linear divergence originates from two sources corresponding to the 
\begin{align}
 \langle \boldsymbol{j}_0 \cdot \boldsymbol{j}_0 \rangle=&
 {\frac{1}{\alpha^2}\Big[}z_1^2\left\langle
 \balpha_1 L_{++}\balpha_1
 \right\rangle
 +z_2^2\left\langle
 \balpha_2 L_{++}\balpha_2
 \right\rangle
 %\right]
 +2z_1z_2
 \left\langle
 \balpha_1 L_{++}\balpha_2
 \right\rangle
 {\Big]}
 \label{eq:Clindet}
\end{align}
separation. The first two terms in Eq.~\eqref{eq:Clindet} are from the instantaneous part of the self-energy divergence, which can be renormalized by the dressed mass arguments of QED \cite{Dy49b}. The last term of Eq.~\eqref{eq:Clindet} is from the dipole approximation of the Breit interaction (in momentum space), which is certainly an unacceptable approximation and turns out to be divergent not surprisingly; hence its subtraction is mandatory. The Breit interaction is accounted for as a direct perturbative correction (without invoking the dipole approximation) or as part of the Dirac--Coulomb--Breit eigenvalue equation.
The only reason it is included here in the dipole approximation (just to be cancelled by a counterterm) is to facilitate the unified description of transverse photon exchange and self-energy.

We obtain a regulated energy correction by subtracting the divergent contributions,
\begin{align}
 \Delta E_K
 =&-\frac{2\alpha^3}{3\pi}
 \left[%
   \int_0^K\mathrm{d}{\kabs}\ {\kabs}J({\kabs})  
   -
   C_\text{lin} K
   -C_{\text{log}}\left(\ln\left(\frac{K}{\Eh}\right)+\ln(2)\right)
 \right] \; ,
 \label{eq:relBlWF0}
\end{align}
\begin{equation}
 C_\text{lin}=
 \langle
 \boldsymbol{j}_0\cdot
 \boldsymbol{j}_0
 \rangle \ ,
 \label{eq:Clinrel}
\end{equation}
\begin{equation}
 C_\text{log}=-
 \langle
 \boldsymbol{j}_0
(H_\text{DC}-E_\text{DC})
 \boldsymbol{j}_0
 \rangle \ .
 \label{eq:Clogrel}
\end{equation}
where it is made explicit that our energy unit is $\Eh$ and the $\ln(2)$ term is included for traditional reasons; it is related to the cancellation of the logarithmic divergence by a similar term from the high-energy part \cite{GrReQEDBook09}.
The $K\rightarrow\infty$ limit can now be freely performed.

It is important to note that the Coulomb self-energy 
$E^{\text{C}(+)}_{\text{SE}}$ is omitted from Eq. (\ref{Eorig0}), since it
gives `only' an infinite but state-independent contribution in the dipole approximation, which can always be discarded by an appropriate shift of the zero energy (see Appendix~\ref{App:Coulombdiv}). It is thus sufficient to consider the transverse contributions only.

The $k$ integration can be performed, and the final result can be expressed in the traditional form
\begin{align}
 \Delta E
  &=
-\frac{2\alpha^3}{3\pi}\left\langle\boldsymbol{j}_0(H_{\text{DC}}-E_{\text{DC}})\boldsymbol{j}_0
  \right\rangle\ln(k_0)
  \ ,
  \label{Eorig}
\end{align}
in terms of 
\begin{align}
  \ln(k_0)
  &=
\frac{1}{\langle \boldsymbol{j}_0(H_\text{DC}-E_\text{DC}) \boldsymbol{j}_0 \rangle}
\lim_{K\rightarrow\infty}
\left[
\int_0^K\mathrm{d}{\kabs}\ {\kabs}J({\kabs})  
-\langle\boldsymbol{j}_0\cdot\boldsymbol{j}_0\rangle K
+\langle \boldsymbol{j}_0(H_\text{DC}-E_\text{DC}) \boldsymbol{j}_0 \rangle
\ln\left(\frac{2K}{E_h}\right)
\right]%+\ln(2) 
\nonumber \\
  &=
  \frac{%
    \left\langle %
      \boldsymbol{j}_0
      (H_\text{DC}-E_\text{DC})\ln\left(2\frac{|H_\text{DC}-E_\text{DC}|}{\Eh}\right)
      \boldsymbol{j}_0
    \right\rangle
  }{%
    \left\langle %
      \boldsymbol{j}_0(H_\text{DC}-E_\text{DC})\boldsymbol{j}_0
    \right\rangle
  }
  %+\ln(2)
  \ ,
  \label{relbl}
\end{align}
the relativistic Bethe logarithm,
in full analogy with its non-relativistic counterpart \cite{KaSa57,Sc61,Ko04,Ko12,FeMa23}.

\subsection{The relativistic Hylleraas functional \label{ch:relHyfun}}
\noindent
While simple and appealing, 
Eqs. (\ref{Eorig}) and (\ref{relbl})
are not suitable for the precise evaluation of the energy correction.
Using them would require knowledge of how the logarithm of $H_{\text{DC}}$ acts on our chosen basis, which is in general not possible. Using the eigenstates of $H_{\text{DC}}$ would circumvent this problem, but also introduce a new problem in the form of very slow convergence towards the complete basis limit due to difficulties associated with the accurate representation of excited states \cite{Sc61}.

Following the ideas of Schwartz about the non-relativistic Bethe-logarithm \cite{Sc61}, it is most convenient to take a step back, and use the regulated integral form of $\Delta E$ in Eq. (\ref{eq:relBlWF0}).
The calculation of $\Delta E$ then boils down to evaluating $J(k)$ accurately for a large (but finite) number of $k$ points,
and perform the ${\kabs}$ integration numerically. Further details about the numerical integration and counterterm subtraction will be given in Sec. \ref{sec:numerical}.

The most crucial point is the accurate
calculation of $J({\kabs})$ for each value of ${\kabs}$.
Similarly to the non-relativistic case \cite{Sc61}, it is possible to define a Hylleraas-like functional for the no-pair relativistic case, 
\begin{equation}
 {\cal{J}}_k[\bos{\Phi}]=
 \langle\bos{\Phi}|H_\text{DC}-E_\text{DC}+{\kabs}|\bos{\Phi}\rangle
 +\iim\langle\bos{\Phi}|\boldsymbol{j}_0|\Phi_\text{DC}\rangle
 -\iim\langle\Phi_\text{DC}|\boldsymbol{j}_0|\bos{\Phi}\rangle \ ,
\end{equation}
which is minimized to obtain auxiliary basis sets which ensure systematic improvement of $J({\kabs})$ for a selected ${\kabs}$ value.

Similarly to the non-relativistic case,  $|\bos{\Phi}\rangle$ has three Cartesian components.
For ground-state computations, $H_{\text{DC}}-E_{\text{DC}}+{\kabs}$ for $k\neq0$ is positive definite, meaning that $(H_{\text{DC}}-E_{\text{DC}}+{\kabs})^{1/2}$ is well-defined, and we can rewrite the functional as
\begin{align}
 {\cal{J}}_k[\boldsymbol{\Phi}]
 &=
 \langle\boldsymbol{\Psi}|\boldsymbol{\Psi}\rangle-J({\kabs})
 \ ,
\end{align}
where
\begin{align}
 |\boldsymbol{\Psi}\rangle
 &=(H_\text{DC}-E_\text{DC}+{\kabs})^{1/2}\left(|\boldsymbol{\Phi}\rangle
 +\iim\frac{1}{H_\text{DC}-E_\text{DC}+{\kabs}}\boldsymbol{j}_0|\Phi_\text{DC}\rangle\right) \ .
\end{align}
This shows that ${\cal{J}}_{{\kabs}}$ is bounded from below, and obtains its minimal value
upon $|\boldsymbol{\Psi}\rangle=|\boldsymbol{\emptyset}\rangle$, that is, when
\begin{equation}
 |\boldsymbol{\Phi}_{\text{min}}\rangle=-\iim
 \frac{1}{H_\text{DC}-E_\text{DC}+{\kabs}}
 \boldsymbol{j}_0|\Phi_\text{DC}\rangle \ .
\end{equation}
We then find $J({\kabs})=-{\cal{J}}_{{\kabs}}[\boldsymbol{\Phi}_{\text{min}}]=i\langle\Phi_\text{DC}|\boldsymbol{j}_0|\boldsymbol{\Phi}_{\text{min}}\rangle$.
To actually find $|\Phi_{\text{min},a}\rangle$ ($a=1,2,3$), we
solve 
\begin{equation}
 (H_\text{DC}-E_\text{DC}+{\kabs})|\boldsymbol{\Phi}_{\text{min}}\rangle=-\iim \boldsymbol{j}_0|\Phi_\text{DC}\rangle
 \label{lineq}
\end{equation}
by expanding $|\Phi_{\text{min},a}\rangle$ in an auxillary basis set $\{X|\Phi'_{\mu q a}\rangle\}$, 
analogously to the case of $|\Phi_\text{DC}\rangle$ (see Appendix \ref{app:DCDCB} for notations):
\begin{align}
 |\Phi_{\text{min},a}\rangle&=X\sum_{\mu=1}^{M'_{\text{b}}}
 \sum_{q=1}^{16}d_{\mu q,a}|\chi'_{\mu q,a}\rangle \ .
 \end{align}
Basis functions are obtained via stochastic optimization, and
coefficients are found by solving the finite basis approximation
of Eq. (\ref{lineq}):
\begin{equation}
 \boldsymbol{M}_a({\kabs})\boldsymbol{d}_a=-\boldsymbol{b}_a \ ,
\end{equation}
where
\begin{align}
 (\boldsymbol{M}_a)_{\mu q,\nu r}({\kabs})
 &=
 \langle{\chi}'_{\mu q,a}|X(H_\text{DC}-E_\text{DC}+{\kabs})X|{\chi}'_{\nu r,a}\rangle \ , \nonumber \\
 &=\tilde{H}'_{\mu q,\nu r,a}-(E_\text{DC}-{\kabs})S'_{\mu q,\nu r,a} \ ,
\end{align}
and
\begin{align}
 (\boldsymbol{b}_a)_{\mu q}
 &=\iim
 \langle{\chi}'_{\mu q,a}|Xj_a|{\Phi}_\text{DC}\rangle \nonumber \\
 &=
 \iim\sum_{\nu=1}^{M_\text{b}}\sum_{r=1}^{16}c_{\nu r}
 \langle{\chi}'_{\mu q,a}|X(\boldsymbol{j}_0)_aX|{\chi}_{\nu r}\rangle \ .
\end{align}
Then, $J({\kabs})$ is given by
\begin{equation}
 J({\kabs})=-\sum_{a=1}^{3}\boldsymbol{b}_a^{\dagger}\boldsymbol{d}_a \ .
 \label{Jcalc}
\end{equation}
The counterterms are evaluated in this basis as
\begin{equation}
 C_{\text{lin}}=
 \sum_{a=1}^{3}\boldsymbol{b}_a^{\dagger}\boldsymbol{S}_a^{-1}\boldsymbol{b}_a
 \ ,  
 \label{linsos}
 \end{equation}
 \begin{equation}
 C_{\text{log}}
 = \sum_{a=1}^{3}\boldsymbol{b}_a^{\dagger}\boldsymbol{S}_a^{-1}\boldsymbol{M}_a(0)\boldsymbol{S}_a^{-1}\boldsymbol{b}_a \ .
  \label{logsos}
\end{equation}

\subsection{{{}Pilot} numerical results and current limitations \label{sec:numerical}}
\noindent
We present {{}pilot} numerical results for the no-pair DC ground state of the helium atom.
\newline
The value of $J({\kabs})$ is calculated according to Eqs.~(\ref{lineq})--(\ref{Jcalc}) at several ${\kabs}$ values, using the analytic matrix elements of $\boldsymbol{j}_0$, Eq.~\eqref{eq:j0def} (Appendix~\ref{ch:velocity}). 
The $k$ integration is performed via Gauss--Legendre quadrature.
For simplicity, we do not consider separate low and medium ranges of the numerical $k$ integration, as it is often carried out in high-precision non-relativistic computations, \emph{e.g.,} Refs.~\cite{Ko19,FeMa22bethe}, but we simply subtract the counterterms at the end of the numerical integration of ${\kabs} J({\kabs})$ integral over the $[0,K]$ interval, Eq.~\eqref{eq:relBlWF0}. 

The fixed `outer basis', which is used for expanding
$|\Phi_{\text{DC}}\rangle$, consists of 400 functions of $^1S$ symmetry (optimized for the non-relativistic ground state of He), while the auxiliary basis set contains 300, 600 or 1200 functions of $^1P$ symmetry.
The relativistic basis space is constructed through the $LS$ coupling scheme, and in this work, we include only the dominant, singlet sector contributions. The triplet sector could be included, also within the $LS$ coupling scheme, to span the entire relativistically relevant space along the lines explained in Ref.~\cite{JeMa22}.
Due to the spherical symmetry of the helium atom, the computations can be made efficient, and we need to explicitly work with only $P_z$-type functions.

The functions of the auxiliary basis of size  $M'_\text{b}$ ($M'_\text{b}=300,600,1200$) are obtained from merging six basis sets of $M'_\text{b}/6$ functions, each of them independently optimized using the non-relativistic Hylleraas functional technique at ${\kabs}=10^n$, $n=0,...,5$. 
In these illustrative computations, the auxiliary basis is taken from non-relativistic Hylleraas function optimizations \cite{FeMa22bethe}, which could be extended along the relativistic scheme outlined in Sec.~\ref{ch:relHyfun}.
The relativistic functional was evaluated using this merged basis.

{{}The auxillary basis spans the positive-energy subspace of the two-particle Hilbert space. The matrices of the positive-energy projection operators (for both the outer and the auxiliary basis sets) were constructed with the `cutting' approach \cite{JeFeMa22} by separating the spectrum of $h_1+h_2$ based on energetic criteria. For low $Z$, this technique gives essentially the same results as the more rigorous alternative based on complex coordinate rotations \cite{JeFeMa22}.}

Regarding the $C_{\text{lin}}$ and $C_{\text{log}}$  counterterms a few comments are in order (with further discussion in relation to the Tables~\ref{table:Jval} and \ref{table:Blval}).
There are efficient techniques to compute the non-relativistic versions,
\begin{equation}
 C_{\text{lin,nonrel}}=\frac{1}{m^2}\langle \bp\cdot\bp\rangle \ ,
 \label{eq:Clinnr}
\end{equation}
and 
\begin{equation}
 C_{\text{log,nonrel}}=-\frac{1}{m^2}\langle \boldsymbol{p}(H_{\text{nonrel}}-E_{\text{nonrel}})\boldsymbol{p}\rangle \ .
 \label{eq:Clognr}
\end{equation}
These techniques rely on operator identities \cite{Ko04}. 
In the relativistic case, an adaptation of the non-relativistic approaches is problematic due {{}to} the presence of the $L_{++}$ positive-energy projector, which is only formally written in an operator form, in practice, its finite-basis matrix representation is constructed in numerical procedures.
This means that we are essentially tied to the sum-over-states representations
in Eqs. (\ref{linsos}), (\ref{logsos}),
which limits the high-precision evaluation of the relativistic counterterms.

Furthermore, the regulated integrand is still a slowly decaying function, whose large ${\kabs}$ part (necessary for high-precision results) is hard to incorporate numerically in the integral. In the non-relativistic case, a semi-analytical approach can be used in this region: the first few terms of the asymptotic expansion of $J({\kabs})$ are known analytically \cite{Sc61,Ko12}, and the coefficients of the higher-order terms can be determined from a numerical fit; the integral on this high ${\kabs}$ region can then be obtained analytically as a function of the fitting parameters.

In the current study, $J({\kabs})$ is evaluated numerically using the above procedure up to ${\kabs}=10^5$ at a large number of $k$ points.
The first numerical results for the example of the ground state of the helium atom are collected in Tables~\ref{table:Jval} and \ref{table:Blval} including $\kabs J(\kabs)$, energy correction, and Bethe logarithm values.

\begin{table}
  \caption{%
    Comparison of relativistic (DC) and non-relativistic (nr, $c\rightarrow\infty$) ${\kabs}J({\kabs})$ values, in atomic unit, for the example of the ground state of the helium atom. 
    See also Table~\ref{table:Blval}.
    \label{table:Jval} 
  }
      \begin{center}
       \begin{tabular}{@{}clcccccc@{}}
         \hline\hline\\[-0.35cm]
         $N_\text{aux}$$^\text{a}$ & ${\kabs}$: & $10^0$ & $10^1$ & $10^2$ & $10^3$ & $10^4$ & $10^5$ \\\hline
         \multicolumn{7}{l}{Non-relativistic, ${\kabs}J_{\text{nr}}({\kabs})$:} \\
          300 &  & $1.788229$  & $4.526040$  & $5.825302$  & $6.086447$  & $6.121174$  &  $6.125015$  \\
          600 & & $1.788274$  & $4.526102$  &  $5.825359$ & $6.086519$  & $6.121266$  & $6.125134$     \\
         1200 &  & $1.788278$ & $4.526110$ & $5.825365$ & $6.086523$ & $6.121270$ & $6.125140$  \\
         \hline\\[-0.4cm]
         \multicolumn{7}{l}{Dirac--Coulomb, ${\kabs}J_{\text{DC}}({\kabs})$:} \\
          300 &  &  $1.788085$ & $4.525428$  & $5.823853$  & $6.084353$  & $6.118804$  & $6.122582$   \\         
          600 &  & $1.788130$ & $4.525491$ & $5.823910$ & $6.084424$ & $6.118889$ & $6.122681$ \\
         1200 &  & $1.788135$ & $4.525499$ & $5.823916$ & $6.084428$ & $6.118894$ & $6.122685$ \\
         \hline\hline\\[-0.4cm]
       \end{tabular}
       \begin{flushleft}
         $^\text{a}$\ $N_\text{aux}$ is the number of auxiliary functions, and the reference state is expanded with 400 ECGs.
       \end{flushleft}
      \end{center}
     \end{table}

\begin{table}
    \caption{%
      Helium atom ground state: self-energy and retardation correction in the dipole approximation, Eq.~\eqref{eq:relBlWF0}, $\Delta E_K$ in $\mu\Eh$, obtained in Dirac--Coulomb (DC) computations and compared with non-relativistic (nr) results. $K=10^5$.
      The effect of the asymptotic correction, $\delta E_\text{as}$ in $\mu\Eh$, is available for the non-relativistic case. 
      The Bethe logarithm, Eq.~\eqref{relbl} is also evaluated.
      %2 x He+ Bethe log járulék:  4/(3*Pi)*Z^4*4.370422/137.035999084^3*10^6 * 2 = 23.0653 microEh
    %
    \label{table:Blval} }
      \begin{center}
       \begin{tabular}{@{}c ll ll cc@{}}
         \hline\hline\\[-0.40cm]
         \multicolumn{1}{c}{$N_\text{aux}$$^\text{a}$} & 
         \multicolumn{1}{c}{$C_{\text{lin}}$$^\text{b}$} & 
         \multicolumn{1}{c}{$C_{\text{log}}$$^\text{b}$} & 
         %\multicolumn{1}{c}{Asym.}&
         \multicolumn{1}{c}{$\Delta E_{K}\ [\mu\Eh]$} & 
         \multicolumn{1}{c}{$\delta E_\text{as}\ [\mu\Eh]$} & 
         $(\lnk0)_{K}$ & 
         $\lnk0$ \\ 
         \hline
         \multicolumn{7}{l}{Non-relativistic \& high-precision counterterms} \\
          300 & $6.125\ 59^\ast$       & $45.501\ 05^\ast$      & $-15.324$ & $-0.131$ & $4.084\ 1$ & $4.118\ 9$ \\
          600 & $6.125\ 59^\ast$       & $45.501\ 05^\ast$      &    $-16.222$       &   $-0.131$       &   $4.323\ 5$         &      $4.358\ 3$      \\
         1200 & $6.125\ 59^\ast$ & $45.501\ 05^\ast$ & $-16.263^\text{c}$ & $-0.131^\text{c}$ & $4.334\ 4$ & $4.369\ 4$ \\
         \hline \\[-0.40cm]         
         \multicolumn{7}{l}{Non-relativistic \& `own-basis' counterterms} \\
          300 & $6.125\ 45$       & $43.747\ 48$      & $-14.683$ & $-0.122$ & $4.070\ 2$ & $4.103\ 8$ \\
          600 &       $6.125\ 58$          &        $44.754\ 13$         &    $-15.542$       &  $-0.128$        &     $4.211\ 3$       &      $4.246\ 1$      \\
         1200 & $6.125\ 59$      & $44.953\ 61$      & $-15.727$ & $-0.129$ & $4.242\ 6$ &  $4.277\ 4$ \\
         \hline \\[-0.40cm]
         \multicolumn{7}{l}{Dirac--Coulomb \& `own-basis' counterterms} \\
          300 & $6.123\ 01$ & $42.839\ 30$ & $-14.079$ & & $3.985\ 4$ & \\
          600 & $6.123\ 11$ & $43.232\ 07$ & $-14.404$ & & $4.040\ 3$ & \\
         1200 & $6.123\ 12$ & $43.265\ 87$ & $-14.434$ & & $4.045\ 6$ & \\
         \hline\hline\\[-0.40cm]
       \end{tabular}
       \begin{flushleft}
         {\footnotesize%
         $^\text{a}$ Number of auxiliary ECGs optimized by minimization of the non-relativistic Hylleraas functional \cite{FeMa22bethe}. The reference state is described with $M_\text{b}'=400$ ECGs. \\
         $^\text{b}$ Linear and logarithmic counterterms. The precise, regularized value of the non-relativistic counterterms is labelled with $^\ast$ and is $C_\text{lin}=6.125\ 587\ 7$, $C_\text{log}=45.501\ 047\ 1$ \cite{KoKo99}. 
         Otherwise, we performed the direct evaluation of Eqs.~\eqref{eq:Clinnr}--\eqref{eq:Clognr} in the non-relativistic computation, and 
         Eqs.~\eqref{eq:Clinrel}--\eqref{eq:Clogrel} by Eqs.~\eqref{linsos}--\eqref{logsos} in the relativistic case. \\
         $^\text{c}$ 
         $\Delta E_K+\delta E_\text{as} ={{}-} 16.395\ \mu\Eh$, 
         which is in good agreement with $\Delta E ={{}-} 16.397\ 348\ \mu\Eh$ calculable, non-relativistic analogue of Eq.~\eqref{eq:relBlWF0}, from the most precise non-relativistic value, $\text{ln}k_0 =4.370\ 160\ 223\ 0703(3)$ \cite{Ko19}. The slight deviation is due to a sub-optimal treatment of the asymptotic part in this work, and could be improved for a smaller $K$ value~\cite{FeMa22bethe}. The large $K$ is used for comparison with relativistic results, where high-precision counterterms and semi-analytic asymptotics are not yet available.
         We used $\alpha^{-1}=137.035\ 999\ 084$ \cite{codata18}.
         \\
         }
       \end{flushleft}
      \end{center}
      \end{table}

First of all, it is interesting to observe in Table~\ref{table:Jval} that the difference between the relativistic and non-relativistic $\kabs J({\kabs})$ values is rather small for small ${\kabs}$, and it becomes more pronounced for large ${\kabs}$, as it can be expected that relativistic contributions are more important for larger momenta. By inspection of the regulated energy correction, Eq.~\eqref{eq:relBlWF0}, and the numerical value of the counterterms (Table~\ref{table:Blval}), it becomes apparent that the divergent terms dominate the $\kabs J(\kabs)$ values already at $k>1000$, and the later significant digits are relevant to the physically relevant energy correction. As it was pointed out earlier, the regulated $kJ{{}(k)}$ function decays slowly, and by choosing the $K=10^5$ cutoff (Table~\ref{table:Blval}), still a non-negligible portion of the energy correction is neglected. This `asymptotic' $\delta E_\text{as}$ contribution to the energy can be calculated for the non-relativistic case in an analytic form \cite{Sc61}. A similar asymptotic analytic expansion is not yet available for the relativistic (Dirac--Coulomb) case, so corresponding $\delta E_\text{as}$ values are not reported in the table, and their magnitude can only be estimated based on the non-relativistic counterpart.

Furthermore, Table~\ref{table:Blval} can be used to assess the effect of the precise value of the 
$C_\text{lin}$ and $C_\text{log}$ counterterms. $C_\text{lin}$ converges as the (auxiliary) basis size is increased, but the $C_\text{log}$ convergence is too slow (`own-basis' counterterms) with a non-negligible effect on the energy correction value. In the non-relativistic case, $C_\text{log}$ can be converged to high-precision \cite{FeMa22bethe} (and references therein) by using operator identities. Similar operator relations are not available in the relativistic case, so we have to rely on the direct evaluation of Eqs.~\eqref{eq:Clogrel} and \eqref{logsos} in the auxiliary basis (`own-basis' counterterms), and estimate the corresponding error based on the non-relativistic `high-precision' vs. `own-basis' energy difference. The effect of the inaccuracy of the counterterm on the energy correction can be calculated from the regulated energy value, Eq.~(\ref{eq:relBlWF0}),
\begin{equation}
  E_{\text{reg}}'
  =
  E_{\text{reg}}
  -
  \frac{2\alpha^3}{3\pi}
  \left[%
    K(C_{\text{lin}}-C_{\text{lin}}')
    +
   \ln\left(\frac{2K}{\Eh}\right)(C_{\text{log}}-C_{\text{log}}')
 \right] \; .
\end{equation}
On the other hand, it is necessary to add that in a numerical procedure, complete cancellation of the divergences for very large $K$ values is ensured only if the same auxiliary basis is used for the evaluation of the counterterms (`own-basis counterterms') as for the evaluation of the $kJ{{}(k)}$ function. But due to the slow convergence of the logarithmic counterterm, this direct evaluation leaves a big numerical error in finite basis, `direct computations'. This property is efficiently circumvented in the non-relativistic case by using the asymptotic expansion of the regulated contribution, which is not yet available in the relativistic case.

All in all, it is difficult to determine the precision of the relativistic result, but in comparison with the non-relativistic data, we estimate that the energy correction, Eq.~\eqref{eq:relBlWF0} is obtained to two significant figures, with numerical uncertainties between 0.5-1~$\mu\Eh$ for the example of the helium atom ground state (Table~\ref{table:Blval}).

\section{Towards a direct evaluation of retardation and one-loop self-energy corrections to the correlated relativistic energy}
\noindent
The question naturally emerges, is it worth improving upon the current limitations of the self-energy plus retardation correction \emph{within the dipole approximation}? To obtain a numerically competitive scheme with existing methodologies, we would like to converge the self-energy (plus retardation) correction to at least a parts-per-billion relative precision (and beyond). At this level of precision, we would like to improve upon also the dipole approximation. 
The dipole approximation is often corrected in the nrQED framework by including higher orders of $\eta_a(\bk)$ 
\begin{equation}
 \eta_a(\bk)=1+\iim\bk\cdot\br_a+... \ ,
\end{equation}
which, however, introduces even more severe divergences, which need to be taken care of by appropriate subtractions \cite{Ko13}.

Alternatively, we can turn to a numerical approach which does not rely on the dipole or any higher approximation of $\eta_a(\bk)$. 
The retardation energy correction, Eqs.~\eqref{eq:trcorr}--\eqref{eq:retcorr}, can be evaluated by direct computation (to be detailed in future work). 
The last section of this paper focuses on the numerical evaluation of the self-energy correction, 
\emph{e.g.,} Eqs.~\eqref{SEsummed}--\eqref{SEpart}, in particular, with a possible numerical treatment of the divergences.

\subsection{On the asymptotic behaviour of the $k$-integrands, and perspectives on renormalization \label{ch:renorm}}
\noindent
The naive form of self-energy (Sec.~\ref{sec:SE}) contains a divergence associated with
the electron mass being modified by its interaction with its own electromagnetic
field.
In order to obtain the finite, observable energy shift, the expressions must
be renormalized by an appropriate subtraction of the divergent part.
The degree of divergence depends on the approximations 
used in the evaluation of the self-energy.
The investigation of the large $k$ limit of the self-energy correction is thus crucial to renormalization.
{We will investigate the large $k$ asymptotics of the positive energy projected transverse and Coulomb self-energies (see Eqs. (\ref{SEpart}) and (\ref{SECoulomb})), and $-$ to some extent $-$ the effect of negative energy parts (see Eqs. (\ref{eq:SEnointTr}) and (\ref{eq:SEnointC})).}

Let us deal with the transverse part of the self-energy first.
To study the large $k$ limit of the positive-energy part of the self-energy correction (Sec.~\ref{sec:SE}), we use the approximation (for large $k$)
\begin{equation}
 \frac{L_{a,+}}{E_\nopair-H_\nopair-{\kabs}} \approx -\frac{\Lambda_{a,+}}{{\kabs}} \ ,
\end{equation}
where $\Lambda_{a,+}$ is the positive-energy part of the free-particle projector, so
\begin{equation}
 \Lambda_\pm(\boldsymbol{p})
 =\frac{1}{2}\left(I\pm\frac{h_0(\bp)}{E_{\bp}}\right)
 \ ,
 \end{equation}
 \begin{equation}
 h_0(\bp)=
  \boldsymbol{\alpha}\cdot\bp+m\beta
 \ \ \ , \ \ \ 
 E_{\bp}=\sqrt{|\bp|^2+m^2} \ .
\end{equation}
For the sake of simplicity, the particle index $a$ will be suppressed.

The $k$ integral in the self-energy is then of the form
(corresponding to the upper sign) 
\begin{align}
 \mp\int\frac{\mathrm{d}^3\bk}{{\kabs}^2}
 \alpha_i
 \eta(-\bk)\Lambda_\pm(\boldsymbol{p})\eta(\bk)
 \alpha_j
 \delta^{\perp}_{ij}(\bk)
 &=\mp
 \alpha_i\alpha_j
 \int\frac{\mathrm{d}^3\bk}{2{\kabs}^2}
 \delta^{\perp}_{ij}(\bk)-
 \int\frac{\mathrm{d}^3\bk}{2{\kabs}^2}
 \alpha_i
 \frac{h_0(\boldsymbol{p}+\bk)}{E_{\bp+\bk}}
 \alpha_j\delta^{\perp}_{ij}(\bk)
 \nonumber \\
 &\approx\mp4\pi {K} I- 
 \int\frac{\mathrm{d}^3\bk}{2{\kabs}^2}
 \alpha_i
 \frac{h_0(\bk)}{E_{\bk}}
 \alpha_j
 \delta^{\perp}_{ij}(\bk)
 \nonumber \\ 
 &=\mp
 4\pi{K} I-
 \frac{4\pi}{3}
 \int^{K}
 \mathrm{d}{\kabs}
 \frac{m}{\sqrt{{\kabs}^2+m^2}}
 \alpha_i\beta\alpha_i \nonumber \\
 &=
{% 
 \mp
 4\pi{K} I+
 4\pi m
 \ln\left[ \frac{K}{m} + \left(1+\frac{K^2}{m^2}\right)^{\frac{1}{2}}\right]\beta
} 
 \nonumber \\
 &\approx\mp
 4\pi{K} I+
 4\pi m \ln\left(2\frac{{K}}{m}\right)\beta \ .
 \label{asymptotic_k}
\end{align}
where we used
that $k\gg p\sim m\alpha$, the fact that
the angular average of $k_i$ and $k_ik_lk_m$ is zero, and introduced a momentum cut-off, $K\gg m$. 
The self-energy restricted to positive-energy states is thus seen to be linearly divergent as {$K\rightarrow \infty$}.

In passing, we note that in the case of transverse photon exchange, the divergence is dampened out by the $\eta$ factors that do not cancel, $\eta_1(\bk)\eta_2(-\bk)=\exp(\iim\bk(\br_1-\br_2))$.

The Coulomb part is investigated similarly, Eq. (\ref{SECoulomb}).
In that case, only replacements $L_\pm\rightarrow\Lambda_\pm$ are necessary:
\begin{align}
 \pm\int\frac{\mathrm{d}^3\bk}{{\kabs}^2}
 \eta(-\bk)\Lambda_\pm(\boldsymbol{p})\eta(\bk)
 &\approx
 \pm2\pi {K} I+
 \int\frac{\mathrm{d}^3\bk}{2{\kabs}^2}
 \frac{h_0(\bk)}{E_{\bk}} \nonumber \\
 &=
 \pm2\pi {K} I+2\pi
 \int^{{K}}\mathrm{d}{\kabs}\frac{m}{\sqrt{{\kabs}^2+m^2}} \beta \nonumber \\
 &\approx 
 \pm2\pi {K} I+2\pi m \ln\left(2\frac{{K}}{m}\right)\beta
 {{} \ .}
 \label{asymptotic_k_coulomb}
\end{align}

In an independent-particle calculation, where the negative-energy self-energy part takes a simple form, Eq.~(\ref{SE0_NES}), the analogous replacements
\begin{equation}
 \frac{L_{a,+}}{\varepsilon-{h}-{\kabs}} \ \rightarrow \  -\frac{\Lambda_{a,+}}{{\kabs}}
 \ \ \ , \ \ \ 
 \frac{L_{a,-}}{\varepsilon-{h}+{\kabs}} \ \rightarrow \  +\frac{\Lambda_{a,-}}{{\kabs}} \ ,
\end{equation}
would result in an asymptotic value of equal magnitude but opposite sign{{}; see}  Eqs.~(\ref{asymptotic_k}) and (\ref{asymptotic_k_coulomb}) with the upper/lower sign.
This means that in the independent-particle approaches to self-energy, where
both positive and negative energy parts are easily included, the linear divergence cancels, and only the usual logarithmic divergence remains, to be taken care of by mass renormalization.
We also note that the next term in the expansion of $1/{\kabs}$ would also contain a logarithmic divergence, which corresponds to the one-potential term ($\sim$ the vertex part) in the many-potential expansion (\emph{e.g.} Ch.~12.3.4 of Ref.~\cite{lindgrenRelativisticManyBodyTheory2011}).
This cancellation in the relativistic self-energy correction
is in agreement with the fact that the unrenormalized non-relativistic self-energy is also `only' logarithmically divergent~\cite{SeMo92};
it is only the often invoked dipole approximation, \emph{i.e.,} $\eta(\bk)\approx 1$,
that renders the expressions linearly divergent.

\subsection{Prospects regarding numerical renormalization schemes}
%\newline
Besides analytic but perturbative renormalization methods, numerical renormalization techniques are also known in the literature, which might
be more suitable for an explicitly correlated multi-electron computation.
The partial wave renormalization (PWR) proposed by Lindgren and co-workers \cite{LiPeSaYn93,PeLiSa93,PeSaSu98} and
by Quiney and Grant \cite{QuGr94,GrQu22} is such a numerical renormalization procedure (developed for (effective) one-particle systems), which produces the finite energy shift
as the difference of two divergent (but term-by-term finite) series. 
We present {some preliminary ideas about this approach for the correlated two-particle case.
For the sake of brevity, let us deal with only one of the terms
from the transverse part of the positive-energy {{} part of} self-energy in Eq.~(\ref{SEpart}), in the Coulomb gauge.
Whether the Coulomb gauge and the positive-energy projection are
good choices for the self-energy problem or can be straightforwardly abandoned in an explicitly correlated relativistic framework remains to be seen (in future work).

After inserting a resolution of unity  in Eq.~\eqref{SEpart} over a complete (bi-)orthogonal two-particle basis set, the first term of the self-energy reads
\begin{align}
 E_{\text{SE++,1}}
 =&
 \frac{z_1^2\alpha}{2\pi^2}\int\frac{\mathrm{d}^3\bk}{2k}
 \langle\Phi_\nopair|\tilde{\balpha}_1\eta_1(-\bk)\frac{L_{1,+}}{E_\nopair-H_\nopair-{\kabs}}\tilde{\balpha}_1\eta_1(\bk)|\Phi_\nopair\rangle \nonumber \\
 =&
 \frac{z_1^2\alpha}{2\pi^2}
 \sum_{p,q}\int\frac{\mathrm{d}^3\bk}{2k}
 \langle\Phi_\nopair|\tilde{\balpha}_1\eta_1(-\bk)|\chi_p\rangle
 R_{pq}({\kabs})
 \langle\tilde{\chi}_q|\tilde{\balpha}_1\eta_1(\bk)|\Phi_\nopair\rangle \nonumber \\
 =&
 \frac{z_1^2\alpha}{2\pi^2}
 \sum_{p,q}\int\frac{\mathrm{d}^3\bk}{2k}
 R_{pq}({\kabs})
 \int\mathrm{d}^3\br_1\mathrm{d}^3\br_2
 \int\mathrm{d}^3\br'_1\mathrm{d}^3\br'_2 \nonumber \\
 &
 \quad\quad
 \Phi^{\dagger}_\nopair(\br_1,\br_2)
 \tilde{\chi}^{\dagger}_q(\br'_1,\br'_2)
 \tilde{\balpha}_1\cdot\tilde{\balpha}_{1'}\eem^{-\iim\bk\cdot(\br_{1}-\br'_{1})}
 \chi_p(\br_1,\br_2)\Phi_\nopair(\br'_1,\br'_2) \ ,
\end{align}
where we introduced 
\begin{equation}
 R_{pq}({\kabs})=
 \langle\tilde{\chi}_p|\frac{L_{1,+}}{E_\nopair-H_\nopair-{\kabs}}|\chi_q\rangle \ ,
\end{equation}
and used $\eta_1(-\bk)\eta_{1'}(\bk)=\eem^{-\iim\bk\cdot(\br_{1}-\br'_{1})}$; $\langle\tilde{\chi}_p|$ is the element of the dual space satisfying $\langle\tilde{\chi}_p|\chi_q\rangle=\delta_{pq}$. 

The angular part of the $\bk$-integral can be calculated, 
\begin{equation}
  \int\mathrm{d}\Omega_k
    \left(%
      \delta_{ij} - \frac{k_i k_j}{{\kabs}^2}
    \right)  
    \eem^{-\iim\bk\cdot(\br_{1}-\br'_{1})} 
  =
  4\pi
  \left(\delta_{ij}-\frac{1}{{\kabs}^2}\partial_{1i}\partial_{1'j}\right)\frac{\sin({\kabs}|\br_{1}-\br'_{1}|)}{{\kabs}|\br_{1}-\br'_{1}|} \ ,
\end{equation}
and expanded using spherical Bessel functions as
\begin{equation}
 \frac{\sin({\kabs}|\br_{1}-\br'_{1}|)}{k|\br_{1}-\br'_{1}|}=\sum_{L=0}^{\infty}(2L+1)j_L({\kabs}r_1)j_L({\kabs}r'_1)
 \boldsymbol{C}_L(\Omega_1)\cdot\boldsymbol{C}_L(\Omega'_1) \ ,
\end{equation}
where
\begin{equation}
 \boldsymbol{C}_L(\Omega_1)\cdot\boldsymbol{C}_L(\Omega'_1)=
 \frac{4\pi}{2L+1}
 \sum_{M=-L}^{+L}Y_L^M(\Omega_1)Y_L^{M \, *}(\Omega'_1) \ .
\end{equation}
The self-energy correction then reads 
\begin{align}
 E_{\text{SE++,1}}
 =&
 \frac{\alpha z_1^2}{\pi}
 \sum_{p,q}\int_0^{\infty}\mathrm{d}{\kabs}\ {\kabs}
   R_{pq}({\kabs}) \nonumber \\
 &\quad\quad %
   \sum_{L=0}^{\infty}(2L+1)
   \Big[%
 \sum_{i=1}^3
 \langle\Phi_\nopair|\alpha_{1i}j_L({\kabs}r_1)\boldsymbol{C}_L(\Omega_1)|
 \chi_p\rangle
 \langle\tilde{\chi}_q|\alpha_{1i}j_L({\kabs}r_1)\boldsymbol{C}_L(\Omega_1)|\Phi_\nopair\rangle
 \nonumber \\
 &\quad\quad %
 -\frac{1}{{\kabs}^2}
\langle\Phi_\nopair|(\balpha_1\cdot\nabla_1j_L({\kabs}r_1)\boldsymbol{C}_L(\Omega_1))|
 \chi_p\rangle
 \langle\tilde{\chi}_q|(\balpha_1\cdot\nabla_1j_L({\kabs}r_1)\boldsymbol{C}_L(\Omega_1))|\Phi_\nopair\rangle 
 \Big] \ ,
 \label{pwSE_unreg}
\end{align}
and the index `$1$'  belongs to particle $1$ in the respective integrals.

The self-energy of the free particle can be represented by a similar partial wave expansion.
During the mass renormalization, the free-electron self-energy contributions
(evaluated with $\Phi_\nopair$) must be subtracted from Eq. (\ref{pwSE_unreg}),
and since both divergent series have finite terms, this subtraction can be carried out term-by-term in a computational procedure. 
A similar expansion and subtraction should be carried out for the Coulomb part
as well.
The renormalized self-energy $E_{\text{SE++}}^{\text{R}}=E_{\text{SE++}}-E_{\text{SE++}}^{\text{free}}$ is calculated numerically term-by-term in
the $L$-expansion. The difficulties associated with this procedure involve
finding a good enough basis and
the calculation of matrix elements of operators $\alpha_{ai}j_L(kr_a)\boldsymbol{C}_L(\Omega_a)$, $\balpha_a\cdot\nabla_aj_L(kr_a)\boldsymbol{C}_L(\Omega_a)$ and $R_{pq}(k)$.

{{}A possible further issue concerns the appearance of further correction terms introduced by the non-covariant regularization scheme. Despite initial conflicting statements over the existence of such terms \cite{PeSaSu98,Ye00},
the conclusion seems to be that no new terms arise for the first-order self-energy when using Feynman gauge and Pauli--Villars regularized momentum integrals \cite{GoPlZsLaSo02}.}
This problem should be re-investigated in the Coulomb gauge, possibly along the lines of Refs. \cite{Ad83,PeSaSu98}.

While the use of the Coulomb gauge introduces many seemingly unnecessary difficulties compared to the Feynman gauge, it 
may be a more natural choice for bound state problems, making it a challenge worth undertaking, as emphasized in pp.~256-257 of Ref. \cite{lindgrenRelativisticManyBodyTheory2011}. We also note that the old nrQED literature proposes a mixed gauge representation, \emph{i.e.,} Coulomb gauge for the interaction, covariant gauge for the radiative terms \cite{sucherPhD1958,DoKr74}. It remains a task to find the most convenient choice for high-precision bound-state computations.

\section{Summary and conclusion}
This work presented formal calculations and initial considerations for the high-precision computation of QED corrections to the correlated relativistic energy of bound two-electron atoms and molecules (and in general, two-spin-1/2 fermion systems with or without an external electrostatic field).
We focused on the lowest-order energy correction due to the single-photon, \emph{i.e.,} retardation and one-loop self-energy processes. 
Throughout the work, the formal connection was elaborated of the present framework to the non-relativistic QED and to the independent-particle QED approaches, which currently provide the state-of-the-art theoretical and computational framework for the low-$Z$ and the high-$Z$ systems, respectively.

{{}Pilot} numerical results were demonstrated for the largest QED contributions to the no-pair relativistic energy; the bulk of the self-energy and retardation corrections were obtained in the dipole approximation for the ground state of the helium atom. For future applications in high-precision computations, alternative theoretical and computational directions were discussed that abandon the dipole approximation and may be best suited for numerical computations by numerical renormalization of the divergent integrals.

\vspace{0.5cm}
\begin{acknowledgments}
\noindent Financial support of the European Research Council through a Starting Grant (No.~851421) is gratefully acknowledged.
\end{acknowledgments}

\appendix
\section{Some properties of the $\eta$ energy-momentum shift operator \label{App:eta}}
\noindent
In the two-particle framework elaborated in this article, the energy-momentum shift operator is defined by its action on a function that depends on the $\bp_1,\bp_2$ momenta and the $\epsi$ relative energy:
\begin{align}
 \eta(\bk,\omega)\Phi(\bp_1,\bp_2,\varepsilon)
 &=
 \eta_1(\bk)\eta_2(-\bk)\eta_\epsilon(\omega)\Phi(\bp_1,\bp_2,\varepsilon) \nonumber \\
 &=
 \Phi(\bp_1-\bk,\bp_2+\bk,\varepsilon-\omega) \nonumber \\
 &=
 \int\mathrm{d}^3\br_1\mathrm{d}^3\br_2
 \int_{-\infty}^{+\infty}\mathrm{d}t\ 
 \Phi(\br_1,\br_2,t)\eem^{\iim[(\varepsilon-\omega)t-(\bp_1-\bk)\cdot\br_1-(\bp_2+\bk)\cdot\br_2]} \nonumber \\
 &=
 \int\mathrm{d}^3\br_1\mathrm{d}^3\br_2
 \int_{-\infty}^{+\infty}\mathrm{d}t
   \left[%
     \eem^{-\iim[\omega t-\bk\cdot(\br_1-\br_2)]}
     \Phi(\br_1,\br_2,t)
   \right]
   \eem^{\iim[\varepsilon t-\bp_1\cdot\br_1-\bp_2\cdot\br_2]} \ ,
\end{align}
and thus, the {action of $\eta$ in} coordinate-space representation is {a simple multiplication:}
\begin{align}
  \eta(\bk,\omega)\Phi(\br_1,\br_2,t)
  =
  \eem^{\iim[\bk \br_{12}- \omega t]}
  \Phi(\br_1,\br_2,t) \; .
\end{align}
This expression shows that $\eta(\bk,\omega)$ is unitary:
\begin{equation}
 \eta^{-1}(\bk,\omega)=\eta^{\dagger}(\bk,\omega)=\eta(-\bk,-\omega) \ .
\end{equation}

\section{Diagram rules for the construction of the irreducible interaction \label{App:diagram}}
\noindent
In this appendix, we summarize the Bethe--Salpeter--Sucher rules \cite{SaBe51,Sa52,sucherPhD1958}
for the irreducible interaction operator:
\begin{itemize}
  \item 
    draw all topologically distinct diagrams distinguishing between all possible time orderings of interactions (both instantaneous and retarded ones);
  \item 
    equip all external and internal lines with appropriate energy and momentum labels by considering the overall conservation of energy-momentum, and with the convention that momentum transfers are directed from particle $2$ to $1$;
  \item 
    starting from the bottom of the diagram and moving to the top, assign the following quantities to the various parts of the diagram in a right-to-left order:
    \begin{itemize}
      \item 
        a factor of $\kappa_\inst(\bk)T_\inst(\bk)/(-2\pi \iim)$ for each instantaneous interaction lines ($\inst$ standing for Coulomb or Coulomb-Breit);
      \item 
        a factor of $\kappa_\text{T}(\bk,\omega)/(-2\pi \iim)$ for each interaction involving transverse photons;
      \item 
        a factor of $\tilde{\alpha}_{a,i}(\bk)$ for the vertex where a transverse photon line starts or ends ($a$ and $i$ being the indices of the particle in question and the Cartesian components, respectively);
      \item 
        a factor of $S_a$ for internal fermion lines of particle $a$ with the appropriate energy-momentum argument;
    \end{itemize}
  \item 
    write a factor of $\eta_1(\bk)\eta_2(-\bk)\eta_\epsilon(\omega)$ to the rightmost of the expression for each interaction line (either instantaneous or retarded); no $\eta$ is assigned to self-contracting lines associated with self-energy or vertex parts; 
  \item 
    assign a factor of $-1$ for each closed fermion loop of the diagram;
  \item 
    integrate over all $\bk$ and $\omega$,
    contract all Cartesian indices of transverse vertices connected by the same photon line.
\end{itemize}

\section{Example calculation with Cauchy's residue theorem and the Sokhotski--Plemelj formula \label{AppC}}
This appendix is about the use of Cauchy's residue theorem and the Sokhotski--Plemelj
theorem for the example,  
\begin{equation}
  {\cal{Q}}(\nu)
  =
  \int_{-\infty}^{+\infty}
    \frac{\mathrm{d}\varepsilon}{-2\pi \iim}
    S_1(\varepsilon)S_2(\nu-\varepsilon) \ .
\end{equation}
We have extensively used Cauchy's residue theorem (in this work and in our earlier work \cite{MaFeJeMa23}) to calculate the following types of integrals, 
\begin{align}
 {\cal{Q}}(\nu)
 &=
 \int_{-\infty}^{+\infty} \frac{\dd\epsi}{-2\pi\iim}
   \left[%
     \frac{L_{1+}}{\frac{E}{2}+\epsi-\ahone +\iim 0^+}
     +
     \frac{L_{1-}}{\frac{E}{2}+\epsi+\ahone -\iim 0^+}
   \right] \nonumber \\
 &\quad\quad\quad\quad\quad\quad
   \left[%
     \frac{L_{2+}}{\frac{E}{2}-\epsi+\nu-\ahtwo +\iim 0^+}
     +
     \frac{L_{2-}}{\frac{E}{2}-\epsi+\nu+\ahtwo -\iim 0^+}
   \right]
 \nonumber \\
 &=
 \frac{L_{++}}{E-h_1-h_2+\nu+\iim 0^+}-\frac{L_{--}}{E-h_1-h_2+\nu-\iim 0^+} \ . 
 \label{eq:S1S2int} 
\end{align}
At the same time, it is often convenient to first re-write the $S_1S_2$ product using the partial fraction decomposition, 
\begin{align}
  S_1 S_2 
  = (S_1^{-1} + S_2^{-1})^{-1} (S_1 + S_2) 
  = (S_1 + S_2) (S_1^{-1} + S_2^{-1})^{-1} \ .
\end{align}
Then, we have
\begin{align}
 {\cal{Q}}(\nu)
 &=
\int_{-\infty}^{+\infty}\frac{\mathrm{d}\varepsilon}{-2\pi \iim}
 \frac{S_1(\varepsilon)+S_2(\nu-\varepsilon)}{S^{-1}_1(\varepsilon)+S^{-1}_2(\nu-\varepsilon)} \nonumber \\
 &=
 \frac{1}{E-{h}_1-{h}_2+\nu+\iim 0^+\left(
 L_{1,+}+L_{2,+}-L_{1,-}-L_{2,-} 
 \right)}
 \int_{-\infty}^{+\infty}\frac{\mathrm{d}\varepsilon}{-2\pi \iim}
 \left[%
 S_1(\varepsilon)+S_2(\nu-\varepsilon)\right]
 \nonumber \\
 &=
\frac{1}{E-{h}_1-{h}_2+\nu+\iim 0^+\left(
 L_{++}-L_{--} 
 \right)} 
\int_{-\infty}^{+\infty}\frac{\mathrm{d}\varepsilon}{-2\pi \iim}
 \left[S_1(\varepsilon)+S_2(\nu-\varepsilon)\right]
 \ ,
\end{align}
where we used
\begin{align}
 S_1^{-1}(\varepsilon)
 &=
 \left[ \frac{E}{2}+\varepsilon-h_1+\iim 0^+\right] L_{1,+}
 +
 \left[ \frac{E}{2}+\varepsilon-h_1-\iim 0^+\right]L_{1,-} \nonumber\\
 &=
 E+\varepsilon-h_1+\iim 0^+\left(L_{1,+}-L_{1,-}\right) \ ,
\end{align}
analogously for $S_2$, 
\begin{align}
 S_2^{-1}(\nu-\varepsilon)
 &=
 \left[ \frac{E}{2}-\varepsilon+\nu-h_2+\iim 0^+\right] L_{2,+}
 +
 \left[ \frac{E}{2}-\varepsilon+\nu-h_2-\iim 0^+\right] L_{2,-} \nonumber \\
 &=
 E-\varepsilon+\nu-h_2+\iim 0^+\left(L_{2,+}-L_{2,-}\right) \ ,
\end{align}
and
\begin{equation}
 L_{1,+}+L_{2,+}-L_{1,-}-L_{2,-}=2(L_{1,+}L_{2,+}-L_{1,-}L_{2,-})=2(L_{++}-L_{--})   \ .
 \label{proj_id}
\end{equation}
{{}We also exploited $2 \, 0^+=0^+$.}
Then, we are left with the integrals of single propagators, which can be processed with the Sokhotski--Plemelj theorem: 
\begin{equation}
 \int_a^b\mathrm{d}x\ \frac{f(x)}{x\pm\iim0^{+}}
 =
 {\cal{P}}\int_a^b\mathrm{d}x\ \frac{f(x)}{x}
 \mp
 \iim\pi\int_a^b\mathrm{d}x\ \delta(x)f(x) \ .
\end{equation}
The principal value terms are all zero
(${\cal{P}}\int(1/x)=0$), and only Dirac delta terms remain.
We get
\begin{align}
 \int_{-\infty}^{+\infty}\frac{\mathrm{d}\varepsilon}{-2\pi \iim}S_1(\varepsilon) 
 =
 \frac{1}{-2\pi \iim}\left(-\iim \pi L_{1,+}+\iim \pi L_{1,-}\right) 
 =\frac{1}{2}\left(L_{1,+}-L_{1,-}\right) \ ,
 \label{eq:S1int}
\end{align}
and
\begin{align}
 \int_{-\infty}^{+\infty}\frac{\mathrm{d}\varepsilon}{-2\pi \iim}S_2(\nu-\varepsilon)
 =
 \frac{1}{-2\pi \iim }\left(-\iim \pi L_{2,+}+\iim\pi L_{2,-}\right) 
 =\frac{1}{2}\left(L_{2,+}-L_{2,-}\right) \ .
 \label{eq:S2int}
\end{align}
Using Eq. (\ref{proj_id}) for the sum of the two terms, 
we end up with
\begin{align}
 {\cal{Q}}(\nu)
 &=
 \frac{L_{++}-L_{--}}{E-{h}_1-{h}_2+\nu+\iim 0^+\left(
 L_{++}-L_{--} 
 \right)} 
 \label{eq:S1S2nu}
 \\
 &=
 \frac{L_{++}}{E-{h}_1-{h}_2+\nu+\iim 0^+}-
 \frac{L_{--}}{E-{h}_1-{h}_2+\nu-\iim 0^+} \ .
\end{align}
The equivalence of the {{}first and second lines} can be simply checked
by expanding the denominators and using $[h_a,L_{\pm\pm}]=0${{}. The result is} of course identical with the result calculated with Cauchy's residue theorem, Eq.~\eqref{eq:S1S2int}.

\section{Example calculation using the Poincaré--Bertrand theorem \label{app:PoBe}}
During the calculation of the Coulomb part of self-energy (and also in the analogous interaction term), it is easy to
run into what is seemingly a contradiction.
The calculation boils down to the evaluation of an integral like
\begin{equation}
 I=\int_{-\infty}^{+\infty}\frac{\mathrm{d}\varepsilon}{-2\pi \iim}
 \int_{-\infty}^{+\infty}\frac{\mathrm{d}\omega}{-2\pi \iim}
 \frac{1}{a+\varepsilon+\omega+\iim0^+}
 \frac{1}{b-\varepsilon+\iim0^+} \ ,
\end{equation}
which is tempting to tackle with a simple two-fold application
of the Sokhotski--Plemelj theorem:
\begin{equation}
 I=\int_{-\infty}^{+\infty}\frac{\mathrm{d}\varepsilon}{-2\pi \iim}
 \frac{-\pi\iim}{-2\pi\iim}
 \frac{1}{b-\varepsilon+\iim0^+}
 =\frac{1}{4} \ \ \ \ \ \  (\text{wrong!}) \ .
\end{equation}
This result is, however, not consistent with the one obtained from
applying the residue theorem for the $\varepsilon$ integral first, 
and then, the Sokhotski--Plemelj theorem for the $\omega$ integral:
\begin{equation}
 I=\int_{-\infty}^{+\infty}\frac{\mathrm{d}\omega}{-2\pi \iim}
 \frac{1}{a+b+\omega+\iim0^+}
 =\frac{1}{2} \ .
\end{equation}
The discrepancy follows from the careless treatment of distribution products in the application of the Sokhotski--Plemelj theorem: freely changing the integration order is not always possible when dealing with singular factors like $1/(x\pm\iim0^+)$.
If one insists on doing the complete double integral with distribution identities, then, the Poincaré--Bertrand formula should be used instead of the Sokhotski--Plemelj theorem:
\begin{equation}
  \left(\frac{1}{x_1-\iim s_1 0^+}\right)
  \left(\frac{1}{x_2-\iim s_2 0^+}\right)
  =
  \left[{\cal{P}}\left(\frac{1}{x_1}\right)+\iim \pi s_1\delta(x_1)\right]
  \left[{\cal{P}}\left(\frac{1}{x_2}\right)+\iim \pi s_2\delta(x_2)\right]
  +\pi^2\delta(x_1)\delta(x_2) \ ,
\end{equation}
where $s_1,s_2=\pm1$.
Setting $x_1=a+\varepsilon+\omega${{},} $x_2=\varepsilon-b$, $s_1=-1$, $s_2=+1$ leads to
\begin{equation}
  I=-\int_{-\infty}^{+\infty}\frac{\mathrm{d}\varepsilon}{-2\pi \iim}
 \int_{-\infty}^{+\infty}\frac{\mathrm{d}\omega}{-2\pi \iim}
 2\pi^2\delta(\varepsilon+\omega+a)\delta(\varepsilon-b)=\frac{1}{2} \ ,
\end{equation}
which is the correct result.

\section{The trivial divergence of the Coulomb self-energy in dipole approximation\label{App:Coulombdiv}}
\noindent
The Coulomb part of the self-energy, Eq.~(\ref{SECoulomb}),
becomes completely divergent in the dipole approximation, 
$\eta_a\approx1$. However, this is only a 
state-independent `constant' divergence that can be absorbed in the redefinition
of zero energy.
Such a constant divergence (the `self-repulsion' of a charged point particle) can already be anticipated from classical electrostatic considerations, 
\begin{align}
 V
 &=
  \frac{1}{8\pi}\int\mathrm{d}^3r\int\mathrm{d}^3r'
  \frac{\rho(\br)\rho(\br')}{|\br-\br'|}
  \nonumber \\
  &=
  \frac{1}{2}\sum_{i,j=1}^{N}\frac{z_iz_j\alpha}{|\br_i-\br_j|}
  \nonumber \\
  &=
  \sum_{\substack{i,j=1 \\ i<j}}^N\frac{z_iz_j\alpha}{|\br_i-\br_j|}
  +\left(\frac{1}{2}\sum_{i=1}^{N}z_i^2\alpha\right)\lim_{|\br|\rightarrow0^+}\frac{1}{|\br|} \ .
 \end{align}
 Indeed, we find the very same divergent term in the low-energy approximation of $\Delta E_{\text{SE}}^{\text{C}(+)}$:
\begin{align}
 \Delta E^{\text{C}(+)}_{\text{SE}}
 =&
 \lim_{|\br|\rightarrow0^+}\Bigg[
\frac{z_1^2\alpha}{2\pi^2}
  \int\mathrm{d}^3\bk\frac{1}{2{{\kabs}}^2}
  \langle\Phi_{\nopair}|\eem^{-\iim\bk\cdot\br_1}
   L_{1,+}\eem^{\iim\bk\cdot(\br_1+\br)}
  |\Phi_{\nopair}\rangle \nonumber \\ &  
  +\frac{z_2^2\alpha}{2\pi^2}\int\mathrm{d}^3\bk\frac{1}{2{{\kabs}}^2}
  \langle\Phi_{\nopair}|\eem^{-\iim\bk\cdot\br_2}
  L_{2,+}\eem^{\iim\bk\cdot(\br_2+\br)}
  |\Phi_{\nopair}\rangle  
  \Bigg]
\nonumber \\
=&
\lim_{|\br|\rightarrow0^+}
  \frac{\left(z_1^2+z_2^2\right)\alpha}{2}
   \int\frac{\mathrm{d}^3\bk}{(2\pi)^3}\frac{4\pi}{{\kabs}^2}\eem^{\iim \bk\cdot\br} \nonumber \\
   &   -
\frac{z_1^2\alpha}{2\pi^2}
  \int\mathrm{d}^3\bk\frac{1}{2{{\kabs}}^2}
  \langle\Phi_{\nopair}|\eem^{-\iim\bk\cdot\br_1}
   L_{1,-}\eem^{\iim\bk\cdot\br_1}
  |\Phi_{\nopair}\rangle \nonumber \\ & 
  -\frac{z_2^2\alpha}{2\pi^2}\int\mathrm{d}^3\bk\frac{1}{2{{\kabs}}^2}
  \langle\Phi_{\nopair}|\eem^{-\iim\bk\cdot\br_2}
  L_{2,-}\eem^{\iim\bk\cdot\br_2}
  |\Phi_{\nopair}\rangle  
 \nonumber \\
 \approx&
  \frac{\left(z_1^2+z_2^2\right)\alpha}{2}
  \lim_{|\br|\rightarrow0^+}
  \int\frac{\mathrm{d}^3\bk}{(2\pi)^3}\frac{4\pi}{{\kabs}^2}\eem^{\iim \bk\cdot\br} 
  \nonumber \\
  =&
  \frac{\left(z_1^2+z_2^2\right)\alpha}{2}
  \lim_{|\br|\rightarrow0^+}\frac{1}{|\br|} \ ,
\end{align}
since $L_{a,+}=1-L_{a,-}$ and $\langle\Phi_\nopair|\eta_a^\dagger L_{a,-}\eta_a|\Phi_{\nopair}\rangle\approx\langle\Phi_\nopair|L_{a,-}|\Phi_{\nopair}\rangle=0$.
The divergence does not depend on the state for which we evaluate the expectation value. This is the reason why we can omit this singular contribution from Eq. (\ref{Eorig0}),
and also why such a term does not appear in the nrQED correction formulae.

\section{A very brief overview of the variational DC/DCB method \label{app:DCDCB}}
\noindent
This appendix serves as a quick overview of the variational solution of
the no-pair DC/DCB equation, mostly to introduce the notations of Sec. \ref{ch:relBl}; see Refs. \cite{JeFeMa21,JeFeMa22,FeJeMa22,FeJeMa22b,JeFeMa22} for details.

The eigenvalue equation of the positive-energy ($++$) sector of the no-pair Hamiltonian, 
Eq.~(\ref{Hnopair}), reads
\begin{equation}
 H^{++}|\Phi_\nopair\rangle=E_\nopair|\Phi_\nopair\rangle \ ,
 \label{nopaireq}
\end{equation}
where
\begin{equation}
 H^{++}=L_{++}H_\nopair L_{++}
\end{equation}
was introduced for computational convenience (as the non-interacting $--$ and $+-$ states of Eq.~(\ref{Hnopair}) are not relevant here).
The eigenfunction is expanded in an overlapping spinor basis as
\begin{equation}
 \Phi_\nopair(\br_1,\br_2)=
 X\sum_{\mu=1}^{M_\text{b}}\sum_{q=1}^{16}c_{\mu q}\chi_{\mu q}(\br_1,\br_2) \ ,
\end{equation}
with the following notations.
The $16\times16$ matrix $X$ is the two-particle kinetic balance matrix,
\begin{equation}
 X=
 \begin{bmatrix}
  I & \emptyset & \emptyset & \emptyset \\
  \emptyset & \frac{(\boldsymbol{\sigma}_2\cdot\boldsymbol{p}_2)}{2m_2} & \emptyset & \emptyset \\
  \emptyset & \emptyset & \frac{(\boldsymbol{\sigma}_1\cdot\boldsymbol{p}_1)}{2m_1} & \emptyset \\
  \emptyset & \emptyset & \emptyset & \frac{(\boldsymbol{\sigma}_1\cdot\boldsymbol{p}_1)(\boldsymbol{\sigma}_2\cdot\boldsymbol{p}_2)}{4m_1m_2} 
 \end{bmatrix} 
 \ .
 \label{kinbalance}
\end{equation}
The basis functions are written as
\begin{equation}
 \chi_{\mu q}(\br_1,\br_2)=
 P_G{\cal{A}}\Theta_{\mu}(\br_1,\br_2)v^{(16)}_q \ ,
\end{equation}
where $P_G$ is the projector onto the desired irrep basis of the given
group, $\cal{A}$ is the operator implementing antisymmetry with respect
to simultaneous permutations in coordinate and spinor space 
(obviously omitted for non-identical particles), and $v^{(16)}_q=e^{(2)}_{\text{L/S},1}\otimes e^{(2)}_{\text{L/S},2}\otimes e^{(2)}_{\text{spin},1}\otimes e^{(2)}_{\text{spin},2}$ is the product of unit vectors
in the large/small component spaces and spin spaces of the particles.
The $\Theta_{\mu}(\br_1,\br_2)$ spatial part of the basis function is taken to be an explicitly correlated Gaussian function (ECG) \cite{SuVa98,MaRe12}:
 \begin{equation}
  \Theta_\mu(\bur)=\exp[-(\bur-\bus_\mu)^T
  \buA_{\mu}(\bur-\bus_\mu)] \ ,
 \end{equation}
 with notations $\bur=(\br_1,\br_2)$, $\bus_\mu=(\bs_1,\bs_2)$
 (understood as block column vectors), and
 \begin{equation}
  \buA_{\mu}=
  \begin{bmatrix}
   a_1 & b \\
   b & a_2
  \end{bmatrix}
  \otimes I^{[3]}
 \end{equation}
 being a positive-definite matrix (with $3\times3$ unit matrix $I^{[3]}$). We also note that the entire relativistic basis space has been efficiently covered and parameterized for low-$Z$ systems using the $LS$ coupling scheme \cite{FeMa23}.
 
 Substituting this basis representation into Eq. (\ref{nopaireq}) gives
 \begin{equation}
  \tilde{\boldsymbol{H}}\boldsymbol{c}=E_{\nopair}\boldsymbol{S}\boldsymbol{c} \ ,
  \label{eq:genereiv}
 \end{equation}
 where 
 \begin{equation}
  \tilde{H}_{\mu q,\nu r}=
  \langle{\chi_{\mu q}}|XH^{++}X|{\chi_{\nu r}} \rangle 
  \end{equation}
  is the matrix of the Hamiltonian in the metric of the kinetic balance,
  and
  \begin{equation}
  S_{\mu q,\nu r}=
  \langle{\chi_{\mu q}}|XX|{\chi_{\nu r}} \rangle
 \end{equation}
 is the overlap matrix.
 Combined with a Pauli-like choice relating large and small 
 components of the wave function \cite{JeFeMa22}, the kinetic balance
 metric guarantees the correct non-relativistic behaviour of the Hamiltonian (and also that of the velocity operator, Appendix \ref{ch:velocity}).
 All required integrals with ECGs are calculated analytically as functions
 of $\buA_\mu$ and $\bs_\mu$.
 The linear basis function parameters $c_{\mu q}$ are determined by solving the generalized eigenvalue problem, Eq.~\eqref{eq:genereiv}, while nonlinear parameters $\buA_\mu$, $\bs_\mu$ are obtained via a stochastic variational method with repeated refinement cycles~\cite{SuVa98,MaRe12}.
 The matrix of the positive-energy projection operator needed to build $H^{++}$ and
 $\tilde{\boldsymbol{H}}$ are typically constructed from the positive-energy solutions of the non-interacting problem (corresponding to $V_\inst=0${{}, while retaining the external field of the nuclei}) \cite{JeFeMa22}.

 \section{Matrix elements of the velocity operator \label{ch:velocity}}
 For the computation of the relativistic Bethe logarithm and corresponding energy correction, Sec. \ref{ch:relBl}, the matrix elements of the `kinetic balance transformed velocity operator' $X\boldsymbol{j}X$ are required, Eqs.~(\ref{velocityop}) and (\ref{kinbalance}). 
 Following the conventions of previous work \cite{JeFeMa22}, 
 \begin{align}
  \boldsymbol{j}={\frac{1}{\alpha}\Big[}z_1\bos{\alpha}\boxtimes I + I\boxtimes z_2\bos{\alpha}{\Big]}  
                                     ={\frac{1}{\alpha}}
  \begin{bmatrix}
   \emptyset & z_2\bsigma_{2} & z_1\bsigma_{1} & \emptyset \\
   z_2\bsigma_{2} & \emptyset & \emptyset & z_1\bsigma_{1} \\
   z_1\bsigma_{1} & \emptyset & \emptyset & z_2\bsigma_{2} \\
   \emptyset & z_1\bsigma_{1} & z_2\bsigma_{2} & \emptyset
  \end{bmatrix}
  \ ,
  \end{align}
  where
  \begin{equation}
   \bsigma_{1}=\bsigma\otimes I \ \ , \ \ 
   \bsigma_{2}=I\otimes\bsigma \ .
  \end{equation}
  Transforming $\boldsymbol{j}$ with the kinetic balance yields
  \begin{align}
  X\boldsymbol{j} X={\frac{1}{\alpha}}
  \begin{bmatrix}
   \emptyset &
   \frac{z_2\bsigma_{2}({\bsigma}_2\cdot\boldsymbol{p}_2)}{2m_2} &
   \frac{z_1\bsigma_{1}({\bsigma}_1\cdot\boldsymbol{p}_1)}{2m_1} &
   \emptyset \\
   \frac{({\bsigma}_2\cdot\boldsymbol{p}_2)z_2\bsigma_{2}}{2m_2} &
   \emptyset &
   \emptyset &
   \frac{p_2^2z_1\bsigma_{1}({\bsigma}_1\cdot\boldsymbol{p}_1)}{8m_1m_2^2} \\
   \frac{({\bsigma}_1\cdot\boldsymbol{p}_1)z_1\bsigma_{1}}{2m_1} &
   \emptyset & 
   \emptyset & 
   \frac{p_1^2z_2\bsigma_{2}({\bsigma}_2\cdot\boldsymbol{p}_2)}{8m_2m_1^2} \\
   \emptyset &
   \frac{p_2^2({\bsigma}_1\cdot\boldsymbol{p}_1)z_1\bsigma_{1}}{8m_1m_2^2} & 
   \frac{p_1^2({\bsigma}_2\cdot\boldsymbol{p}_2)z_2\bsigma_{2}}{8m_2m_1^2} &
   \emptyset
  \end{bmatrix}
  \ . \label{jtransform}
  \end{align} 
  The calculation of matrix elements between spinor basis functions is then
  seen to boil down to the calculation of the following matrix elements
  between ECGs:
  \begin{align}
   \langle\Theta_{\mu}|\partial_{ai}\Theta_{\nu}\rangle=
   S_{\mu\nu}\Large(\underline{\boldsymbol{e}}_{ai}^{T}
   \underline{\boldsymbol{q}}_{\mu\nu}\Large)
   \ ,
  \end{align}  
  \begin{align}
   \langle\Theta_{\mu}|\partial_{ai}\partial_{bj}\partial_{ck}\Theta_{\nu}\rangle=&
   S_{\mu\nu}
   \Large(\underline{\boldsymbol{e}}_{ai}^{T}\underline{\boldsymbol{q}}_{\mu\nu}\Large)
   \Large(\underline{\boldsymbol{e}}_{bj}^{T}\underline{\boldsymbol{q}}_{\mu\nu}\Large)
   \Large(\underline{\boldsymbol{e}}_{ck}^{T}\underline{\boldsymbol{q}}_{\mu\nu}\Large) %\\
    -
   S_{\mu\nu}\left[
   \Large(\underline{\boldsymbol{e}}_{ai}^{T}
   \underline{\hat{\boldsymbol{A}}}_{\mu\nu} 
   \underline{\boldsymbol{e}}_{bj}\Large)
   \Large(\underline{\boldsymbol{e}}_{ck}^{T}\underline{\boldsymbol{q}}_{\mu\nu}\Large)
   \right] \nonumber \\
   &-
   S_{\mu\nu}\left[
   \Large(\underline{\boldsymbol{e}}_{ck}^{T}
   \underline{\hat{\boldsymbol{A}}}_{\mu\nu} 
   \underline{\boldsymbol{e}}_{ai}\Large)
   \Large(\underline{\boldsymbol{e}}_{bj}^{T}\underline{\boldsymbol{q}}_{\mu\nu}\Large)
   \right]-
   S_{\mu\nu}\left[
   \Large(\underline{\boldsymbol{e}}_{bj}^{T}
   \underline{\hat{\boldsymbol{A}}}_{\mu\nu} 
   \underline{\boldsymbol{e}}_{ck}\Large)
   \Large(\underline{\boldsymbol{e}}_{ai}^{T}\underline{\boldsymbol{q}}_{\mu\nu}\Large)
   \right] \ ,
  \end{align} 
  where $a,b,c=1,2$ and $i,j,k=1,2,3$ denote particle and Cartesian indices, respectively, and
  \begin{equation}
   \underline{\boldsymbol{q}}_{\mu\nu}=
   -\underline{\hat{\boldsymbol{A}}}_{\mu\nu}(\underline{\boldsymbol{s}}_{\mu}-\underline{\boldsymbol{s}}_{\nu})
   \ ,
  \end{equation}
  \begin{equation}
   \underline{\hat{\boldsymbol{A}}}_{\mu\nu}
   {{}=2\underline{\boldsymbol{A}}_{\nu}
   (\underline{\boldsymbol{A}}_{\mu}+\underline{\boldsymbol{A}}_{\nu})^{-1}\underline{\boldsymbol{A}}_{\mu}}
  \end{equation}
  The only non-zero component of unit vector $\boldsymbol{e}_{ai}$ is component $3(a-1)+i$. The overlap of the ECGs reads
  \begin{equation}
   S_{\mu\nu}=\langle\Theta_{\mu}|\Theta_{\nu}\rangle=\sqrt{\frac{\pi^{3N}}{\det({{}\underline{\boldsymbol{A}}_{\mu}+\underline{\boldsymbol{A}}_{\nu}})}}
   \exp\left(-\frac{1}{2}
   (\underline{\boldsymbol{s}}_{\mu}-\underline{\boldsymbol{s}}_{\nu})^{T}
   \underline{\hat{\boldsymbol{A}}}_{\mu\nu}(\underline{\boldsymbol{s}}_{\mu}-\underline{\boldsymbol{s}}_{\nu})\right) \ .
  \end{equation}  

{
\section{On atomic units \label{app:AU}}
\noindent
Let us convert the last, retarded term of Eq. (\ref{eq:dEretinst}),
\begin{equation}
  \Delta{\cal{E}}
  =
  \frac{\alpha^3}{2\pi^2}
  \int\mathrm{d}^3\bk\frac{1}{2{\kabs}^2}\left\langle
    \boldsymbol{j}^{\dagger}(\bk)
    \left[%
    \frac{E_{\text{DC}}-H_{\text{DC}}}{E_{\text{DC}}-H_{\text{DC}}-{{\kabs}}}
    \right]
    \boldsymbol{j}(\bk)
   \right\rangle 
\end{equation}
into atomic units $(m=e=\hbar=4\pi\varepsilon_0=1)$ by expressing everything in terms of the $a_0=1/(m\alpha)$ Bohr radius and the $E_\text{h}=m\alpha^2$ Hartree energy. This corresponds to the scaling
\begin{equation}
 \br=a_0\br' \ \ , \ \ 
 \bk=\frac{\alpha}{a_0} \bk' \ \ , \ \
 \bp=\frac{1}{a_0}\bp' \ \ , \ \
 E_\text{DC}=E_\text{h} E'_\text{DC} \ , \ \
 H_\text{DC}=E_\text{h} H'_\text{DC} \ ,
 \label{auscaling}
\end{equation}
where primed quantities are dimensionless (the norm of the wave function is also rescaled).
Note that we deliberately scale an extra $\alpha$ out of $\bk$ (thereby, rescaling the integration variable), and thus, we formally have a factor of $E_\text{h}=\alpha/a_0$ in front of $\bos{k}'$.
Then, we find
\begin{equation}
  \Delta{\cal{E}}=E_h
  \frac{\alpha^3}{2\pi^2}
  \int\mathrm{d}^3\bk'\frac{1}{2{\kabs'}^2}\left\langle
    \boldsymbol{j}^{\dagger}(\alpha\bk')
    \left[%
    \frac{E'_{\text{DC}}-H'_{\text{DC}}}{E'_{\text{DC}}-H'_{\text{DC}}-{{\kabs}'}}
    \right]
    \boldsymbol{j}(\alpha\bk')
   \right\rangle \ . \label{tr_au}
\end{equation}
The suppression of the exponent of $\eta_a$ by an extra factor of $\alpha$ once again motivates the dipole approximation $\eta_a\approx1$, $j_i(\alpha\bk')\approx\delta^{\perp}_{il}(\bk')(\boldsymbol{j}_0)_{l}$ (see Eq. (\ref{eq:j0def})), leading to (after the same steps as in Sec. \ref{ch:relBl}, and again discarding the logarithmically diverging part)
\begin{equation}
\Delta E
=
-E_\text{h} 
\frac{2\alpha^3}{3\pi}\left\langle\boldsymbol{j}_0(H'_{\text{DC}}-E'_{\text{DC}})\boldsymbol{j}_0
  \right\rangle\ln(k_0) \ . 
\end{equation}
Of course, this transition to atomic units could have been performed directly after Eq. (\ref{relbl}), but in this way, we can see the non-trivial $\alpha$-dependence of Eq. (\ref{tr_au}) when not resorting to the dipole approximation. By using Eqs. (\ref{auscaling}) and (\ref{jtransform}), we can see that the contribution of $\boldsymbol{j}_0$ to the expectation value is ${\cal{O}}(\alpha^0)$, so the leading term in $\Delta E$ is indeed ${\cal{O}}(\alpha^3E_h)={\cal{O}}(m\alpha^5)$, as we know from nrQED.
}

\vspace{0.5cm}
%\bibliography{references}

\end{document}